\documentclass[fleqn,usenatbib]{mnras}

\usepackage{newtxtext,newtxmath}

\usepackage[T1]{fontenc}

\DeclareRobustCommand{\VAN}[3]{#2}
\let\VANthebibliography\thebibliography
\def\thebibliography{\DeclareRobustCommand{\VAN}[3]{##3}\VANthebibliography}

\usepackage{graphicx}	
\usepackage{amsmath,nccmath}
\usepackage{mhchem}
\usepackage{threeparttable}
\usepackage{subcaption}
\usepackage{multirow}
\usepackage{hyperref}

\newcommand{\solar}{$_\odot$}
\newcommand{\solarperyr}{$_\odot \, \textrm{yr}^{-1}$}
\newcommand{\bse}{\texttt{bse}}
\newcommand{\binaryc}{\texttt{binary\_c}}
\newcommand{\Mchand}{$M_{\rm Ch}$}
\newcommand{\Mwd}{$M_{\rm WD}$}
\newcommand{\Mhsh}{$M_{\rm H \ shell}$}
\newcommand{\Mhesh}{$M_{\rm He \ shell}$}
\newcommand{\Mig}{$M_{\rm ig}$}
\newcommand{\Mdot}{$\dot{M}$}

\newcommand*\diff{\mathop{}\!\mathrm{d}}

\title[Population synthesis of H and He novae]{Population synthesis of accreting white dwarfs: \\ Rates and evolutionary pathways of H and He novae}

\author[A. Kemp et al.]{
Alex. J. Kemp$^{1,7}$\thanks{E-mail: alexander.kemp@monash.edu},
Amanda I. Karakas$^{1,7}$,
Andrew R. Casey$^{1,7}$,
Robert G. Izzard$^{2}$,\newauthor
Ashley J. Ruiter$^{3,8}$,
Poojan Agrawal$^{4,8}$,
Floor S. Broekgaarden$^{5}$,
Karel D. Temmink$^{6}$
\\
$^{1}$School of Physics \& Astronomy, Monash University, Clayton 3800, Victoria, Australia\\
$^{2}$Astrophysics Research Group, University of Surrey, Guildford, Surrey GU2 7XH, UK\\
$^{3}$School of Science, University of New South Wales, Australian Defence Force Academy, Canberra, ACT 2600, Australia\\
$^{4}$Centre for Astrophysics and Supercomputing, Swinburne University of Technology, Hawthorn, VIC 3122, Australia\\
$^{5}$Harvard-Smithsonian Centre for Astrophysics, 60 Garden Street, Cambridge, MA, 02138, USA\\
$^{6}$Department of Astrophysics/IMAPP, Radboud University Nijmegen, P.O. Box 9010, 6500 GL Nijmegen, The Netherlands\\
$^{7}$Centre of Excellence for Astrophysics in Three Dimensions (ASTRO-3D), Melbourne, Victoria, Australia\\
$^{8}$OzGrav: The ARC Centre of Excellence for Gravitational Wave Discovery, Melbourne, Victoria, Australia\\
}

\date{Accepted XXX. Received YYY; in original form ZZZ}

\pubyear{2020}

\begin{document}
\label{firstpage}
\pagerange{\pageref{firstpage}--\pageref{lastpage}}
\maketitle

\begin{abstract}
Novae are some of the most commonly detected optical transients and have the potential to provide valuable information about binary evolution.
Binary population synthesis codes have emerged as the most effective tool for modelling populations of binary systems, but such codes have traditionally employed greatly simplified nova physics, precluding detailed study.
In this work, we implement a model treating H and He novae as individual events into the binary population synthesis code \binaryc. This treatment of novae represents a significant improvement on the `averaging' treatment currently employed in modern population synthesis codes.
We discuss the evolutionary pathways leading to these phenomena and present nova event rates and distributions of several important physical parameters.
Most novae are produced on massive white dwarfs, with approximately 70 and 55 per cent of nova events occurring on O/Ne white dwarfs for H and He novae respectively.
Only 15 per cent of H-nova systems undergo a common-envelope phase, but these systems are responsible for the majority of H nova events. All He-accreting He-nova systems are considered post-common-envelope systems, and almost all will merge with their donor star in a gravitational-wave driven inspiral.
We estimate the current annual rate of novae in M31 (Andromeda) to be approximately $41 \pm 4$ for H novae, underpredicting the current observational estimate of $65^{+15}_{-16}$, and $0.14\pm0.015$ for He novae. When varying common-envelope parameters, the H nova rate varies between 20 and 80 events per year.

\end{abstract}




\section{Introduction}

White dwarfs (WDs) are the inert, degenerate cores of stars left behind once a star's envelope is blown away in the asymptotic giant branch (AGB) phase or stripped by a binary companion. Carbon-oxygen (C/O) WDs and oxygen-neon (O/Ne) WDs are typically formed from AGB progenitors with initial masses from 0.8--10 M\solar\ \cite[see][]{herwig2005evolution, karakas2014} while the lower mass He WDs are formed through binary stripping prior to the cessation of core helium burning \citep{kamath2014Optically, kamath2016binary, kamath2016newly, kamath2019}.
Once the natal WD emerges from the ionised gas of its surrounding planetary nebula \citep{paczynski1971}, it does little other than cool and undergo chemical stratification \citep{koester1990,horowitz2010,tremblay2019core}.

In the context of binary stellar evolution however, a WD can interact with its companion star in a myriad of complex ways \citep{demarco2017} that can result in energetic transient phenomena such as novae and type Ia supernovae. Consider the evolution of, for example, a binary containing a 6 M\solar\ primary (initially more massive star) and a 2 M\solar\ secondary (initially less massive star). The primary evolves through the AGB phase, loses its outer layers (possibly with the assistance of its companion) and forms a WD, all while the secondary is on the main sequence. The less evolved companion star eventually expands as it approaches the end of its own life. It may then fill its Roche lobe and transfer material onto its degenerate companion; in this configuration the secondary is labelled the `donor', and the primary the `accretor', and the donor star is said to be undergoing `Roche lobe overflow' \cite[RLOF, see][]{paczynski1971}. During stable mass transfer, material from the donor star is able to be transferred to the surface of the WD where it may undergo nuclear processing. This period of stable mass transfer is fundamental to explaining observations of novae, type Ia supernovae, and populations of compact objects.

Detailed models that study WD transients, such as novae \citep{henze2018,izumi2019} or type Ia supernovae \citep{seitenzahl2016,townsley2019,delosreyes2020manganese}, are ideal tools for reproducing the details of individual observations, such as elemental abundances derived from spectroscopy and light curve decay profiles, providing a crucial link between observations and the fundamental physics of these events. However, the computationally expensive nature of detailed models makes it difficult to explore the effect of many aspects of uncertain physics relevant to binary evolution such as mass transfer, tides, and stripping episodes.

Alternatively, binary population synthesis (BPS) models can be used to calculate the rates and distributions of these events. Binary population synthesis involves computing the evolution of a large population of binary systems using approximate methods such as fits to detailed stellar evolution models and analytic approximations to calculate stellar properties and drive evolutionary decisions \citep[e.g.,][]{tout1997,hurley2000,hurley2002,izzard2004,stevenson2017formation}. Modelling stellar evolution in this way has the advantage of being extremely fast, a quality that is especially important when considering the large parameter space surrounding binary stars.

In the past, population synthesis studies have treated novae as average events \citep[e.g.,][]{han1995formation,hurley2002,ruiter2014,claeys2014}, with no way to check whether a nova eruption actually occurred, when it occurred, or how many occurred in a given binary system. This made accurate and direct estimation of the rates and distributions of nova properties impossible. \cite{chen2016} inferred nova rates in M31 based on the system properties at the onset of mass transfer to the WD accretor using the detailed stellar evolution code MESA \citep{Paxton2013} and the BPS code \bse\ \citep{hurley2002}. This method necessarily decouples the novae from subsequent evolution of the binary. Furthermore, many competitor BPS codes are available with different specialisations and capabilities e.g., \binaryc\ \citep{izzard2004,claeys2014}, \texttt{BPASS} \citep{stanway2016,stanway2018}, \texttt{COMPAS} \citep{stevenson2017formation,barrett2018}, \texttt{SeBa} \citep{zwart1996,toonen2012supernova}, \texttt{STARTRACK} \citep{belczynski2008compact,ruiter2014}.

In this work, we present results from our population synthesis study of accreting C/O and O/Ne WDs using \binaryc. We report rates, distributions, and pathways to H and He novae having updated the evolutionary behaviour of accreting C/O and O/Ne WDs using results from recent detailed models. This work aims to provide a description of the distributions of physical parameters surrounding novae, with a focus on the evolutionary pathways that these systems may travel. In particular, this work distinguishes itself from earlier population synthesis efforts by considering the individual nova eruptions consistently within the BPS model.

In Section \ref{sec:wdphysics}, we describe physics relevant to accreting WDs, and compare with the model implemented in \binaryc . Section \ref{sec:methodology} summarises the numeric and physical details of the grids computed. The results are presented in Section \ref{sec:results}. We discuss the results in the context of observations and uncertain common-envelope  physics in Section \ref{sec:discussion}. We summarise our findings in Section \ref{sec:conclusion}.

\section{Physics of accreting WDs}
\label{sec:wdphysics}

Here we describe aspects of WD accretion physics relevant to novae, with a focus on the input physics used in our calculations. In our work, we consider C/O and O/Ne WDs accreting either H- or He-rich material. Consequently, our model WDs have (potentially) three masses associated with them: their total mass \Mwd , the mass of a H shell, \Mhsh, and the mass of a He shell, \Mhesh. Upon formation of the WD, the mass of H and He shell are both assumed to be zero. This is certainly not generally true in nature, but is assumed to prevent material present at the birth of these WDs giving rise to spurious nova systems. This assumption is expected to have little effect on the true nova rate due to the repeating nature of novae.

\subsection{Accretion Limits}

The behaviour of accreting WDs varies with the rate of accretion, and can be classified into different \textit{accretion regimes}. The bounds of these regimes are in principle dependent on a number of physical properties, including the mass of the WD, the composition of the accreted material, and the temperature of the WD.

\begin{figure*}
\centering
\begin{subfigure}{1\columnwidth}
\centering
\includegraphics[width=\textwidth=1]{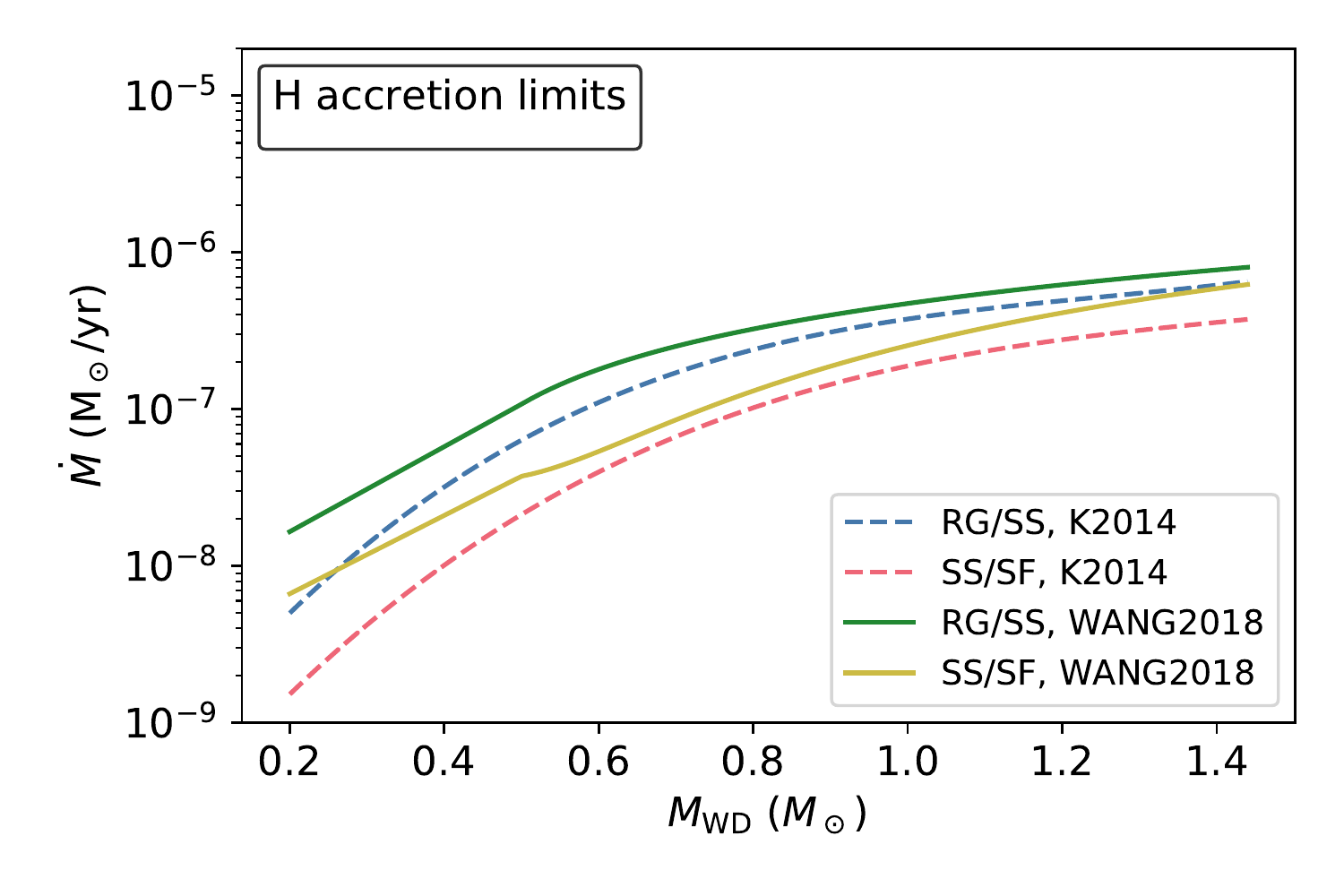}
\end{subfigure}%
\begin{subfigure}{1\columnwidth}
\centering
\includegraphics[width=\textwidth=1]{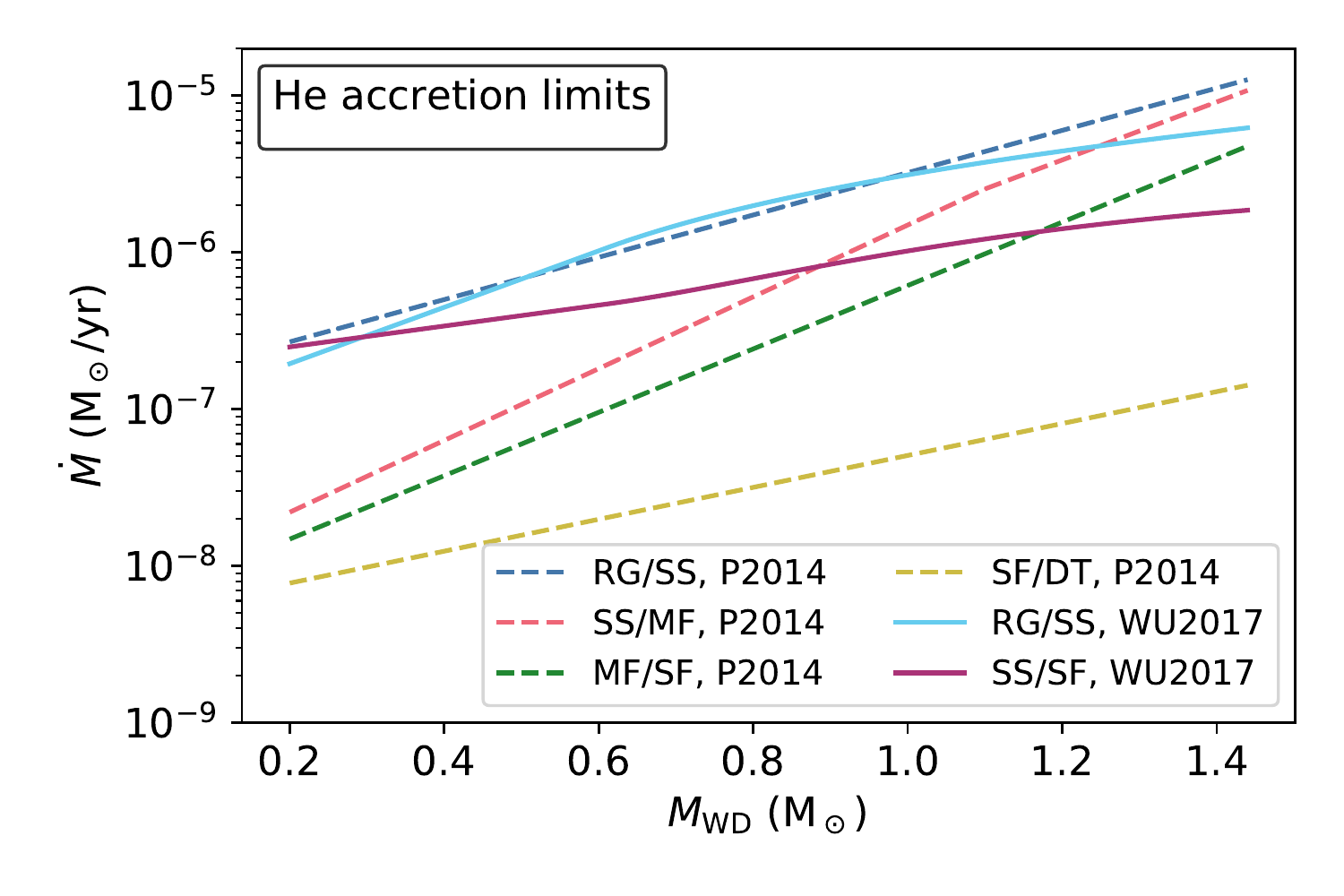}
\end{subfigure}
\caption{Limits for H and He accretion onto different white dwarf masses. Solid lines represent the models used in this work, WANG2018 \protect \citep{wang2018} and WU2017 \protect \citep{wu2017}, while the dotted lines represent K2014 \citep{kato2014} and P2014 \citep{piersanti2014} accretion limits for comparison. Key: \textit{RG: giant regime. SS: steady burning regime. MF: mild flash regime. SF: strong flash (nova) regime. DT: detonation regime.}}
\label{fig:wdaccretionlimits}

\end{figure*}

Detailed modeling works of novae typically consider only the effect of variations in the WD mass in pre-computed grids. The works of \cite{yaron2005}, which varied the temperature of the isothermal WD core, and \cite{piersanti2000,starrfield2000,chen2019}, which varied the metalicity of accreted H-rich material, are rare examples of studies that also quantified the impacts of other physical properties. In this work, we use accretion limits from \cite{wang2018} and \cite{wu2017} in our standard physics case for H and He accretion, respectively, onto C/O and O/Ne WDs, shown in Figure \ref{fig:wdaccretionlimits}.

In our model the accretion regime is determined by the WD mass, the rate of mass transfer, and whether the accreted material is H- or He-rich. The accretion regime dictates how the mass transfer onto the accretor is treated. If the WD is accreting at a rate above the maximum steady burning limit, then in the case of H accretion the core accretes and processes material at the maximum steady accretion rate. The processed material is stored in a He shell and the accretor becomes a born-again giant\footnote{See also \cite{bertolami2016} for the single-stellar channel of forming born-again giants.} \citep{nomoto1982, kato1994, tauris2013}. In the case of He accretion, the processed material is added directly to the core and the accretor becomes a born-again He giant. 

If the system is in the `steady burning' or `mild flash' \citep[see][]{piersanti2014} regime, then the material is accreted and processed at the accretion rate and stored in a He shell (or added directly to the core in the case of He accretion), and the accretor may be observed as a super-soft x-ray source \citep{hoshi1973}. Finally, if the system is in the `unsteady burning' regime, then material is allowed to accumulate in a H or He shell, as appropriate, until sufficient material has been accreted for a nova eruption.

In the case of He accretion, the unsteady regime can be sub-divided into the deflagration and detonation regimes, depending on the mode of the explosion \citep{shen2009,piersanti2014}. This distinction is relevant to double-detonation type Ia supernovae \citep{piersanti2014,ruiter2014}. In this work, we neglect this channel and treat both the detonation and deflagration regimes as viable regions of the parameter space for He novae.

\subsection{Novae}

Here we discuss how we determine when individual nova eruptions occur, and the impact these eruptions have on the system.

The decision to trigger a nova eruption is made by comparing the mass of the H or He shell (as appropriate) with the corresponding critical ignition mass, \Mig . If, at a given timestep, the mass of either of the layers is greater than \Mig , then the conditions for nuclear burning at the base of the accreted layer are assumed to have been met and a nova eruption is triggered. Mass is lost from the WD according to the accretion efficiency $\eta$, and orbital parameters are recalculated.

The critical ignition mass is sensitive to the WD mass and the accretion rate. At higher masses, the surface gravity is stronger, resulting in greater heating of the shell due to increased compression of the accreted mass shell and higher infall energy, resulting in a lower \Mig. At higher accretion rates, more gravo-thermal is released per unit time resulting once again in a higher degree of compression, a higher temperature shell, and therefore a lower \Mig.  Note that when increasing \Mdot\ there is a two-fold effect on the recurrence time (the period of time between subsequent nova eruptions). In addition to reaching a given \Mig\ sooner due to material being added to the shell faster, a system with higher \Mdot\ will also have a lower \Mig, further reducing the recurrence time. 

Critical ignition masses for He shells are significantly higher than those for H, due to the more extreme conditions required to initiate nuclear burning of He, which occurs through the triple-$\alpha$ nuclear process, compared to that of H, which occurs through the hot-CNO cycle \citep[see, for example, ][]{jose1998,iliadis2007}.

In our model, \Mig \ is calculated by interpolating between fits to the results of \cite{kato2014} for H accretion, and \cite{piersanti2014} and \cite{kato2018} for He accretion. The interpolation curves are piecewise polynomial fits in $\log_{10}(M_{\rm ig}) - \log_{10}(\dot{M}) - M_{\rm WD}$ space. Beyond the limits of these detailed models, the broad trends of the data are extrapolated using linear fits, preventing any runaway behaviour due to higher order extrapolation. The functions used to calculate \Mig\ are included as supplementary material \footnote{For code, see \href{https://zenodo.org/record/4415007\#.YA47mSZxU5k}{DOI:10.5281/zenodo.4415007}}.

These detailed models were selected based on factors such as the scope and resolution of the grids computed, the recency of the work, and the physical assumptions employed. Scope and resolution in particular were emphasised to reduce the degree of extrapolation necessary where the input parameters from \binaryc\ fall out of the scope of the parameter space of the input models. The effect of selecting different input models for \Mig, of which there are several in the literature \citep[e.g.,][]{yaron2005,chen2019}, will be explored in a future work.

\begin{figure*}
\centering
\begin{subfigure}{0.95\columnwidth}
    \centering
    \includegraphics[width=\textwidth]{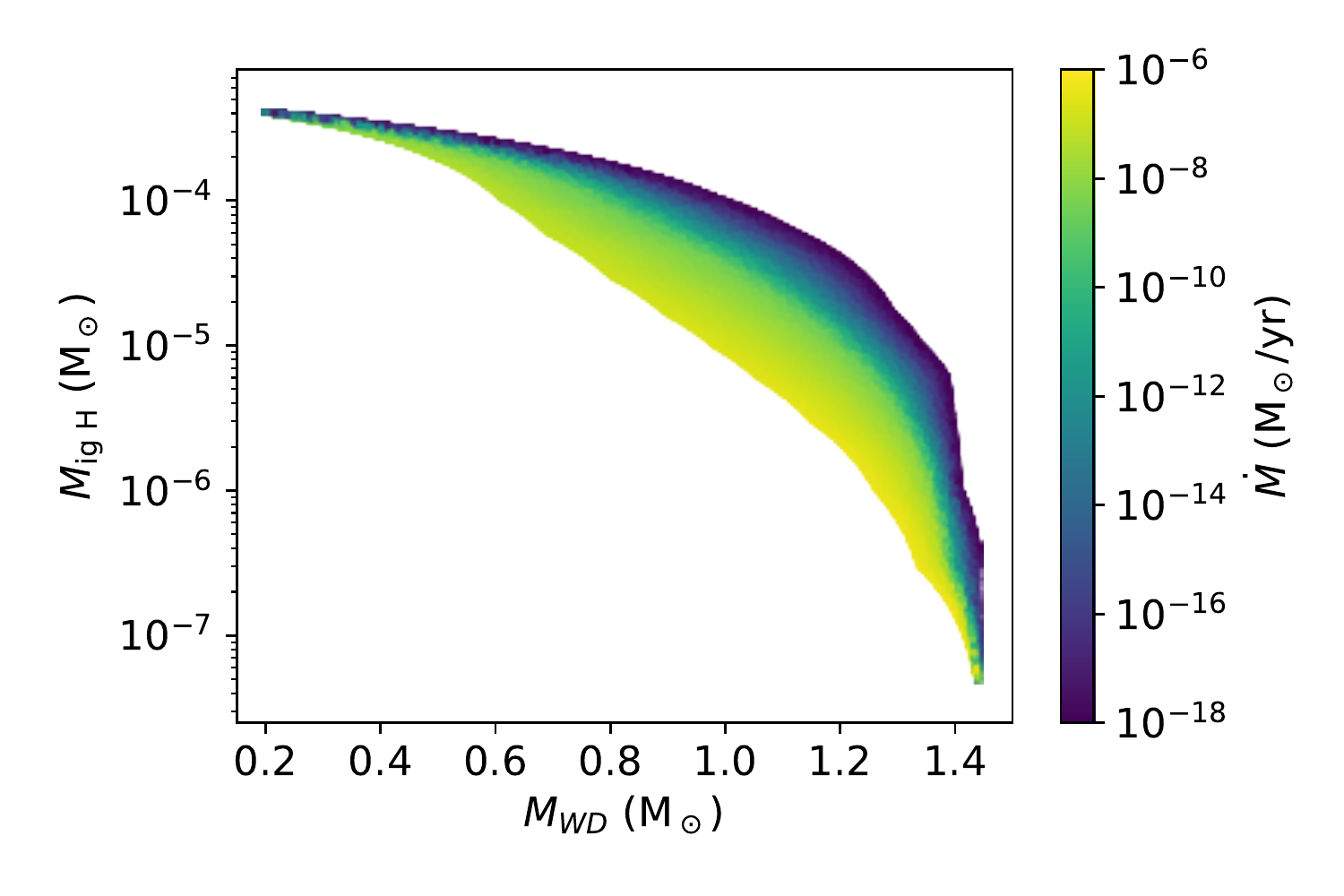}
\end{subfigure}%
\begin{subfigure}{0.95\columnwidth}
    \centering
    \includegraphics[width=\textwidth]{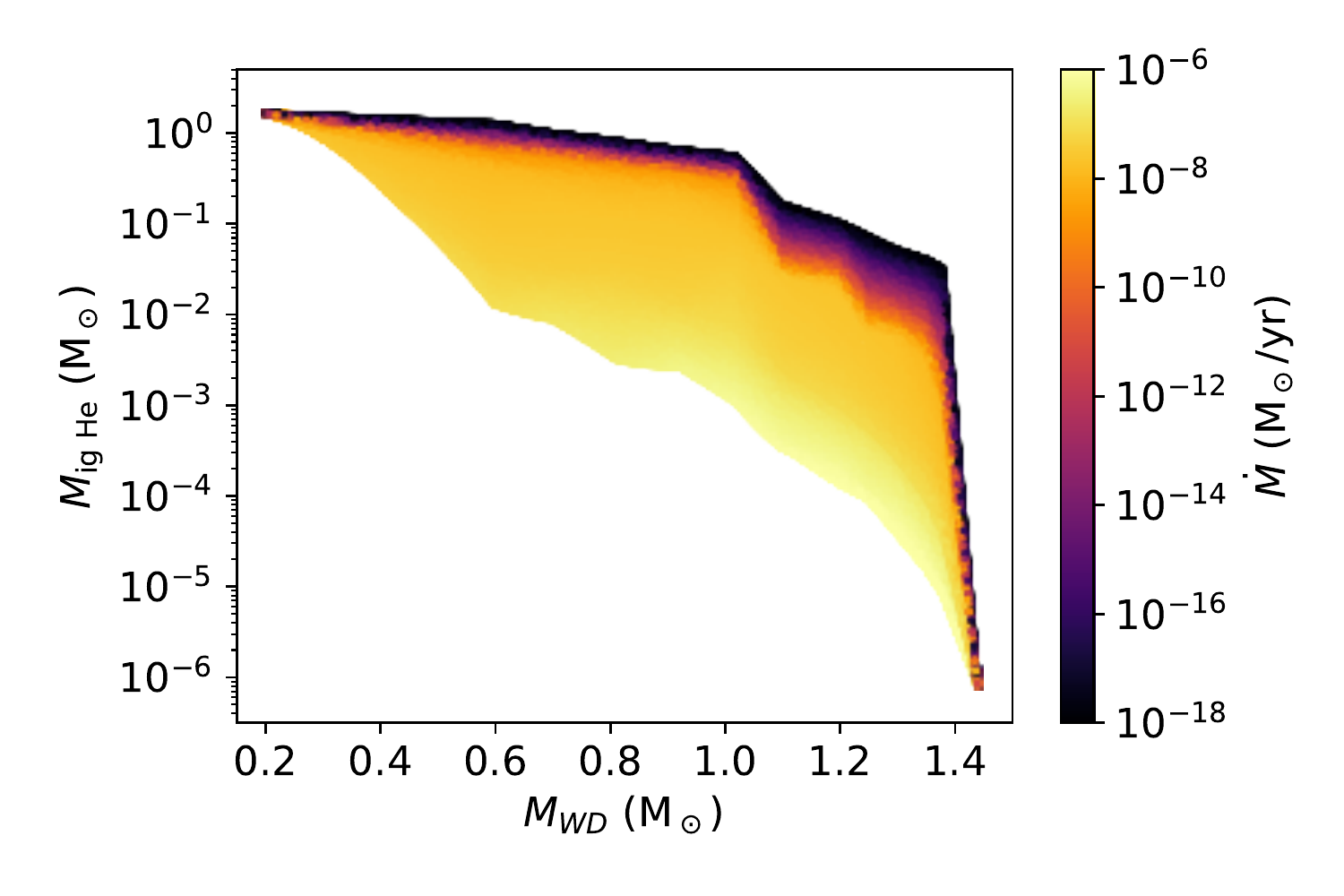}
\end{subfigure}
\caption{Critical ignition masses as a function of white dwarf mass and accretion rate. Ignition masses for H novae (left) are calculated using fits to data from \protect \cite{kato2014}, while data from \protect \cite{piersanti2014} and \protect \cite{kato2018} are used for He novae (right).}
\label{fig:interpfitstuffmig}

\end{figure*}

Figure \ref{fig:interpfitstuffmig} summarises the outputs of the \Mig\ functions when \Mwd\ and \Mdot\ are allowed to vary over a wide range of values, representing the values that \Mig\ \textit{could} potentially take. In practice, it is determined by the WD masses and accretion rates obtained in the simulated binary systems.
It should be noted that the inputs in Figure \ref{fig:interpfitstuffmig} are limited to values which are permissible for novae to occur, i.e., fall within the unsteady burning regime (Figure \ref{fig:wdaccretionlimits}).

As the nova event is triggered only once $M_{\rm accreted}>M_{\rm ig}$, where $M_{\rm accreted}$ is the mass of the shell at the time of the nova eruption, there will necessarily be some excess material $M_{\rm excess} = M_{\rm accreted} - M_{\rm ig}$. This excess material is used to calculate a weighting factor correcting for this overshoot: 

\begin{equation}
F_{\rm nova}=1+\frac{M_{\rm excess}}{M_{\rm ig}}=\frac{M_{\rm accreted}}{M_{\rm ig}}.
\end{equation}

By weighting each eruption by $F_{\rm nova}$, we approximate the effect of adding the excess material back to the H or He shell, as appropriate, and thereby allowing it to contribute toward the next eruption. This correction is applied to all results pertaining to individual nova events, including nova recurrence times. The recurrence time $T_{\rm rec,i}$ for a nova eruption logged at time $t_{\rm i}$ where the next recorded nova eruption occurs at $t_{\rm{i}+1}$ is be computed as

\begin{equation}
    T_{\rm rec,i}=\frac{t_{\rm{i}+1}-t_{\rm i}}{F_{\rm nova,i}}.
\end{equation}

When a nova event is triggered, some fraction of the mass accreted by the WD is lost during the nova, and some fraction is retained. This fraction is characterised by the accretion efficiency $\eta$, defined such that the mass lost is

\begin{equation}
M_{\rm lost} = (1-\eta) M_{\rm ig}.
\end{equation}

When $\eta$ = 1 no mass is lost during the eruption, and all accreted mass is lost when $\eta$ = 0. For situations where $\eta$ < 0, the WD loses \textit{more} mass than was accreted due to material from the surface of the core being dredged up into the shell during, or prior to, the eruption. 

Mass loss through novae is assumed to be spherically symmetric, carrying away angular momentum from the binary. We further assume that no material is re-accreted onto the companion.

In similar fashion to the critical ignition mass \Mig , $\eta$ is calculated by interpolating between fits to the results of \cite{wang2018} for H novae and \cite{wu2017} for He novae.

Assessing the physical soundness of computed accretion efficiencies of detailed models of novae is difficult. The results of detailed models are sensitive both to numeric decisions, such as the number of nova cycles computed and the spatial grid resolution, as well as the physical choices made, particularly the choice of mass-loss mechanism (an assumption necessary due to numeric difficulties which arise as the envelope expands during outburst). \cite{kato2017} compare the results of several codes from different groups, and find significant discrepancies between the results due primarily to the above considerations.

The models of \cite{wang2018} and \cite{wu2017} were computed by the same research group using the same code, and so can be considered in some sense consistent. This allows a better motivated comparison between the WD growth rates due to H and He accretion.
However, studies deriving accretion efficiencies from observations of chemical abundances in nova ejecta suggest negative accretion efficiencies are prevalent, particularly in novae exhibiting significant enrichment of elements heavier than He \citep{fujimoto1992}. \cite{wang2018} and \cite{wu2017}, the works on which this study bases its accretion efficiency calculations on, do not find any significant parameter space with negative accretion efficiencies.

Negative accretion efficiencies have also been found to be common in several theoretical works dedicated to novae (\citep[e.g.,]{yaron2005}. Not all detailed modelling works on novae publish accretion efficiency data, making it somewhat difficult to get a consistent picture of the state of the field, but in general it appears that works making use of the \cite{prialnik1984,kovetz1985} code tend to find negative accretion efficiencies to be common, though not ubiquitous \citep{prialnik1995errosionofHe,prialnik1995anextendedgrid,yaron2005,epelstain2007,idan2013}, while the use of other codes such as MESA \citep{newsham2014,wu2017,wang2018,chen2019} and the `Saio' code \citep{kato2004,kato2017amillenium} tends to result in higher accretion efficiencies. It may be that these discrepancies are more closely related to physical choices such as the assumed degree of mixing increasing at the WD-shell interface, which has been shown to dramatically effect accretion efficiencies \citep{starrfield2012,chen2019}, than the numeric quirks of each code.

\begin{figure*}
\centering
\begin{subfigure}{0.95\columnwidth}
    \centering
    \includegraphics[width=\textwidth]{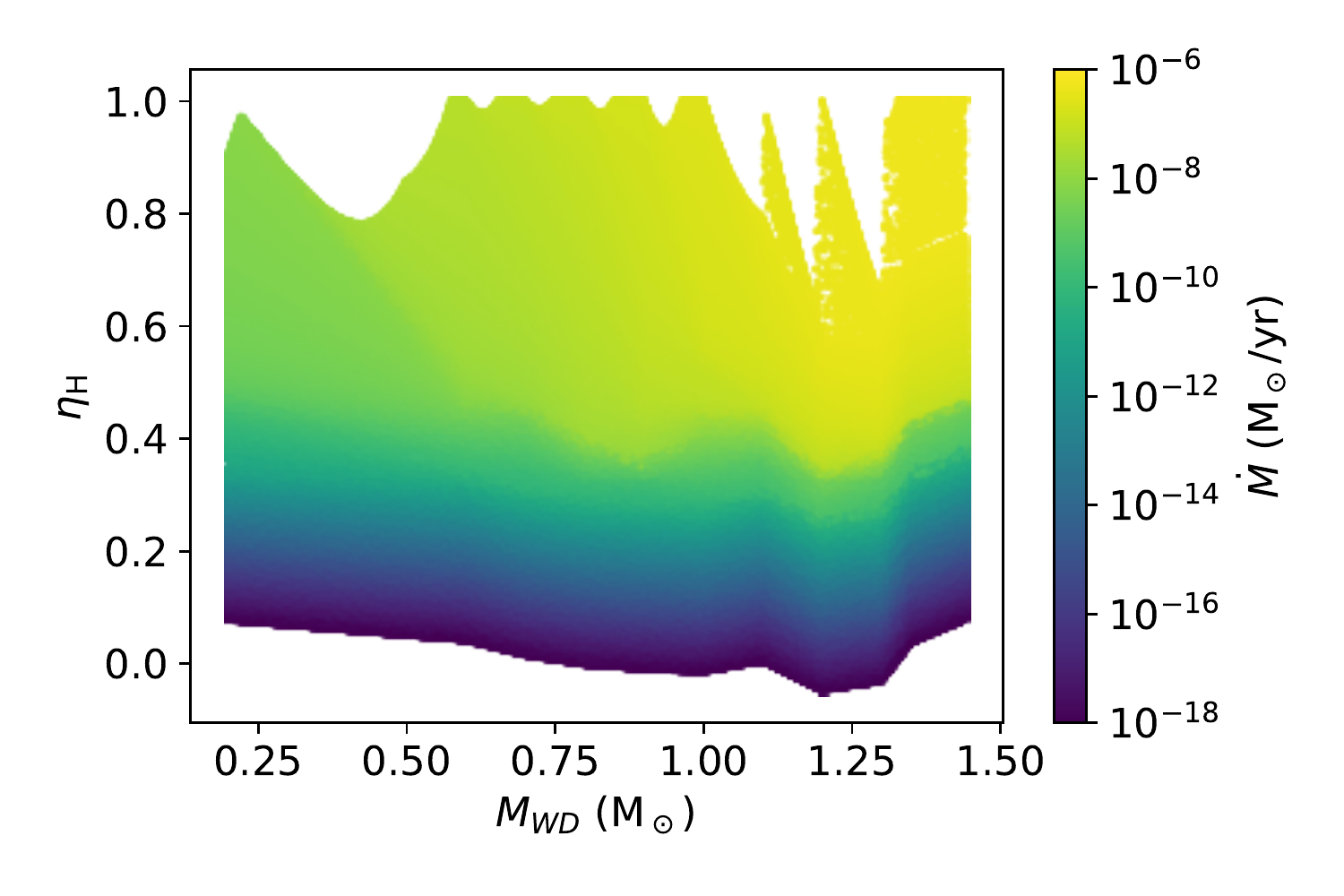}
\end{subfigure}%
\begin{subfigure}{0.95\columnwidth}
    \centering
    \includegraphics[width=\textwidth]{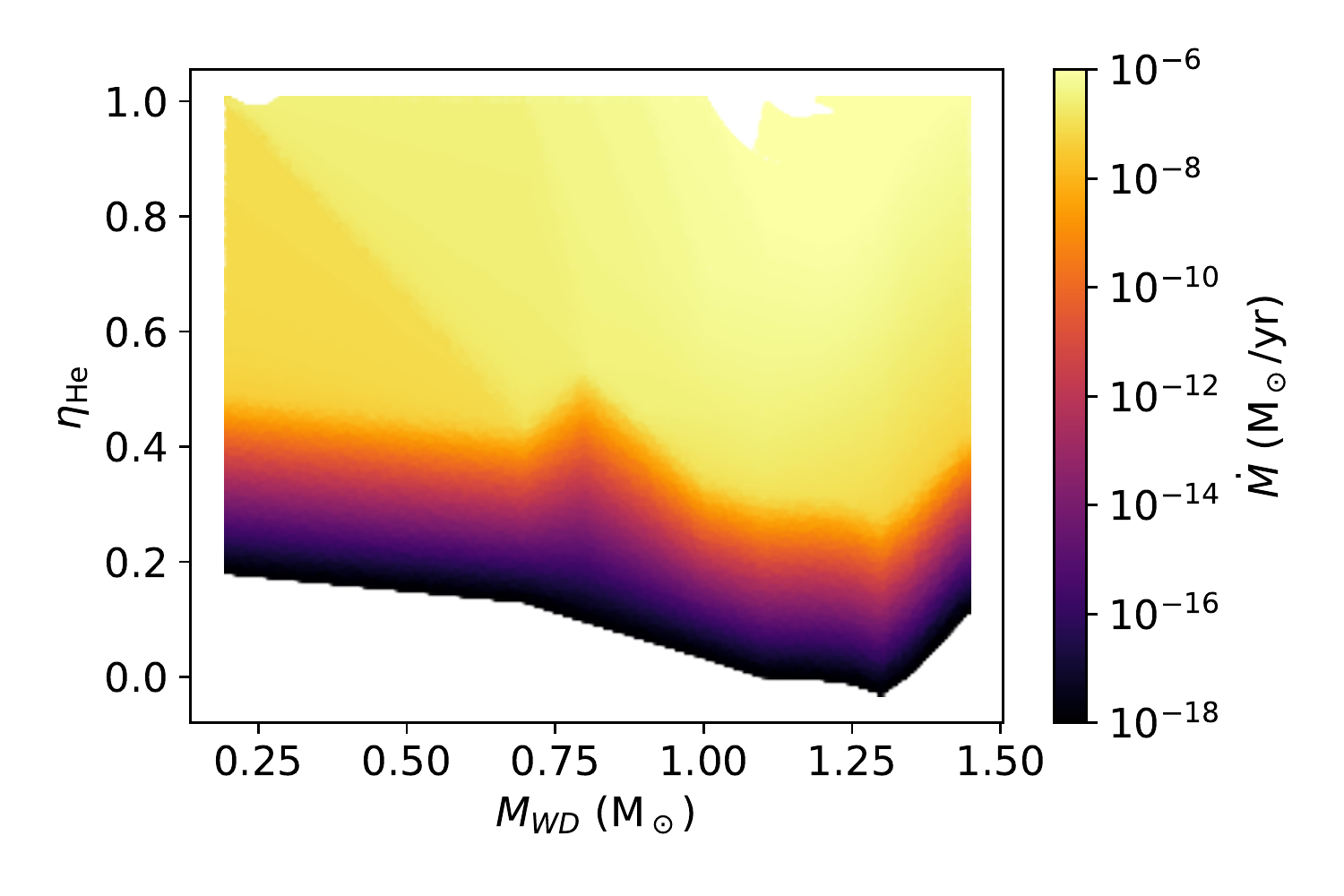}
\end{subfigure}
\caption{Accretion efficiencies as a function of white dwarf mass and accretion rate. The accretion efficiencies for H novae (left) are calculated using fits to data from \protect \cite{wang2018}, while data from \protect \cite{wu2017} are used for He novae (right).}
\label{fig:interpfitstuffeta}

\end{figure*}

As with \Mig , $\eta$ is determined by \Mwd \ and the accretion rate \Mdot, and Figure \ref{fig:interpfitstuffeta} summarises the outputs of the interpolation functions for $\eta$ in a similar fashion to Figure \ref{fig:interpfitstuffmig}. The behaviour with varying \Mdot \ is relatively intuitive; reduced \Mdot \ leads to reduced gravo-thermal heating of the layer which increases the total layer mass and degree of degeneracy at the base of the layer at ignition. This leads to a more powerful explosion which in turn leads to more material lost from the WD, lowering $\eta$ for a given \Mwd . The behaviour of $\eta$ when \Mwd \ is varied is more complicated, however, and the intuitive idea that more powerful explosions cause more mass loss no longer applies. Ultimately, material is lost through winds and $\eta$ can be considered to be the result of competition between the rate of nuclear processing of material at the surface and the rate of wind mass loss \citep{kato2004}. When \Mwd\ is less than approximately 0.8 M\solar\ the winds are too weak to effectively remove mass from the WD, increasing $\eta$, while at the highest WD masses (\Mwd $\gtrsim 1.25$ M\solar) the rate of nuclear burning dominates as the high surface gravity impedes wind mass loss, causing $\eta$ to increase once again.

\subsection{Quiescent carbon burning}
\label{sec:qcb}
A potential evolutionary channel leading to a single-degenerate, Chandrasekhar mass (\Mchand) type Ia supernova involves accreting He-rich material onto a C/O WD at a rate such that the He is burnt steadily at the surface \citep{solheim2005}, directly depositing C and O onto the core. This has the advantage of potentially growing the WD at close to the maximum steady accretion limit for He (`RG/SS' in the right panel of Figure \ref{fig:wdaccretionlimits}), as mass loss due to novae is avoided when accreting in the steady burning regime.

However, at extremely high He accretion rates, it is possible that C burning may be initiated at the surface, resulting in a C burning front that propagates through the WD \citep{nomoto1991,brooks2016,wang2017}. This has the effect of converting the C/O WD to a O/Ne WD, likely preventing it from producing a type Ia supernova in the event it reaches \Mchand. This effect is included in our standard physics set (Table \ref{tab:gridphysics}) of our simulations by changing the stellar type from C/O to O/Ne if the He accretion rate becomes higher than $2.05\times 10^{-6}$ M\solarperyr\ \cite[the threshold computed by][]{wang2017}.

\subsection{Unstable mass transfer}

Whether or not mass transfer proceeds in a stable manner depends on the response of the stellar radii and Roche lobes to mass loss, described by the mass-radius exponent $\zeta \equiv \frac{\diff \ln R}{\diff \ln M}$ \cite[see e.g.][]{hjellming1987, ge2010,ge2015, ge2020}. 
If the radius of the donor star persistently increases relative to its Roche lobe in response to mass loss, the resulting increase in the mass transfer rate causes a runaway situation to develop that ultimately results in a common envelope that engulfs both stars.
Conversely, if the donor star remains roughly constant in size or shrinks relative to its Roche lobe in response to mass loss, mass transfer is stable. Note that stability also depends on the reaction of the accretor to mass gain. If the accretor can structurally re-adjust on timescales shorter than the mass transfer timescale, mass transfer remains stable. If it cannot, then the transferred material quickly piles up on the accretor in a high-entropy envelope, overflowing its Roche lobe and producing a common envelope surrounding both stars.

Without access to detailed information about stellar structure, population synthesis codes typically rely on greatly simplified prescriptions for mass transfer stability. The most common approximation used is to compare the mass ratio $q=M_{\rm donor}/M_{\rm accretor}$ to a set of predefined critical values. This method is motivated by the fact that the mass transfer rate and changes in Roche lobe size can be described as a function of $q$  \citep[see e.g.,][]{soberman1997}.

In \binaryc, the stability of all mass transfer episodes is estimated in this way, henceforth referred to as the `$q_{\rm crit}$ method'. Alternative prescriptions relying on the calculation of $\zeta$ may incorrectly predict unstable mass transfer in the event that the Roche lobe $\zeta$ only briefly exceeds that of the donor, introducing spurious common envelope events that can dramatically affect the evolution of the binary. The $q_{\rm crit}$ method does not suffer from this issue, as the critical ratios are typically determined by simulating the evolution of mass-transferring binaries up to the onset of instability. However, a disadvantage of this method is that any set of critical mass ratios is only truly valid under a single set of assumptions regarding mass transfer efficiency and angular momentum loss. The critical values of $q$ depend on the phase of evolution, structure, and composition of the donor star, as well as whether or not the accretor is degenerate. For an overview of the critical mass ratios and calculation of the mass transfer rates and efficiencies employed in \binaryc, see \cite{claeys2014}. 

If mass transfer becomes unstable, a common-envelope (CE) is formed that engulfs both stars. This CE is not co-rotating with the binary, and drag forces inside the CE cause the binary components to spiral into tighter orbits. The CE phase ends either when the envelope is ejected from the system, producing a more compact (`hard') binary with a stripped former-donor star, or when the two stars merge, producing a single merger remnant. Despite the importance of the CE phase in binary evolution and significant effort from the community \cite[e.g.,][]{demarco2011alpha,ivanova2013common,ohlmann2016,chamandy2018accretion}, this phase remains poorly understood. Therefore, population synthesis studies often employ simplistic prescriptions to determine the outcome of CE evolution, and estimate the impact of uncertainties in CE evolution by testing a number of different treatments.

The canonical model for the CE phase is the $\alpha$--$\lambda$ prescription, which is based on the energy budget of the binary system \citep{paczynski1976ce,webbink1984,livio1988,dekool1990common}. In this prescription $\alpha_{\rm CE}$ parameterises the fraction of orbital energy which is spent in ejecting the envelope

\begin{equation}
    E_{\rm bind,i}=\alpha_{\rm CE}(E_{\rm orb,f}-E_{\rm orb,i}),
\end{equation}

where subscripts `i' and `f' denote properties before and after common envelope, $E_{\rm bind}$ is the binding energy of the system, and $E_{\rm orb}$ is the orbital energy of the system. We can also write $E_{\rm bind, i}$ as 

\begin{equation}
E_{\rm bind,i}=-\frac{G}{\lambda_{\rm CE}} \Bigg(\frac{M_1 M_{\rm env1}}{R_1} + \frac{M_2 M_{\rm env2}}{R_2}\Bigg),
\end{equation}

where $M_{\rm env}$ is the envelope mass for each star. This allows us to define $\lambda_{\rm CE}$ as a parameter inversely proportional to the binding energy of the envelope. Note that depending on the phase of evolution (e.g., for the case of a WD remnant), there may be only one envelope and one of the terms reduces to zero. For a more detailed discussion, see \cite{hurley2002}, from which we have reproduced the above equations.

Uncertainty in common-envelope physics is explored through varying $\alpha_{\rm CE}$ between 0.5 and 2. The energy required to eject the envelope depends on the mass distribution of the envelope and is parameterised by $\lambda_{\rm CE}$. It is therefore expected that this parameter is sensitive to the stellar mass and phase of evolution \citep{ivanova2011conference,demarco2011alpha}. We vary $\lambda_{\rm CE}$ between 0.05 and 1, and additionally use two prescriptions to calculate $\lambda_{\rm CE}$ based on the stellar type of the donor. The first, included in our standard physics model, uses fits provided in \cite{wang2016binding}, and the other, formulated by \cite{claeys2014}, is based on the data of \cite{dewi2000}.

The critical values of $q$ that determine whether unstable mass transfer occurs for a given binary configuration are taken from \cite{hurley2002}.
This treatment is likely to be a somewhat poor approximation, particularly for the case of giant donors. However, it does have the advantage of being widely used and easily described. A truly satisfactory treatment of mass transfer stability remains elusive in BPS codes, although work in this area continues. Examining the effect of employing a more complex approximation for the stability of mass transfer is deferred to a future work where it can be given the attention it deserves.

\section{Methodology}
\label{sec:methodology}
\subsection{Our Standard Model}

The results presented in this work are computed using \binaryc\ \citep{izzard2004}. \binaryc\ is a BPS code built on the backbone of the \cite{hurley2000,hurley2002} stellar fits and binary evolution models. The code features many updates and improvements, including the introduction of an updated treatment of AGB stars \citep{izzard2004}, a synthetic nuclear yields module \citep{izzard2006}, wind-RLOF \citep{abate2013wind}, improved stellar rotation algorithms \citep{demink2013}, updated stellar lifetimes \citep{schneider2013}, and RLOF corrections \citep{claeys2014}.

Table \ref{tab:gridphysics} summarises the model parameters used to compute our results, referred to as our `standard physics case'. Other model variables are set to the \binaryc\ defaults, and a complete set of model parameters may be obtained on request from the corresponding author.

In our models the effect of magnetic braking is modelled following \cite{hurley2000}, which does not include magnetic braking effects on remnant stellar types, including WDs. The effect of magnetic braking on our accreting WDs is therefore neglected, although it is taken into account for the donor stars. Magnetic braking has been demonstrated \citep{belloni2020} to be important in reproducing period distributions of synchronous magnetic cataclysmic variables (polars), although its effect on nova rates and properties remains unknown.

\begin{table}
\begin{tabular}{p{0.45\columnwidth}p{0.5\columnwidth}}
\textbf{Variable   / physics} & \textbf{Standard} \\ \cline{1-2}
Metallicity $Z$ & 0.02 \\
Simulation time (Myr) & 15000 \\
RLOF model & \cite{claeys2014} \\
$\alpha_{\rm CE}$ & 1 \\
$\lambda_{\rm CE}$ & \cite{wang2016binding} \\
Hachisu disk wind & ON \\
Eddington limited accretion from WD to remnant & OFF \\
Eddington limited steady accretion to WD & OFF \\
wind-RLOF: & $q$ dependent \cite{mohamed2007wind,abate2013wind} \\
Angular momentum loss through winds & Spherically symetric \cite{mohamed2007wind,abate2013wind} \\

Magnetic braking & \citep{hurley2000} \\

H and He accretion limits onto WDs & H: \cite{wang2018}, He: \cite{wu2017} \\
Nova accretion efficiencies & H: \cite{wang2018}, He: \cite{wu2017} \\
Nova critical ignition masses & H: \cite{kato2014} He: \cite{piersanti2014,kato2018} \\
Minimum He shell mass for Hybrid COWD (M\solar) & 0.001 \\
Quiescent carbon burning limit & $2.05\times 10^{-6}$ M\solarperyr\ \citep{wang2017} \\
Chandrasekhar mass \Mchand\ & 1.38 M\solar \\
Critical mass ratios for mass transfer stability $q_{\rm crit}$ & \cite{hurley2002} \\
$M_{\rm max\ NS}$ & 2.5 \\
Remnant mass scheme (NS/BH) & Fryer rapid, \citep{fryer2012}
\end{tabular}
\caption{Table summarising selected input physics in our standard physics case.}
\label{tab:gridphysics}
\end{table}

\subsection{Grids}
\label{sec:grids}

\begin{table}
\begin{threeparttable}[b]
\begin{tabular}{lllr}
 & Bounds \tnote{a} & Spacing function & Resolution \\ \cline{2-4}
\multicolumn{1}{l|}{$M_{1,\rm init}$ (M\solar)} & (0.8, 20) & $\Delta M_{1, \rm init}=\rm const$ & 80 \\
\multicolumn{1}{l|}{$q_{\rm init}$} & ($\frac{0.1}{M_{1, \rm init}}$,1) & $\Delta (q_{\rm init})=\rm const$ & 50 \\
\multicolumn{1}{l|}{$a_{\rm init}$ (R\solar)} & (3, $1\mathrm{e}{6}$) & $\Delta\ln(a_{\rm init})=\rm const$ & 60
\end{tabular}
\begin{tablenotes}
\item[a] (min, max)
\end{tablenotes}
\end{threeparttable}
\caption{Grid bounds, spacing between grid points, and resolution ($80\times50\times60$) used to model H-nova systems.}
\label{tab:gridH}
\end{table}

\begin{table}
\begin{threeparttable}[b]
\begin{tabular}{lllr}
 & Bounds \tnote{a} & Spacing function & Resolution \\ \cline{2-4}
\multicolumn{1}{l|}{$M_{1,\rm init}$ (M\solar)} & (1.2, 12) & $\Delta\ln(M_{1, \rm init})=\rm const$ & 100 \\
\multicolumn{1}{l|}{$q_{\rm init}$} & ($\frac{1}{M_{1, \rm init}}$,1) & $\Delta (q_{\rm init})=\rm const$ & 50 \\
\multicolumn{1}{l|}{$a_{\rm init}$ (R\solar)} & (10, $3\mathrm{e}{4}$) & $\Delta\ln(a_{\rm init})\rm const$ & 50
\end{tabular}
\begin{tablenotes}
\item[a] (min, max)
\end{tablenotes}
\end{threeparttable}
\caption{Grid bounds, spacing between grid points, and resolution ($100\times50\times50$) used to model He-accreting He-nova systems.}
\label{tab:gridHe}
\end{table}

Results for H- and He-nova systems are computed from two different grids of binary input parameters, summarised in Tables \ref{tab:gridH} and \ref{tab:gridHe}. These grids were formulated by optimising the selection of bounds, spacing functions determining the spacing of grid points, and resolution of each of the input parameters in terms of the number of resolvable unique evolutionary sequences of H and He novae, while not leaving any systems outside the bounds of the respective grids. The exception to this constraint is the initial secondary mass, $M_{\rm 2 \ init}$, for the H-nova grid, for which we impose a minimum mass of 0.1 M\solar\ despite the fact that, according to our simulations, systems with secondaries with initial masses as low as 0.01 M\solar\ are found to be capable of producing H novae.

We impose the 0.1 M\solar\ limit on the grounds that these stars are already below the 0.5 M\solar\ minimum mass that the original \cite{pols1998} grid of stellar models used in \binaryc. Stars with masses lower than 0.1 M\solar\ are approaching the brown dwarf regime, for which the binary frequencies are uncertain and likely to be significantly lower than their more massive counterparts \citep{farihi2005,grether2006}. The inclusion of systems with secondaries with initial masses below 0.1 M\solar\ increases predicted nova rates by approximately 5 per cent under the generous assumption of a 50 per cent binary fraction.

\section{Results}
Throughout this work we make use of the \cite{hurley2000,hurley2002} stellar types, in addition to further classifying each of the stellar types as either an unevolved, evolved, or remnant phase of evolution. This classification scheme is summarised in Table \ref{tab:evotags}.

All subsequent figures present results normalised according to theoretical birth distributions of the primary mass, secondary mass, and orbital separation for each system. The normalisation process is described in detail in Appendix \ref{sec:normalisationprocess}.
Figures \ref{fig:barCEeventsnovae}-\ref{fig:delaytimedist} represent a scenario where a burst of star formation occurs at time $t=0$, and an observer is able to count all events over the entire 15 Gyr duration of the simulation. These figures provide insight into fundamental nova properties, and are not expected (or intended) to reproduce distributions of nova properties derived from observations \citep[e.g.,][]{shara2018,fuentes2020}. These figures are normalised per solar mass of star forming material, $\rm M_{\rm \odot SFM}$.

\begin{table}
\begin{tabular}{lll}
Stellar Type & Description & Classification \\ \hline
LMMS & low-mass main sequence & Unevolved \\
MS & Main sequence & Unevolved \\
HG & Hertzsprung gap & Evolved \\
FGB & First giant branch & Evolved \\
CHeB & Core He burning & Evolved \\
EAGB & early asymptotic giant branch & Evolved \\
TPAGB & Thermally pulsing asymptotic giant branch & Evolved \\
HeMS & He main sequence & Unevolved \\
HeHG & He Hertzsprung gap & Evolved \\
HeGB & He giant branch & Evolved \\
HeWD & He white dwarf & Remnant \\
COWD & C/O white dwarf & Remnant \\
ONeWD & O/Ne white dwarf & Remnant \\
NS & Neutron star & Remnant \\
BH & Black hole & Remnant \\
MR & Massless remnant & Remnant
\end{tabular}
\caption{Summary of the \protect\cite{hurley2000,hurley2002} stellar types and terminology used throughout this work.}
\label{tab:evotags}
\end{table}

\label{sec:results}

\subsection{Typical Evolutionary Sequences for H and He Novae}

There are a plethora of different evolutionary sequences leading to systems which may undergo H or He novae. We can quantify this diversity if we consider a binary evolution sequence to be unique if it is comprised of a distinct combination of single stellar evolution phases \citep[or `stellar types', see][and Table \ref{tab:evotags}]{hurley2000,hurley2002}. In our standard physics case we can distinguish 167 unique evolutionary sequences involving at least one H nova (where we have 11521 distinct H-nova systems and 80 million eruptions). Performing the same assessment for He nova, we obtain 143 unique evolutionary sequences (where we have 8730 distinct He-nova systems and 1.75 million eruptions). In truth, these figures only hint at the true extent of the complexity surrounding these systems. Some of these evolutionary sequences contain within them significant spreads in the initial masses and orbital separations which can dramatically change the number and nature of novae a system will experience over its lifetime.

Despite this complexity, it is instructive to describe in broad terms a typical evolutionary sequence for H- and He-nova systems. Facets of other evolutionary pathways will be discussed in subsequent sections as they become relevant.

A typical binary evolution sequence involving H novae proceeds as follows. The primary evolves similarly to a single star, ending its life as either a C/O or O/Ne WD. Mass transfer commences as the secondary evolves off the main sequence and ascends the giant branch, with mass transfer ceasing as the star contracts due to the initiation of core He burning before restarting upon the ascent of the asymptotic giant branch (AGB). Here it loses its H envelope to increasingly strong winds on the AGB, leaving the star in its final remnant state as a C/O or O/Ne WD. The binary then remains in this double WD configuration for the remainder of the simulation. H novae may occur during any of the periods of mass transfer onto the WD, most frequently during the FGB and TPAGB evolutionary phases, when the donor star is at its largest radial extent and the mass transfer rate is typically highest.

While this is far from the most complex evolutionary sequence involving H novae, it is by far the most common. More exotic sequences involving H novae can include interesting manifestations of binary physics, such as stripping episodes, Algol systems, CE events, mergers, or type Ia supernovae. 

The life stories of He-nova systems are far more complex. It is also more difficult to neatly describe a single `typical' channel for their production, so we shall instead outline two of the most important.

The first evolutionary sequence, which is most common when the initial separation is greater than approximately 300 R\solar\ and involves two CE phases, begins with the primary evolving normally until it leaves the main sequence. As it expands, it begins to lose material through stellar winds, a small fraction of which is accreted by the companion, with the remainder carrying away angular momentum from the binary. This situation continues throughout the HG, the FGB, and CHeB phases until it reaches the AGB, having formed a C/O or O/Ne core. During its ascent of the AGB the first CE of the system occurs, significantly hardening the binary and ejecting the remaining envelope of the star and leaving it as a C/O or O/Ne WD. The system then remains quiescent until the secondary evolves off the main sequence. The expansion of the secondary results in a second CE event as it ascends the FGB, leaving the binary, now consisting of the previously formed WD and a new-born HeMS star, with a sub-hour orbital period. If the orbital separation is too large for mass transfer to commence immediately, angular momentum loss through gravitational radiation hardens the binary further, until mass transfer is initiated from the HeMS star onto the WD. At this point in the evolution of the binary, the WD is usually the more massive of the two stars, so that the effect of mass transfer is to widen the binary. However, despite the binary having a mass ratio $q<1$, it is still too large for the effect of mass transfer from the He star to the WD to overcome angular momentum losses through gravitational-wave emission. Tides hasten the inspiral as angular momentum from the orbit is transferred into the spins of the two stars to keep them tidally locked \cite[the Darwin instability][]{darwin1879}. As the system loses mass and yet more angular momentum through He novae, the fate of the system becomes inevitable. The HeMS star and the WD merge, forming a single He giant, and the system typically ends its life as a single WD. The inspiral phase typically lasts tens of Myrs once mass transfer is initiated, during which time the system experiences He novae.

The second sequence, involving a single CE phase, is favoured for systems with initial separations less than approximately 300 R\solar. The primary transfers material to the companion at a non-negligible rate (typically $\gtrsim 10^{-9}$ M\solarperyr) upon leaving the main sequence, typically beginning mass transfer in the HG, and continuing to do so as it ascends the giant branch. This mass transfer is initially non-conservative, with greater than 50 per cent of material being ejected from the binary due to the orbit being too wide. The effect of this initial phase of mass transfer is to tighten the binary, making the mass transfer more rapid and more conservative in the process. However, soon the primary becomes less massive than the secondary and subsequent mass transfer widens the binary as the mass transfer rate stabilises. This situation persists until the primary loses all of its remaining envelope on the FGB and transitions to the HeMS. The net effect of the mass transfer up to this point is to widen the binary, typically by a factor of a few. The primary evolves along the HeMS unperturbed until it reaches the HeHG, at which point a final episode of stable mass transfer occurs leaving it as a C/O or O/Ne WD and widening the binary further. The secondary continues to evolve along the main sequence before undergoing a CE event on the FGB, leaving the binary in a sub-hour period orbit with a WD and a HeMS star. It is again in this configuration that He novae are produced. Gravitational-wave radiation hardens the binary until mass transfer occurs, but the mass ratio is too large for the effect of mass transfer to overcome the orbital decay due to gravitational radiation. The binary merges, once again tending to produce a single WD remnant.

Note that both of the above channels involve He novae produced in systems with HeMS donor stars. We find that HeMS donor systems utterly dominate populations of He novae, producing 99.4 per cent of all He novae (excluding He novae produced during periods of H accretion). The remaining contributions are HeHG donors (0.385 per cent), HeGB donors (0.215 per cent) and a tiny contribution from HeWD donors (less than 0.01 per cent).

\subsubsection{He novae from H donors}

Apart from the two (dominant) channels of He novae which occur under conditions of He accretion discussed above, He novae can also be produced through H accretion. In our model, we allow a He layer to be built up through successive novae -- provided the accretion efficiency is positive -- or through steady H burning, provided the WD mass and H accretion rate do not place the system in the steady He burning regime. A He nova eruption is triggered if, through successive H nova eruptions or a period of steady H burning, the He layer grows more massive than $M_{\rm ig \ He}$. However, two important caveats should be raised at this point.

Firstly, the models on which we base our calculations of the He burning regimes and critical ignition masses are He-accreting models. Therefore, at a fundamental level we are using a model for one scenario and hoping that it produces a result that is somewhat relevant for another, and that physics such as heating due to the nuclear burning of H is relatively unimportant to the underlying physics of this channel. Secondly, a He nova that results from the accretion of H material may not actually be able to be positively identified as a He nova due to the presence of at least some H in the ejecta. Distinguishing a He nova driven by H accretion from a H nova that has dredged up a large amount of He-rich material from the surface of the WD is likely to be impossible observationally. It is of interest to compute and compare light-curves for the two scenarios, but that is beyond the scope of this work.

With these caveats in mind, we may proceed to discuss the relevance of this channel. If we consider all channels that result in at least one He nova to be equal in importance, then the channel is of paramount importance, making up 94 per cent of all He nova systems produced per burst of star formation. However, the channel accounts for very few He novae, with only around 10 per cent of all He novae produced per burst of star formation occurring during H accretion. The situation can become complicated when considering a real star forming environment due to the delay-times of some of these events, but we defer this discussion to Section \ref{sec:novarates}.

We find that the most common scenario for producing He novae from H accreting systems is as follows. After the primary becomes a WD, the donor star expands as it evolves up the FGB, filling its Roche lobe and depositing material onto the WD, producing novae. In this phase, the accretion rate is relatively low, with a correspondingly high $M_{\rm ig \ He}$, often on the order of a tenth of a solar mass. The FGB phase ends, and appreciable mass transfer ceases as the donor star enters the CHeB phase. As the star reascends the giant branch after completing CHeB, mass transfer resumes on the AGB, growing the He layer further as the system produces further H nova eruptions. As the donor star approaches the end of the TPAGB its radius expands further, significantly driving up the mass accretion rate onto the WD such that the system ceases to produce H novae as it is pushed into the steady H-burning regime. This increase in the mass accretion rate can reduce $M_{\rm ig \ He}$ by almost two orders of magnitude. It is this reduction, as much as the increased growth rate of the He layer due to the regime transitioning to steady H burning, that allows the He novae to occur. Typically, between 1 to 5 He novae occur during the brief time before the accretion rate increases to the degree where it is accreting material beyond the limit for steady He burning to occur, or beyond the maximum steady burning limit for H burning. Regardless of which limit is reached first, soon after steady burning commences the WD accretion regime transitions into the `giant' accretion regime, where material cannot be processed sufficiently rapidly at the surface and instead builds up in a high-entropy envelope. The white dwarf then becomes a born-again giant star and typically remains in this phase until the donor star evolves off the TPAGB and becomes a WD remnant. The system ends its life as a double white dwarf.

From Figure \ref{fig:interpfitstuffeta} onwards, figures relating to He novae will specify whether they include or neglect He novae from H accretion. Where the channel is neglected from the figure in the interests of demonstrating features of the He-accretion channels, which in some cases the presence of the H accreting systems obscure, any noteworthy features of the H accretion channel will be discussed in the main body of text.

\subsubsection{The importance of CE events on novae}

\begin{figure*}
\centering
\begin{subfigure}{0.8\columnwidth}
    \centering
    \includegraphics[width=\textwidth=1]{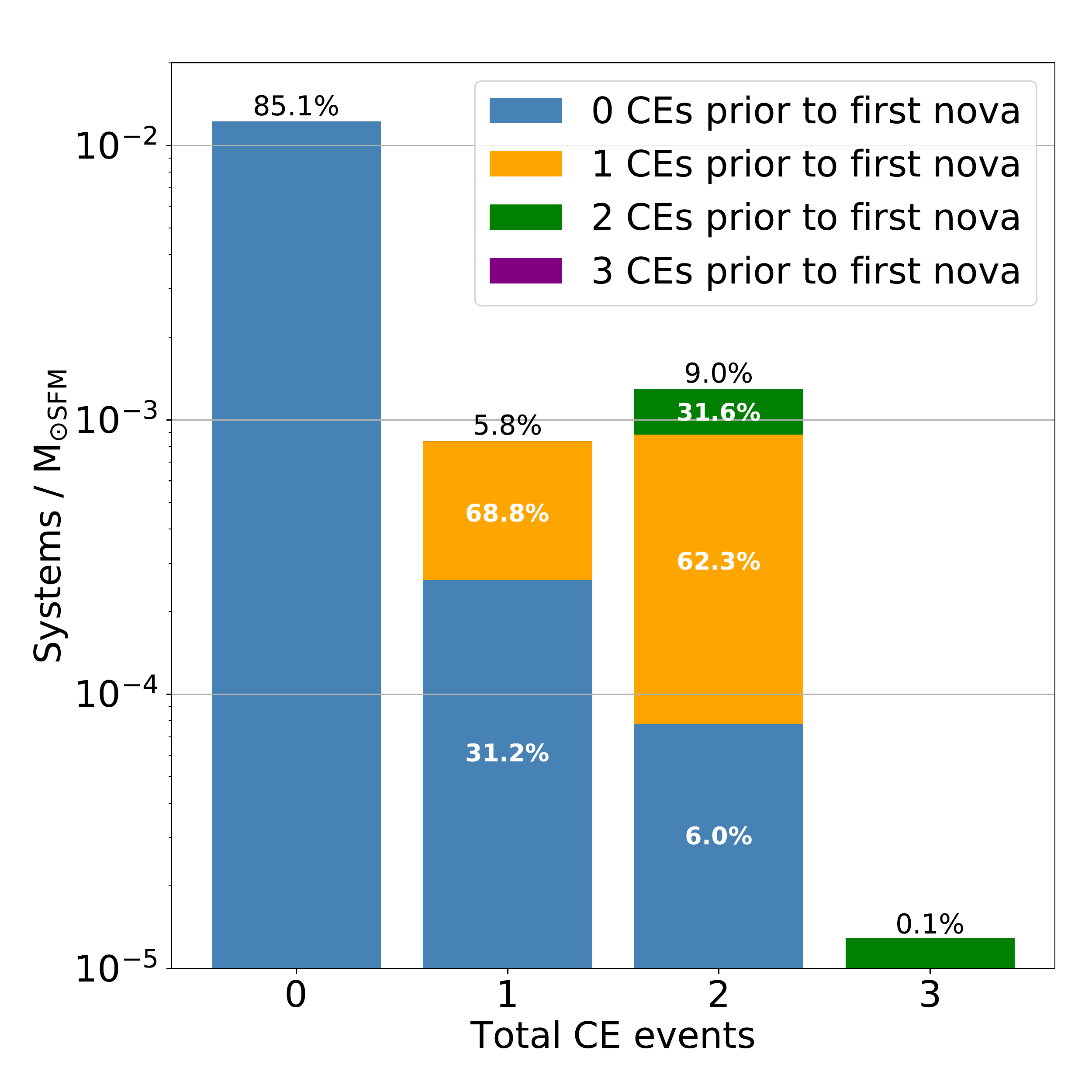}
    \caption{H Novae: system-weighted}
    \label{fig:barCEeventsnovaeH}
\end{subfigure}%
\begin{subfigure}{0.8\columnwidth}
    \centering
    \includegraphics[width=\textwidth=1]{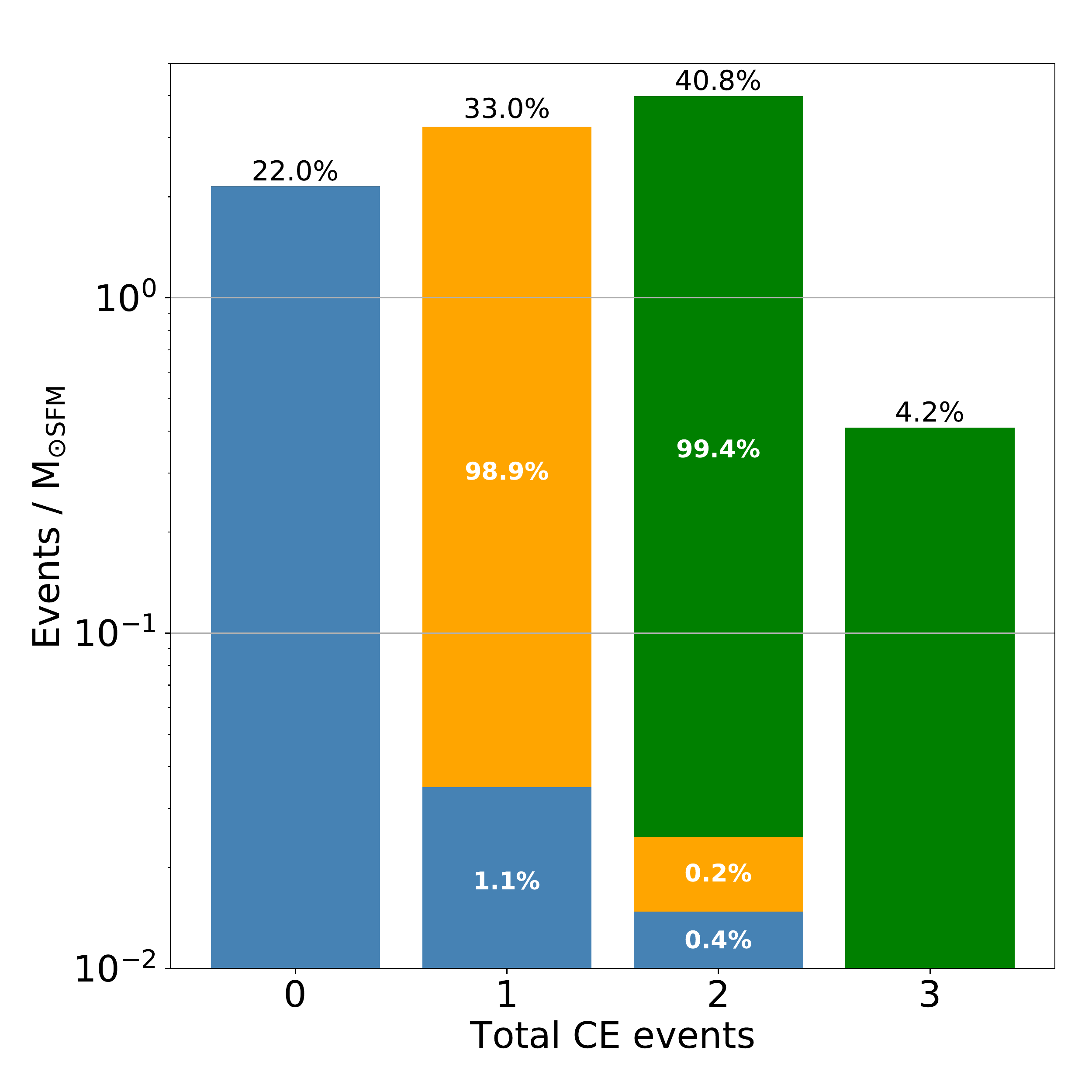}
    \caption{H Novae: event-weighted}
    \label{fig:barCEeventsnovaeHEXP}
\end{subfigure}

\begin{subfigure}{0.8\columnwidth}
    \centering
    \includegraphics[width=\textwidth=1]{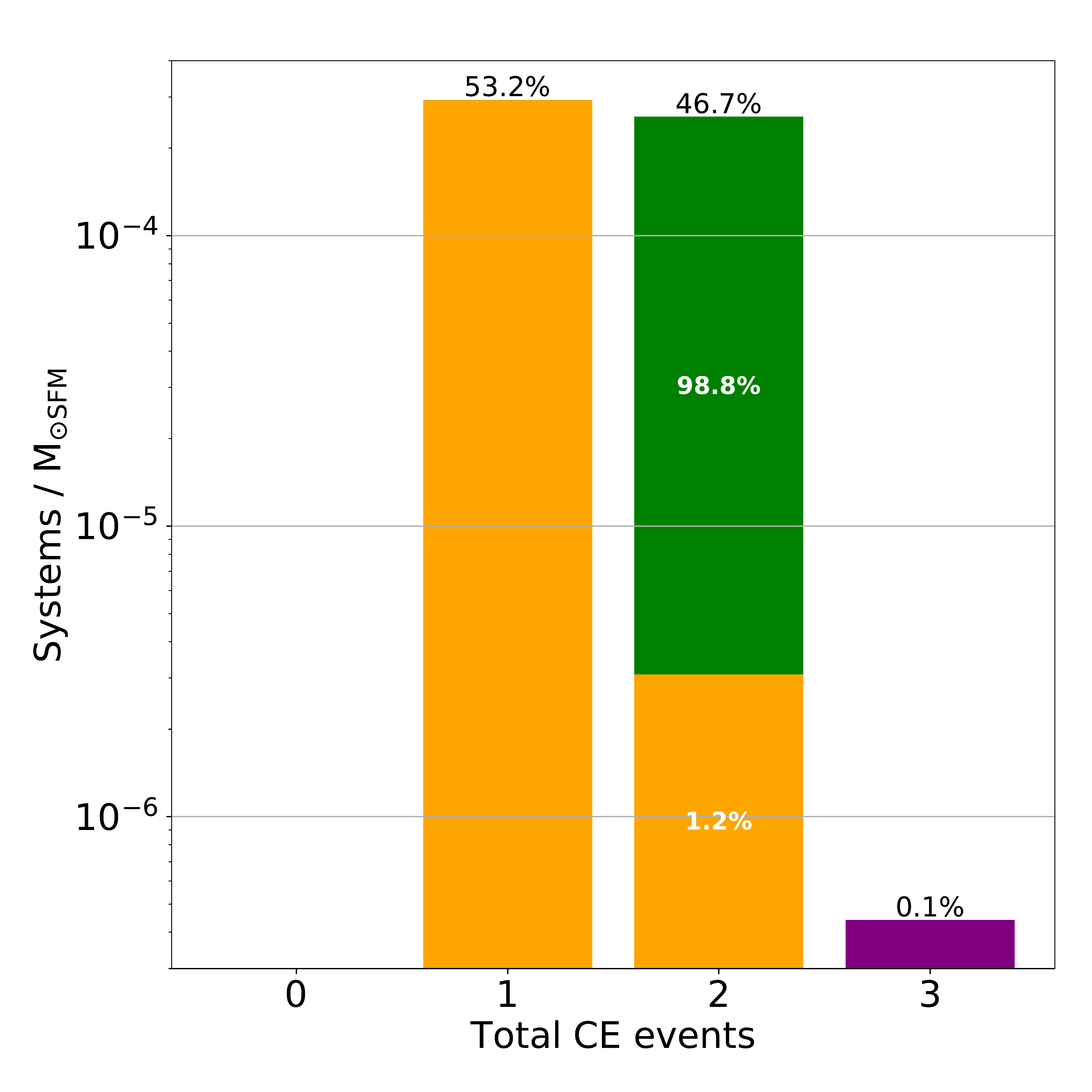}
    \caption{He Novae: system-weighted, excluding H donors}
    \label{fig:barCEeventsnovaeHe_ignoringHdon}
\end{subfigure}%
\begin{subfigure}{0.8\columnwidth}
    \centering
    \includegraphics[width=\textwidth=1]{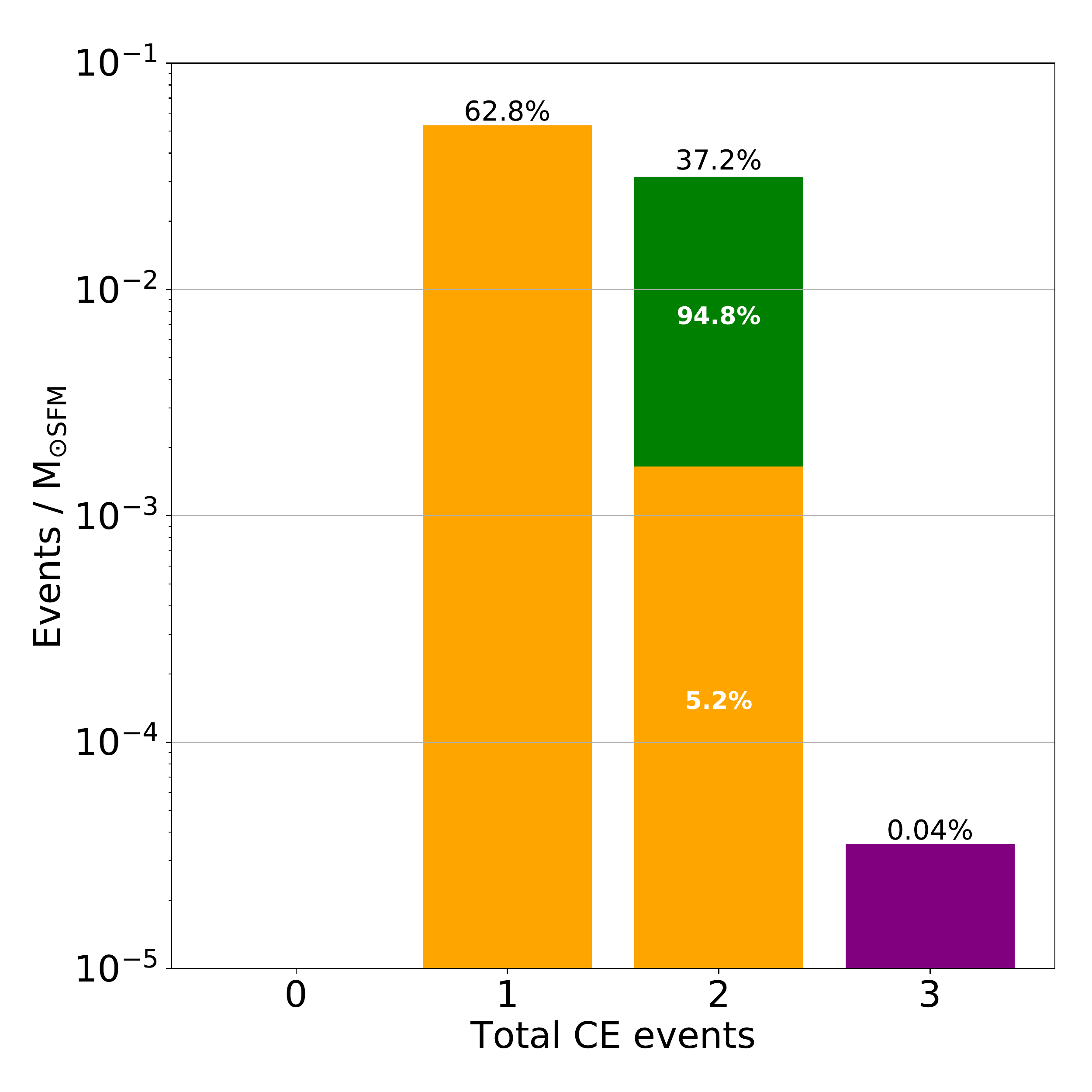}
    \caption{He Novae: event-weighted, excluding H donors}
    \label{fig:barCEeventsnovaeHeEXP_ignoringHdon}
\end{subfigure}

\begin{subfigure}{0.8\columnwidth}
    \centering
    \includegraphics[width=\textwidth=1]{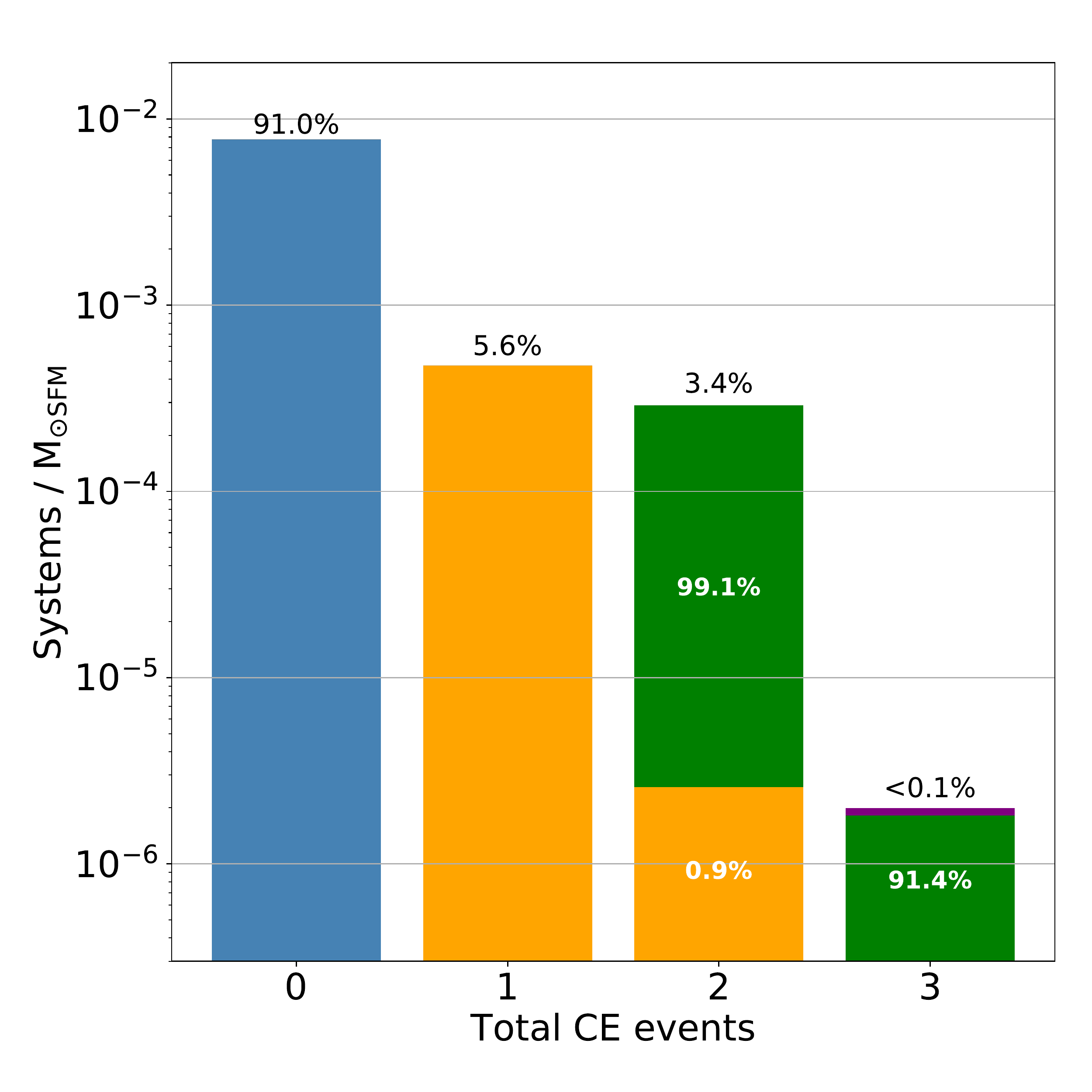}
    \caption{He Novae: system-weighted,  including H donors}
    \label{fig:barCEeventsnovaeHe}
\end{subfigure}%
\begin{subfigure}{0.8\columnwidth}
    \centering
    \includegraphics[width=\textwidth=1]{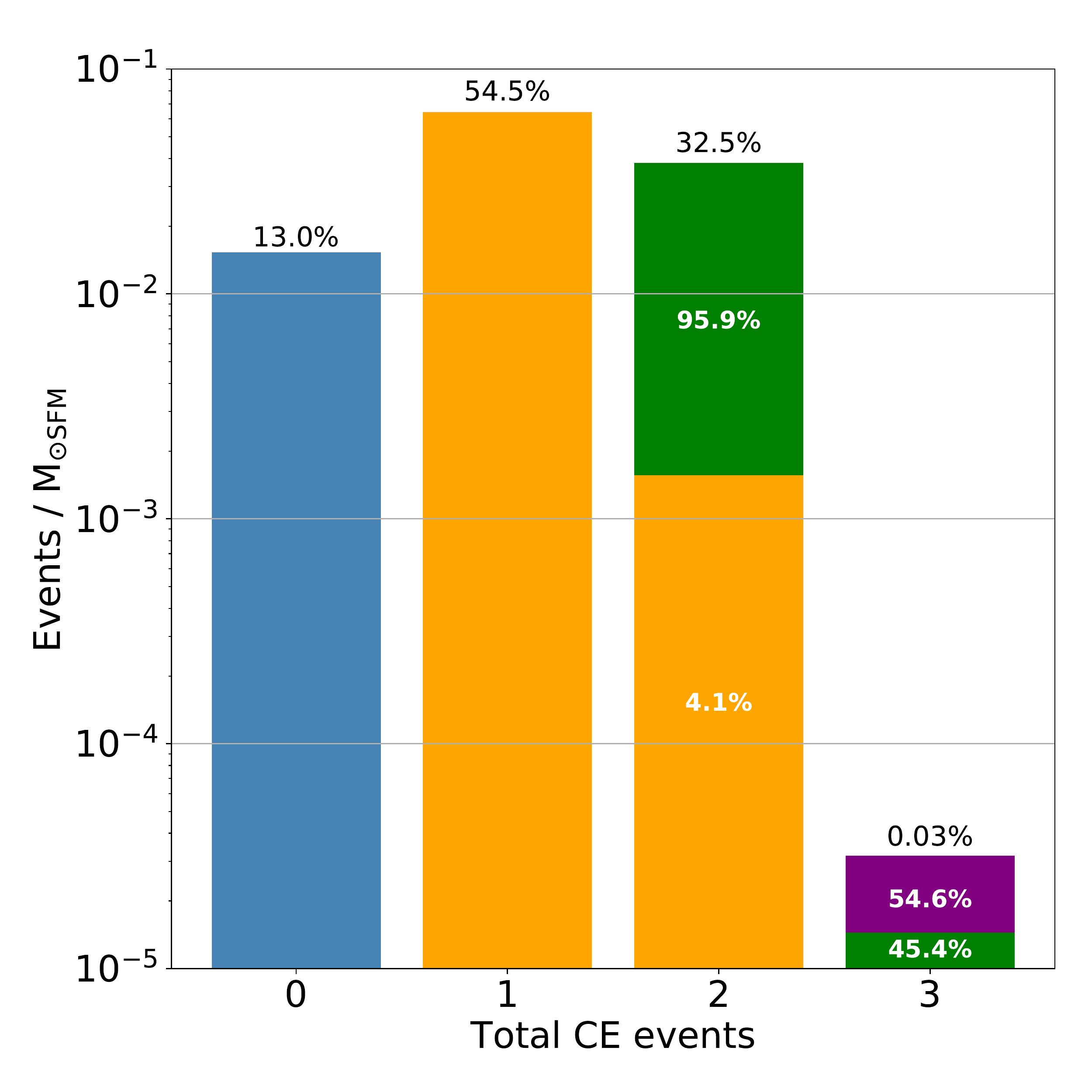}
    \caption{He Novae: event-weighted, including H donors}
    \label{fig:barCEeventsnovaeHeEXP}
\end{subfigure}

\caption{H and He nova counts of common-envelope (CE) events, coloured by the number of CEs that occur prior to the first nova event. The system-weighted figures weight each system equally regardless of the number of nova events that system produces, while the event-weighted figures weight each system according to the total number of novae produced over the systems lifetime.}
\label{fig:barCEeventsnovae}
\end{figure*}

Although both H and He novae are, fundamentally, nuclear explosions on the surfaces of WDs in binaries, the evolutionary channels which govern populations of these systems are distinct. One measure of the importance of binary physics to novae is the number of CE events, shown in Figure \ref{fig:barCEeventsnovae}, which shows the occurrence of CE events weighting each system equally (a `system weighting', Figures \ref{fig:barCEeventsnovaeH}, \ref{fig:barCEeventsnovaeHe_ignoringHdon} and \ref{fig:barCEeventsnovaeHe}) and also according to the total number of novae that occur in each system's lifetime (an `event weighting', Figures \ref{fig:barCEeventsnovaeHEXP}, \ref{fig:barCEeventsnovaeHeEXP_ignoringHdon}, and \ref{fig:barCEeventsnovaeHe}).

When considering the importance of different channels leading to H novae, the distinction between these two weightings is important. Systems which never undergo any CE events dominate the population of nova systems; Figure \ref{fig:barCEeventsnovaeH} shows that only around 12 per cent of all H-nova systems have a CE event occurring prior to the first H nova. However, when considering the number of eruptions each system produces the picture changes substantially, as shown in Figure \ref{fig:barCEeventsnovaeHEXP}. Almost 80 per cent of all H nova eruptions originate from systems which undergo at least one CE event prior to the first H nova eruption, implying that the overall H nova rate can be expected to be quite sensitive to CE physics.

Returning to Figure \ref{fig:barCEeventsnovaeH}, the small ($\approx 15$ per cent) fraction of H-nova systems that do undergo a CE event at some point in their lives experience either 1 or 2 CEs, with a 0.1 per cent minority undergoing 3. Of those that undergo exactly 1 CE, roughly 70 per cent do so prior to the first H novae; therefore only 70 per cent of these `one CE' systems produce H novae which are potentially influenced by CE physics. The systems which undergo 2 CEs are more complex. A small fraction (6 per cent) have both CEs occurring after the first H nova, implying complete insensitivity to CE physics, while around 60 per cent undergo 1 CE prior before the first H nova and their second CE after, and the remaining 30 per cent have both CEs occurring prior to the first H nova.

Under the assumption that CE events eject the entirety of either a H or He envelope, which may not be true in general \citep{yoon2017,gotberg2017}, we can make some comments about the evolutionary sequences surrounding the information shown in Figure \ref{fig:barCEeventsnovae}. In order for two CEs to occur prior to the first H nova, the primary must lose (and survive the loss of) its H and its He envelope. Therefore the primary must have its first CE on either the HG, FGB, CHeB or EAGB\footnote{The He layer in TPAGB stars is assumed to be too thin to form a true He envelope, and a CE event in this evolutionary phase instead produces a WD remnant.} such that it is stripped to become a He star. Its second CE occurs when the He star evolves off the HeMS after forming a C/O or O/Ne core, leaving it as a WD that can then undergo H novae. This is the only permissible evolutionary channel (assuming total stripping of the H envelope in the first CE), as the secondary cannot undergo its first CE event until after the first H nova has occurred, or it would be left without H to transfer onto a WD companion to produce H novae. This requirement of a hydrogen atmosphere on the donor star during H novae leads us to the expectation that at most 2 CEs can occur prior to H novae in isolated binary systems. This expectation is supported by the tiny fraction of H-nova systems that undergo 3 CEs, all of which have 2 CEs occurring prior the first H nova, with their third CE phase occurring subsequently.

CE phases are found to be ubiquitous in He-nova systems where the white dwarf is accreting material from a He star (Figures \ref{fig:barCEeventsnovaeHe_ignoringHdon} and \ref{fig:barCEeventsnovaeHeEXP_ignoringHdon}. We find that all such He-nova systems can be considered post-CE systems, without exception. Slightly over half of these systems experience one CE event prior to the first eruption, and the vast majority of the remainder experience two. This is unsurprising, as we require at least one stripping episode in order to produce a He-donor star. Furthermore, He stars tend to be far smaller than their H cousins \citep{divine1965,woosley1995,pols2002,dewi2002,laplace2020}, and so require tighter orbits for mass transfer to occur.

Only a very small percentage of these He-nova systems ($\lesssim$ 0.5 per cent) undergo a CE event after He novae have occurred. The minuscule population ($\approx$ 0.1 per cent) of systems which undergo three CE events undergo all three prior to the first He nova.

When considering the event-weighted importance of the different channels towards He novae where the donor is a He star, there are only minor differences from the system-weighted case. Therefore, we conclude that whether a He-nova system undergoes one or two CE events prior to the first He nova is not the dominant factor determining the number of He novae that the system produces.

Figures \ref{fig:barCEeventsnovaeHe} and \ref{fig:barCEeventsnovaeHeEXP} show the numbers of common-envelope events across all He nova systems, including those where He novae occur while accreting material from H donors. We find that over 90 per cent of all systems never undergo a common envelope, reflecting the distribution of H-nova systems, which greatly outnumber the He-accreting He-nova systems. When considering the event-weighted distributions, there are only relatively minor discrepancies between the distribution found when considering only He-accreting He-nova system, reflecting the fact that H-accreting white dwarfs produce relatively few He novae ($\approx 15$ per cent).

Despite the majority of all H- and He-nova systems never undergoing a CE event, both H- and He-nova rates are expected to be sensitive to CE physics. This makes both nova rates potentially useful in ongoing efforts to understand these highly uncertain phases of binary evolution in the mass range that contributes to novae.

\subsection{Initial System Properties}

\begin{figure*}
     \centering

\begin{subfigure}{0.85\columnwidth}
    \centering
    \includegraphics[width=\textwidth=1]{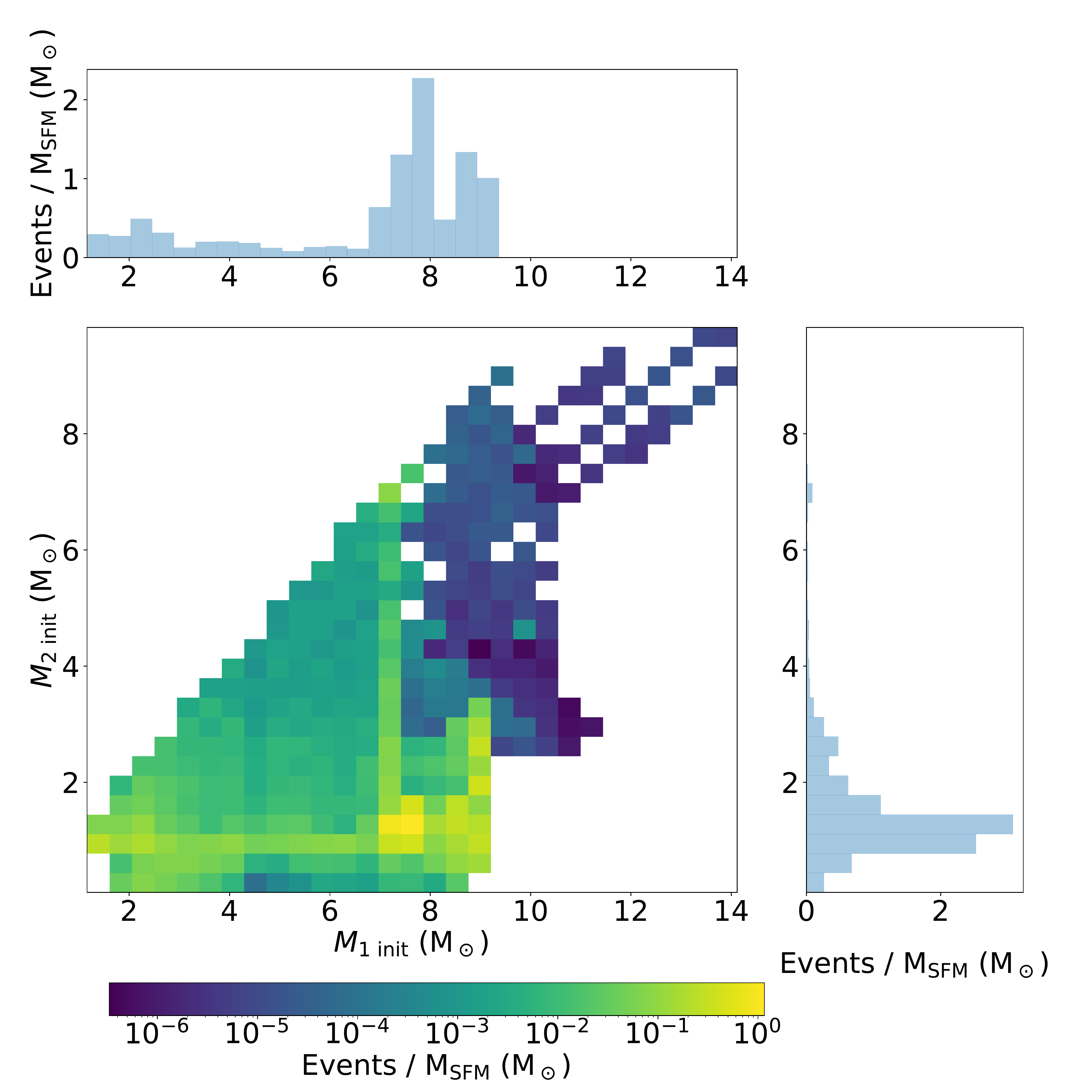}
    \caption{H novae: initial primary mass vs initial secondary mass}
\end{subfigure}%
\begin{subfigure}{0.85\columnwidth}
    \centering
    \includegraphics[width=\textwidth=1]{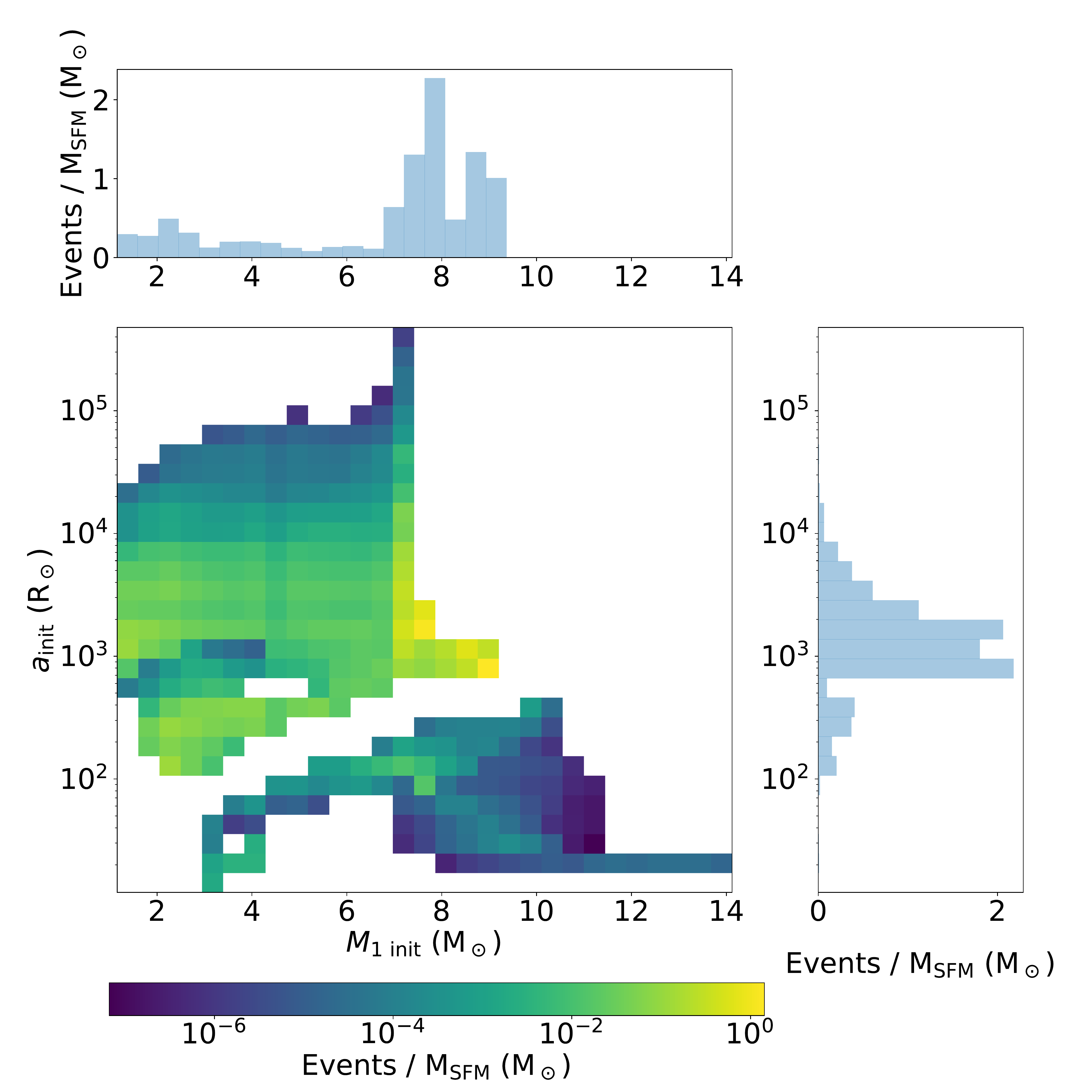}
    \caption{H novae: initial primary mass vs initial separation}
    \label{fig:hist2DnovainitialpropertiesHnovam1initvsainit}
\end{subfigure}

\begin{subfigure}{0.85\columnwidth}
    \centering
    \includegraphics[width=\textwidth=1]{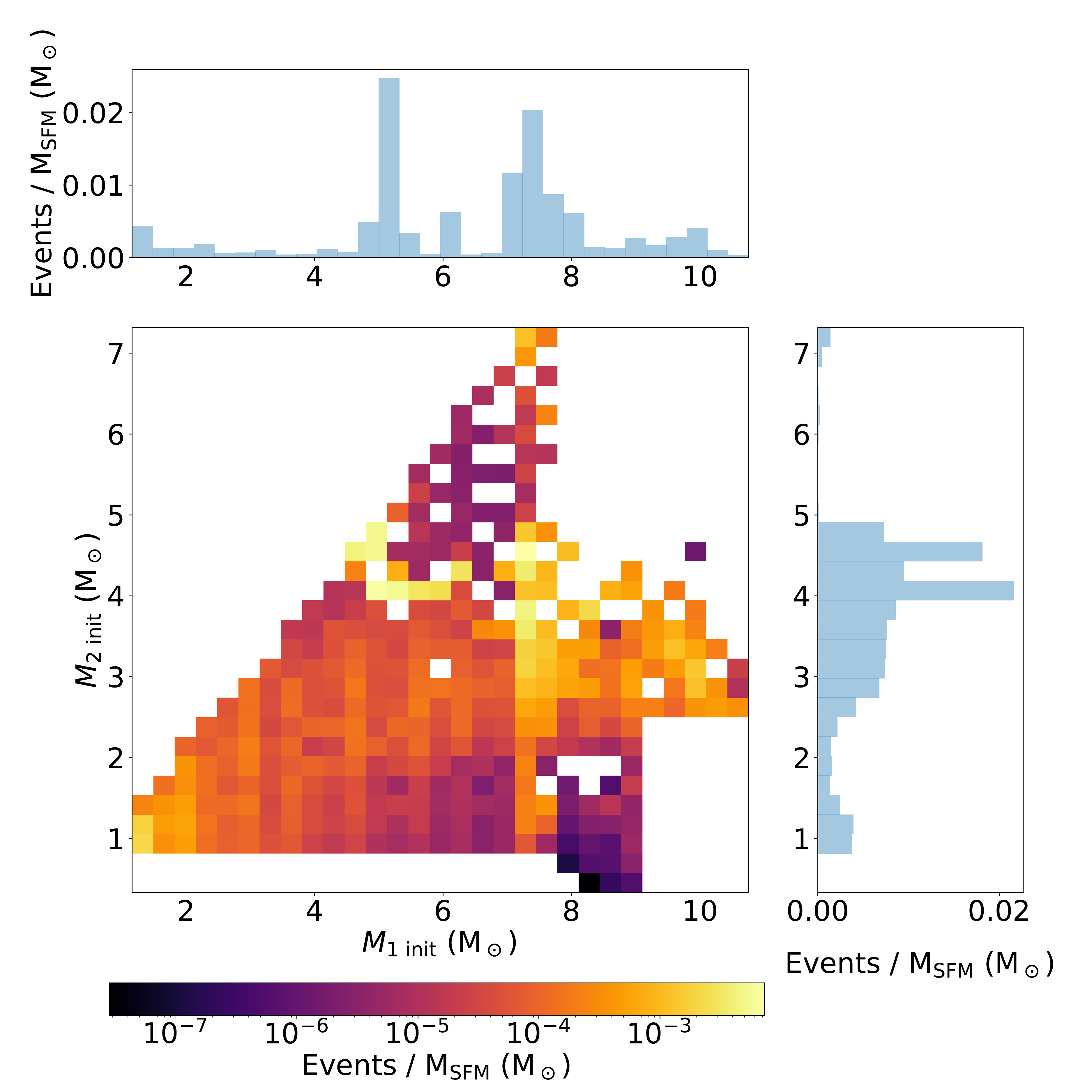}
    \caption{He novae$^{\rm *}$: initial primary mass vs initial secondary mass}
\end{subfigure}%
\begin{subfigure}{0.85\columnwidth}
    \centering
    \includegraphics[width=\textwidth=1]{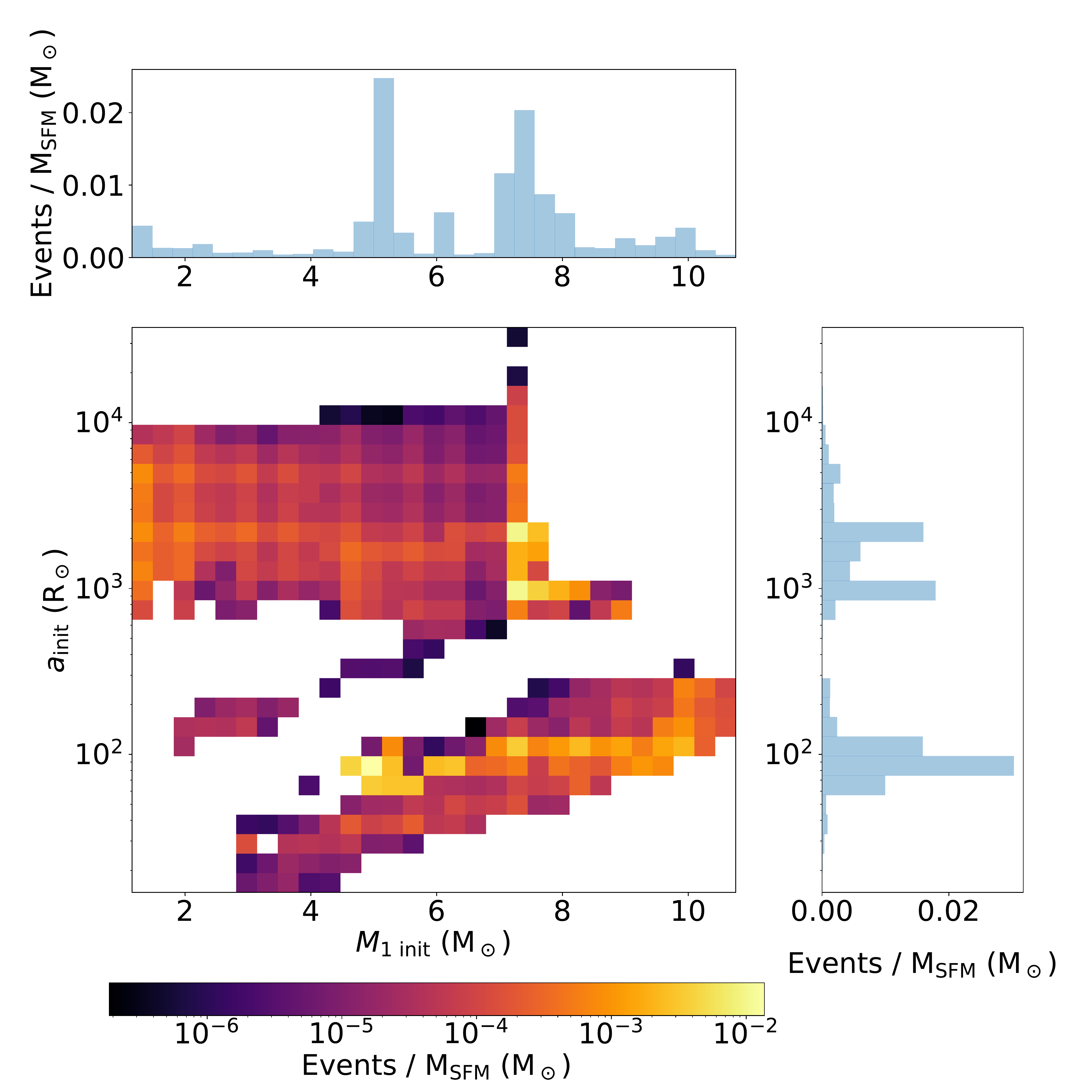}
    \caption{He novae$^{\rm *}$: initial primary mass vs initial separation}
    \label{fig:hist2DnovainitialpropertiesHenovam1initvsainit}
\end{subfigure}

\begin{subfigure}{0.85\columnwidth}
    \centering
    \includegraphics[width=\textwidth=1]{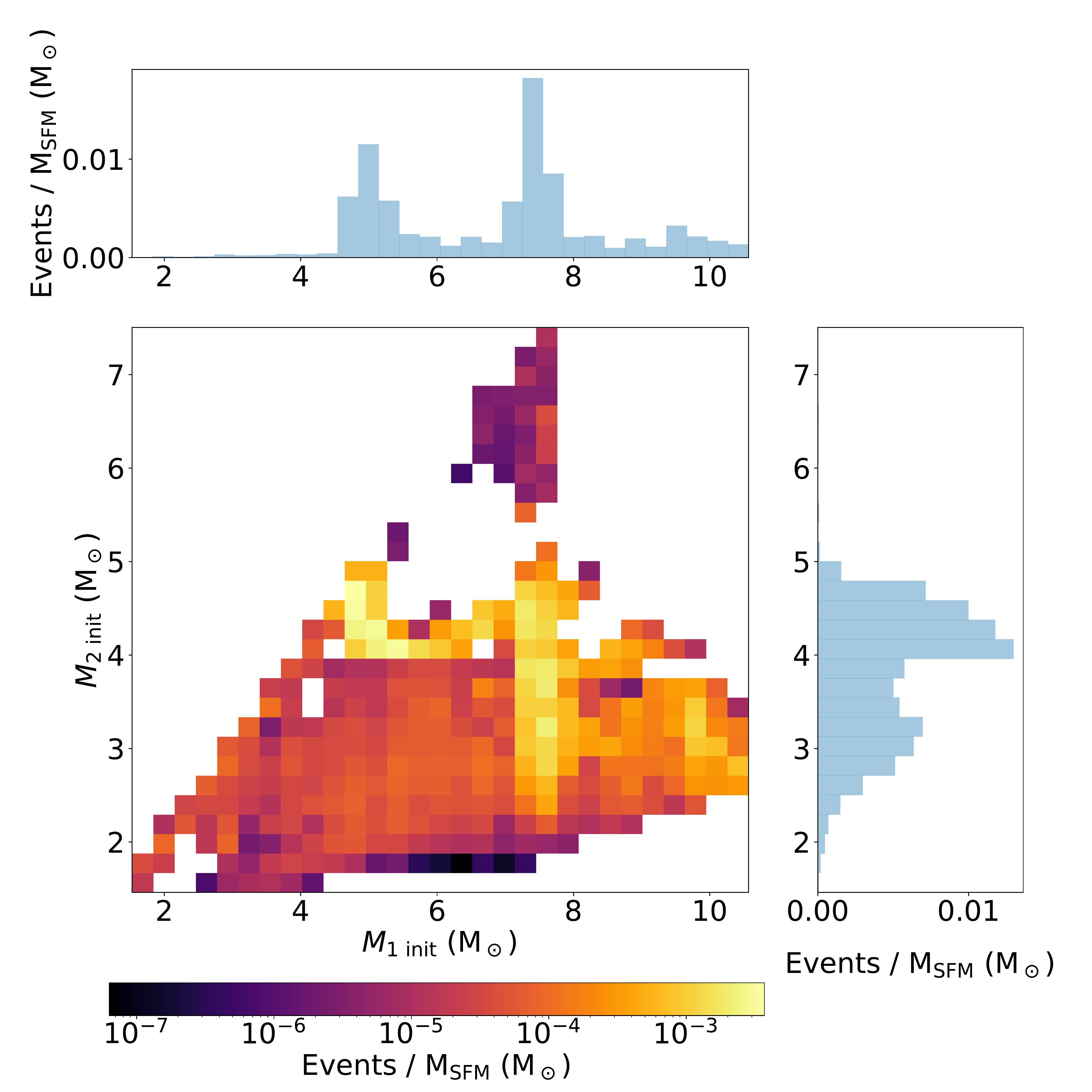}
    \caption{He novae$^{\rm **}$: initial primary mass vs initial secondary mass}
\end{subfigure}%
\begin{subfigure}{0.85\columnwidth}
    \centering
    \includegraphics[width=\textwidth=1]{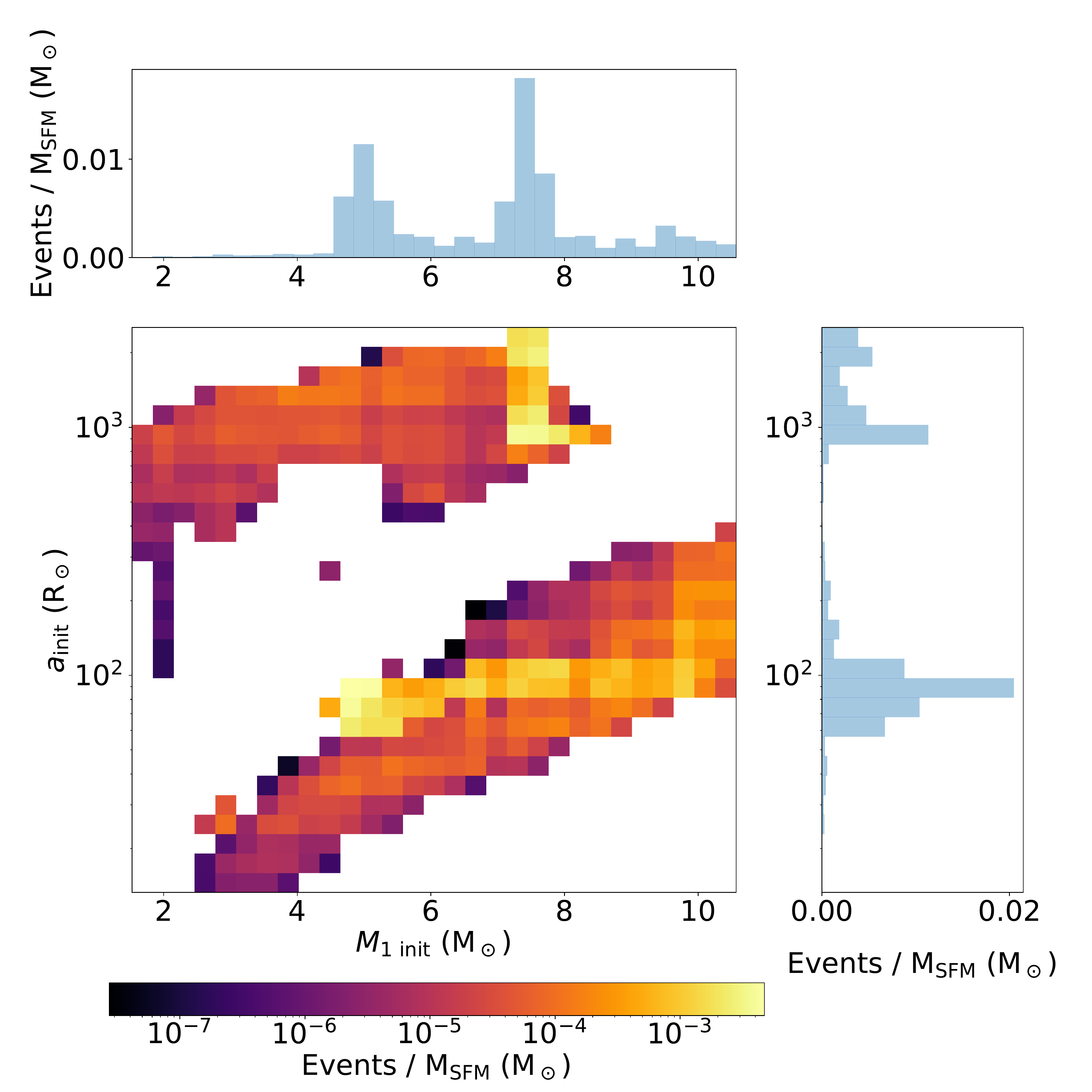}
    \caption{He novae$^{\rm **}$: initial primary mass vs initial separation}
    \label{fig:hist2DnovainitialpropertiesHenovam1initvsainit_ignoringHdon}
\end{subfigure}
\caption{Distributions of initial secondary masses and orbital separations plotted against initial primary masses for systems undergoing H and He novae. Systems are weighted according to the number of novae each system produces over the 15 Gyr simulation time. "$^{\rm *}$" / "$^{\rm **}$": including / excluding H donors.} 
\label{fig:hist2Dnovainitialproperties}
\end{figure*}

The distributions of initial binary system parameters which produce H and He novae are shown in Figure  \ref{fig:hist2Dnovainitialproperties}, weighted by the total number of nova eruptions produced by these systems.

The distribution of masses reveals most H nova eruptions originate from relatively massive primaries with initial masses in the range of of 7-9 M\solar\ and lower mass ($\lesssim 2.5$ M\solar) secondaries.
The high initial primary mass peak around 7-9 M\solar\ is largely driven by systems which undergo a CE event caused by the expansion of the primary, usually as the star approaches the end of the TPAGB, which produces a hardened binary with a massive WD and a MS companion. This WD is then able to accrete material from its binary companion as it leaves the main sequence, with the first H novae from this channel typically produced by either MS or HG donors. The combination of a hardened binary with a massive WD accretor allows novae in these systems to occur with very low recurrence times (see Section \ref{sec:rectimes}). It is the large number of novae these systems produce that overwhelms the effect of the initial mass function (IMF), which favours the birth of lower mass systems. This population of systems is readily identifiable in Figure \ref{fig:hist2DnovainitialpropertiesHnovam1initvsainit}.

In conventional single stellar evolution, stars more massive than around 9-10 M\solar\ are not expected to form WD remnants, but this is not necessarily the case when such a star is placed in a binary. A small number of novae are produced from primaries initially more massive than 10 M\solar\ when $a_{\rm init}$ is small. In these systems, the primary experiences significant mass transfer on the main sequence, reducing the final mass of the He core sufficiently to produce a C/O or O/Ne WD which goes on to accrete from its companion star and produce nova eruptions.

The distribution of the initial primary masses of He-nova systems reveals a bimodal distribution with one peak around $M_{1\rm \ init}$ of  5-5.5 M\solar\ driven by high initial mass ratio binaries ($M_{2\rm \ init}>4$ M\solar), and the other around 7.5 M\solar, driven by a wider range of lower mass secondaries (2.5-5 M\solar). This distribution changes little when comparing between Figure Examination of Figures \ref{fig:hist2DnovainitialpropertiesHenovam1initvsainit} and \ref{fig:hist2DnovainitialpropertiesHenovam1initvsainit_ignoringHdon} reveals these two peaks to be driven by two distinct channels, one on each side of the gap in viable initial separations.

The H accreting He nova channel primarily manifests itself in Figure \ref{fig:hist2Dnovainitialproperties} through the introduction of progenitors of low initial primary (<4 M\solar) and secondary (<1.5 M\solar) masses, as well as opening up nova progenitor systems with higher initial orbital separations ($\gtrsim 2.5 \times 10^3$ R\solar).

The origin of the 7.5 M\solar\ peak is similar to the channel for the previously described peak in the H nova distribution, where a massive WD is formed from an initially massive star which ends its life in a CE event on the TPAGB, leaving the massive WD in close proximity to its future donor star. The main difference is that a second CE event occurs as the secondary ascends the first giant branch, leaving a HeMS donor star in an extremely tight orbit with a massive WD. This configuration provides ideal conditions for He novae.

The second peak from 5-5.5 M\solar\ is driven by systems which avoid the first CE event by losing the primary's envelope predominantly through stable, conservative mass transfer shortly after leaving the main sequence, leaving it as a relatively massive He star which eventually produces a WD of approximately 1-1.1 M\solar. This episode of mass transfer widens the binary significantly, but also increases the mass of the secondary. When it eventually undergoes CE on the FGB to produce the HeMS donor star for the system, it has a sufficiently massive H envelope to tighten the orbit again so that the HeMS donor may fill its Roche lobe and accrete onto its companion WD.

Figures \ref{fig:hist2DnovainitialpropertiesHnovam1initvsainit}, \ref{fig:hist2DnovainitialpropertiesHenovam1initvsainit}, and to a lesser extent \ref{fig:hist2DnovainitialpropertiesHenovam1initvsainit_ignoringHdon}, show a well defined `desert' region dividing the distribution of initial separations ($a_{\rm init}$) of H and He-nova systems that is related to the initial mass of the primary. Close examination of H-nova systems above the desert reveals that systems with $a_{\rm init}\lesssim10^3$ R\solar\ commonly undergo a CE event prior to H novae. Above this threshold no CE events occur and H novae are produced almost exclusively from WDs accreting from the winds of TPAGB donor stars. When systems below this threshold undergo a CE event they will either merge or survive to form a hardened binary which may go on to experience nova eruptions. The upper bound of the desert marks the minimum orbital separation for which the binary can survive and go on to produce novae, while the lower bound of the desert marks the beginning of populations of H-nova systems which do not undergo a CE phase prior to H novae by avoiding this fate through earlier episodes of mass transfer. The similar `desert' region which exists for He-nova systems is also due to the threshold beyond which CEs caused by the primary's expansion become inevitable and unsurvivable, and below which this CE phase may be avoided.

Finally, it should be mentioned that when not weighting by the nova count of each system (i.e., applying a system weighting rather than an event weighting), the distributions of both H and He novae are heavily skewed towards low mass primaries, as expected due to the IMF favouring the birth of lower mass systems.

\subsection{Remnant States}
\subsubsection{H novae}

\begin{figure*}
     \centering
\begin{subfigure}{1\textwidth}
    \centering
    \includegraphics[width=0.8\textwidth]{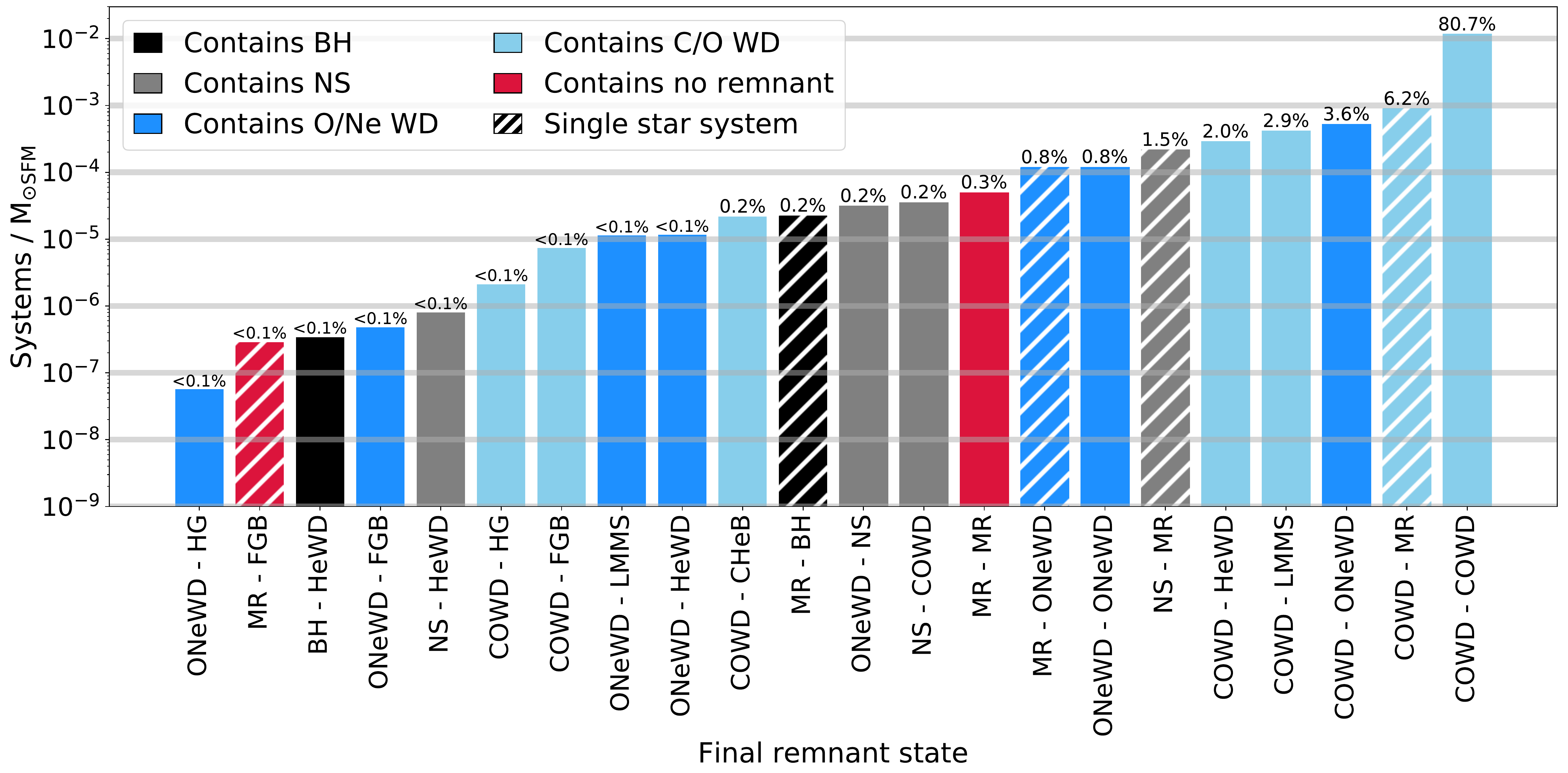}
    \caption{H-nova systems}
    \label{fig:finalbinarystateHnova}
\end{subfigure}

\begin{subfigure}{1\textwidth}
    \centering
    \includegraphics[width=0.8\textwidth]{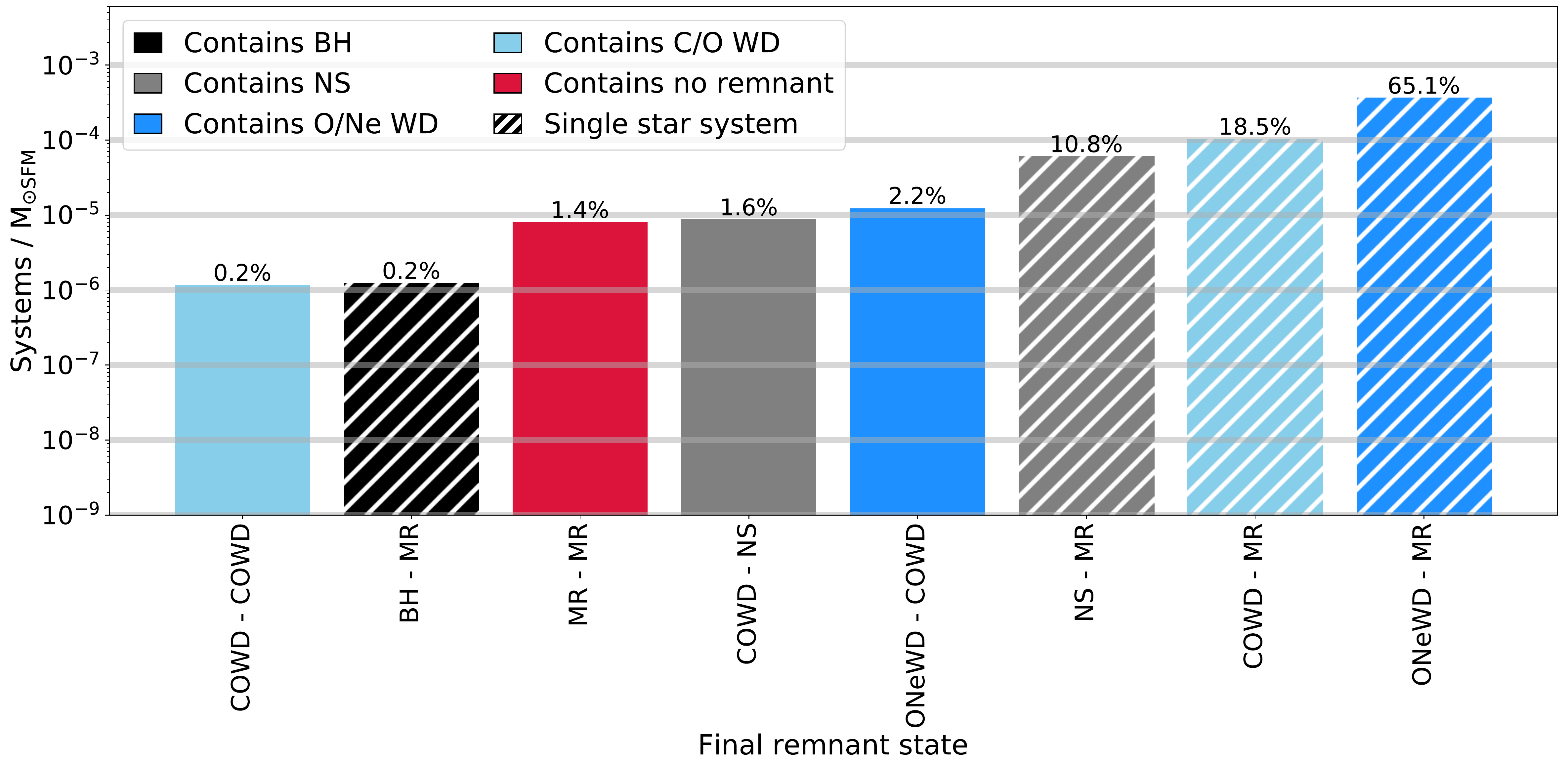}
    \caption{He-nova systems, excluding H donors}
    \label{fig:finalbinarystateHenova}
\end{subfigure}
     \caption{Final remnant states of all systems which experienced at least one H or He nova. Each bar is coloured according to the most exotic star in the binary, with massless remnants (MR) given the lowest priority. Diagonally striped bars denote the systems where the final remnant is a single star, having undergone either a supernova or a merger event.
}
\label{fig:finalbinarystate}
\end{figure*}

There are diverse final states in H-nova systems, as shown in Figure \ref{fig:finalbinarystateHnova} which shows the final evolutionary states of all systems which underwent at least one H nova. By a full order of magnitude the most common binary state is a double C/O WD binary, with over $10^{-2}$ systems created per M\solar\ of stars formed. Significant numbers ($\gtrapprox 10^{-4}$ systems per M\solar) of C/O WD-O/Ne WD, C/O WD-He WD, LMMS-C/O WD, and double O/Ne WD binaries are also formed, in addition to single NSs and C/O WDs. The population of LMMS-C/O WD binaries in particular are noteworthy as they represent the most common binary configuration which will continue to produce novae beyond the 15 Gyr scope of the simulation.

Single O/Ne WDs are formed at a comparable rate to single black holes, WD-NS binaries, and binaries where both stars have been destroyed in a double-degenerate type Ia supernova. The single black hole population is dominated by the gravitational-wave driven merger of NS-WD binaries, supplemented by contributions from Algol systems where the WD primary merges via a final CE with its companion star, which has grown to approximately 20 M\solar\, and  subsequently forms a black hole in a stripped supernova event.

\subsubsection{He novae}

\begin{figure}
    \centering
    \includegraphics[width=0.8\columnwidth]{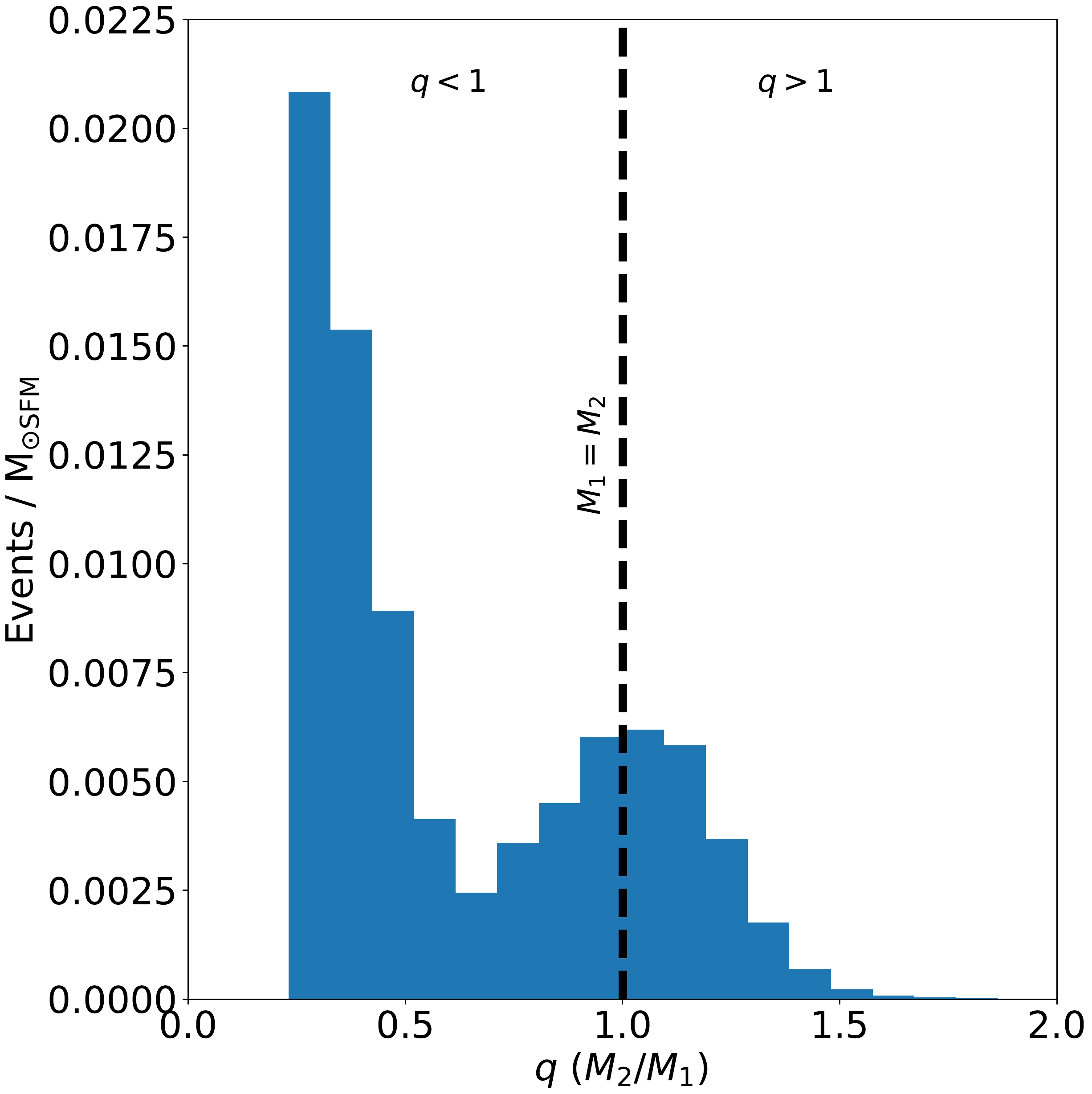}    
    \caption{Distribution of the mass ratio $q=M_2/M_1$ at the time of each He nova, excluding H donors. When q<1, the effect of conservative mass transfer is to widen the binary. However, sources of orbital decay such as significant gravitational-wave emission may overcome this effect even when the system is actively mass transferring, depending on the orbital separation and the mass ratio of a system.}
    \label{fig:qhistatHenova}
\end{figure}

Figure \ref{fig:finalbinarystateHenova} shows the distribution of binary remnant states of all systems which undergo at least one He nova under conditions of He accretion. When considering all He-nova systems irrespective of whether the nova occurs under conditions of H or He accretion, the distribution roughly mirrors that of H novae, being dominated by double-WD systems with a selection of more exotic binaries being produced in small numbers. Subsequent discussion in around the remnant states of He novae will be restricted to He-accreting systems.

When considering only the He-accreting channel, however, an interesting feature of these systems becomes apparent. The vast majority of systems which undergo He novae end their lives as single systems. This is surprising, as He novae do not rely on either the merger of two stars or the destruction of one of the stars, as type Ia supernovae do. The reason behind this result is complex. Here we outline some of the key physics determining whether a given He-nova system ends its life in a binary or single star configuration.

Almost all He novae are found to be produced by accreting material from a HeMS donor star. An obvious explanation of the tendency of these systems to end their lives as single stars is that they undergo a final CE event when the HeMS donor star evolves off the main sequence. Because of the close orbital separations required for mass transfer in these systems, this final CE event would result in a merger of the donor star and the accretor. In our simulations this scenario only occurs in very few systems, demonstrated in Figure \ref{fig:barCEeventsnovae} by the very low number of systems that undergo a CE event after the first He nova. Further, \cite{halabi2018} demonstrate that stable mass transfer can occur for a wide range of binary parameters in He-shell burning, post-CE binaries. The answer, then, lies elsewhere.

Whether an orbit widens or shrinks as mass is transferred conservatively from the donor to the accretor is determined by the mass ratio $q=M_{\rm donor}/M_{\rm accretor}$. If $q<1$, then the effect of mass transfer is to widen the orbit, and if $q>1$, the effect is to shrink the orbit. Figure \ref{fig:qhistatHenova} shows the distribution of $q$ at the time of He nova for all nova eruptions. Most of the He novae occur in systems with $q<1$, therefore acting to drive the two stars apart. This effect is responsible for the predicted `bounce' of AM CVn binaries, where mass transfer from a low mass (typically 0.2-0.26 M\solar\footnote{AM CVn binaries are typically only observed post-bounce, with He donor masses 0.01-0.1 M\solar.} at the time of the bounce) He star accreting onto a WD causes the binary to widen after a minimum period is reached \citep{kraft1962,yungelson2008,solheim2010,neunteufel2020}.

In the simulated He-nova systems, however, this bounce does not occur, and instead the binary merges. The difference between these two scenarios is the mass of the donor star. The lowest mass donor star involved in a He nova in our standard physics case is 0.32 M\solar, significantly higher than typical AM CVn binaries. In this scenario, angular momentum loss through gravitational-wave emission overcomes the effect of mass transfer and the binary inspirals. 

As it does so and the orbital frequency increases, tides transfer more orbital angular momentum into the spins of the stars, hastening the inspiral. He nova outbursts transfer material away from the system (we assume no interaction occurs between the nova ejecta and the companion star) and with it angular momentum, further acting to drive the binary towards merger.

Even in systems where orbital shrinkage does not occur, the binary state of the system is still not safe. For the binary to survive, it must avoid growing a C/O WD accretor to \Mchand. The fact that O/Ne WDs instead collapse into a NS upon growing to \Mchand\ is the reason that the C/O WD-NS remnant population is so large, representing approximately half of all He-nova systems which survive in a binary remnant state. Further, the system may destroy itself entirely in a double-degenerate type Ia supernova, producing the double massless remnant (MR-MR) populations in Figure \ref{fig:finalbinarystate}. It is noteworthy that in our simulations, more double-degenerate type Ia systems are produced from post-H-nova systems than post-He-nova systems.

With all this acting to prevent the survival of He nova binaries, it may be wondered how those few systems that do remain in a binary configuration survive. The accretion induced collapse of an O/Ne WD into a neutron star is primarily responsible for the population of the survivors with NS elements; the remainder, comprised of C/O WD-O/Ne WD pairings, is more complicated. 

Of this population, almost all involve He novae occurring in systems involving mass transfer from an evolved He star. The importance of this is that, unlike the HeMS, which often lasts long enough to either merge the stars through inspiral or grow them to \Mchand, this phase of evolution typically does not. Under the competing influences of core growth and envelope mass loss, these systems often stop mass transferring relatively quickly \citep{halabi2018}, leaving the system in a binary configuration. Further, these systems tend to have wider orbital separations, greatly reducing angular momentum loss through gravitational waves. However, it should be noted that mass transfer from an evolved He star does not guarantee the survival of a binary, it only makes it far more probable.

\subsection{Nova Physics and System Properties at Nova} 

\subsubsection{White Dwarf Mass}

\begin{figure}
\centering
\begin{subfigure}{0.8\columnwidth}
    \centering
    \includegraphics[width=\textwidth=1]{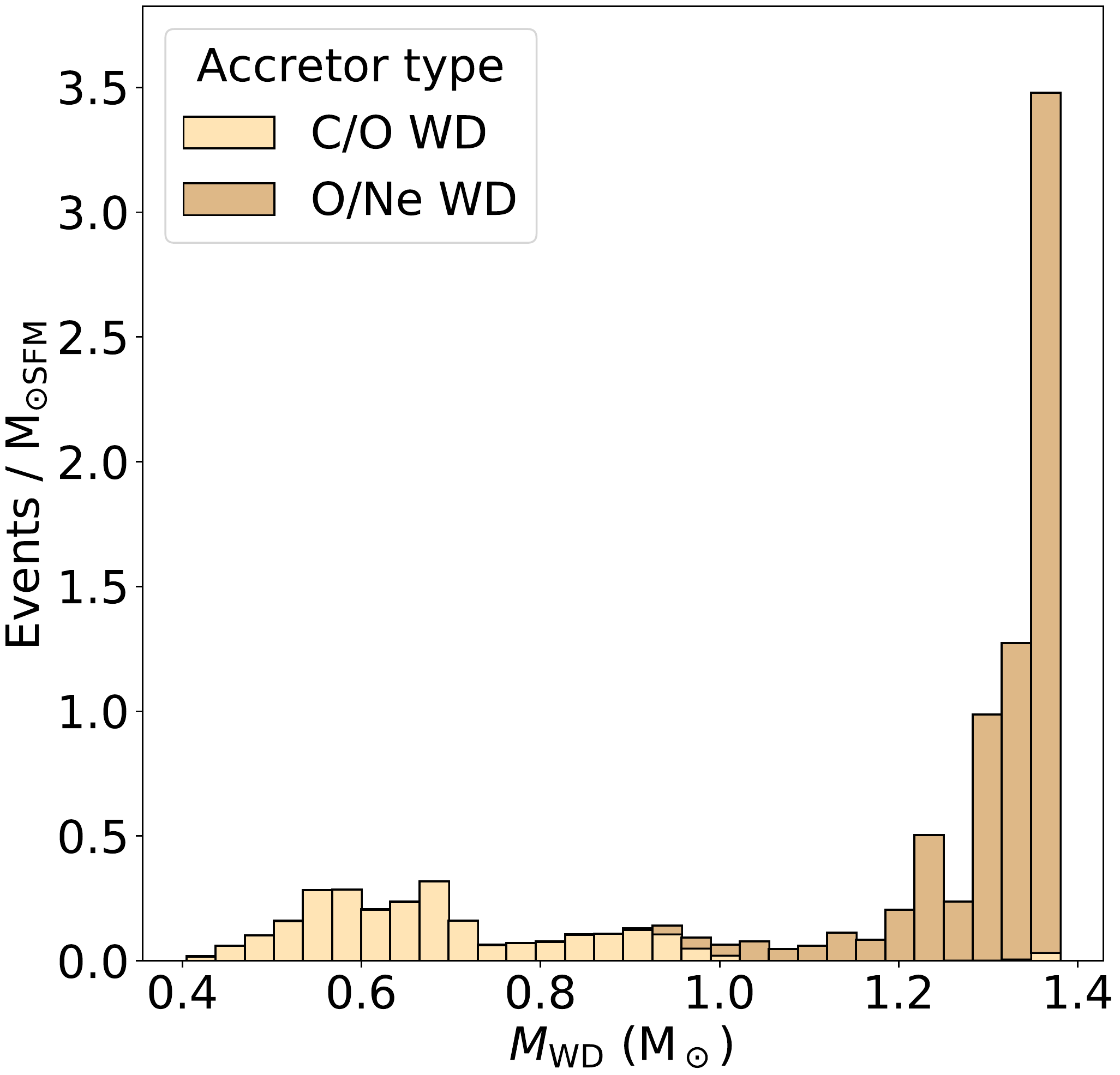}
    \caption{H novae}
\end{subfigure}

\begin{subfigure}{0.8\columnwidth}
    \centering
    \includegraphics[width=\textwidth=1]{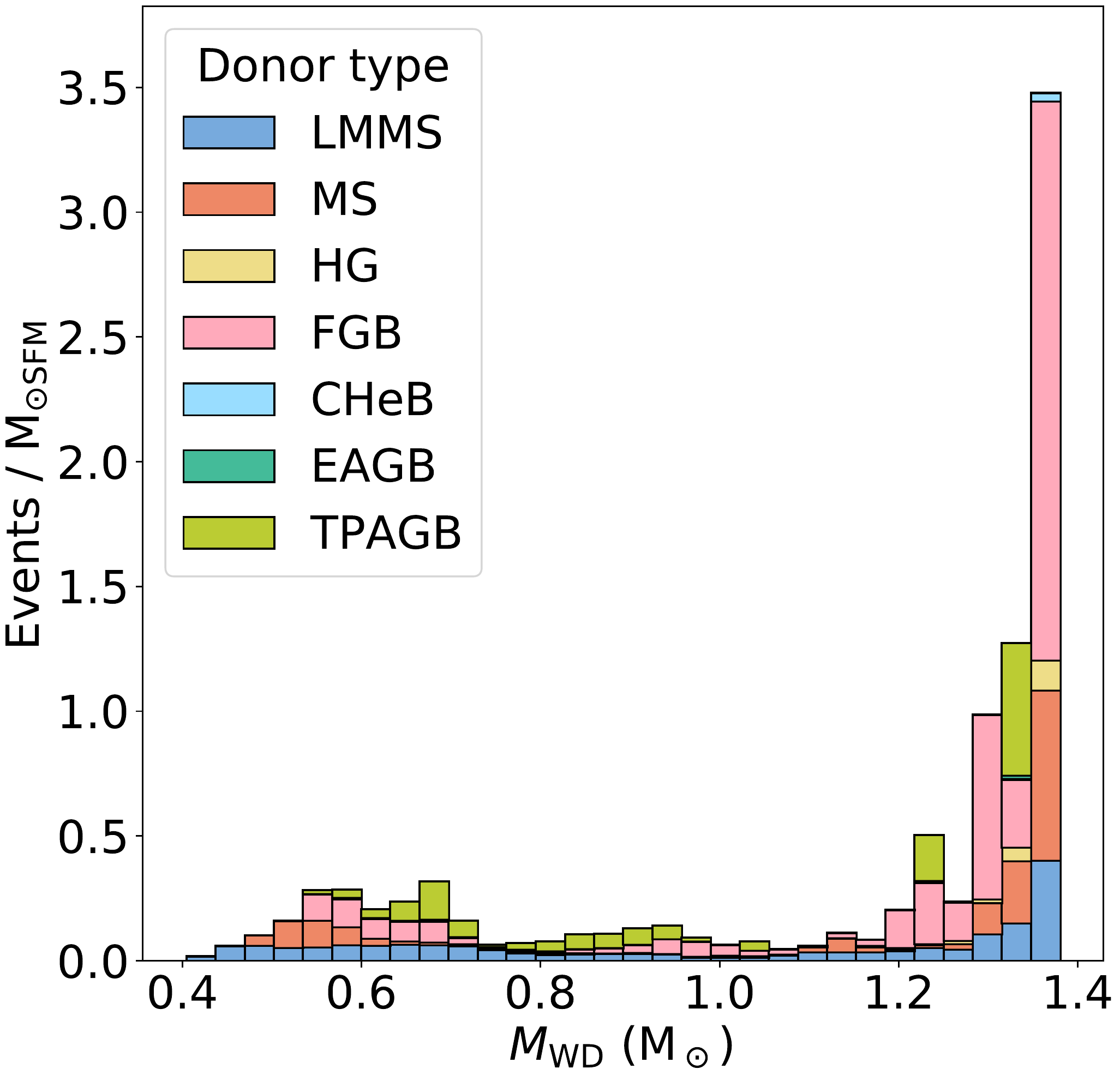}
    \caption{H novae}
\end{subfigure}

\begin{subfigure}{0.8\columnwidth}
    \centering
    \includegraphics[width=\textwidth=1]{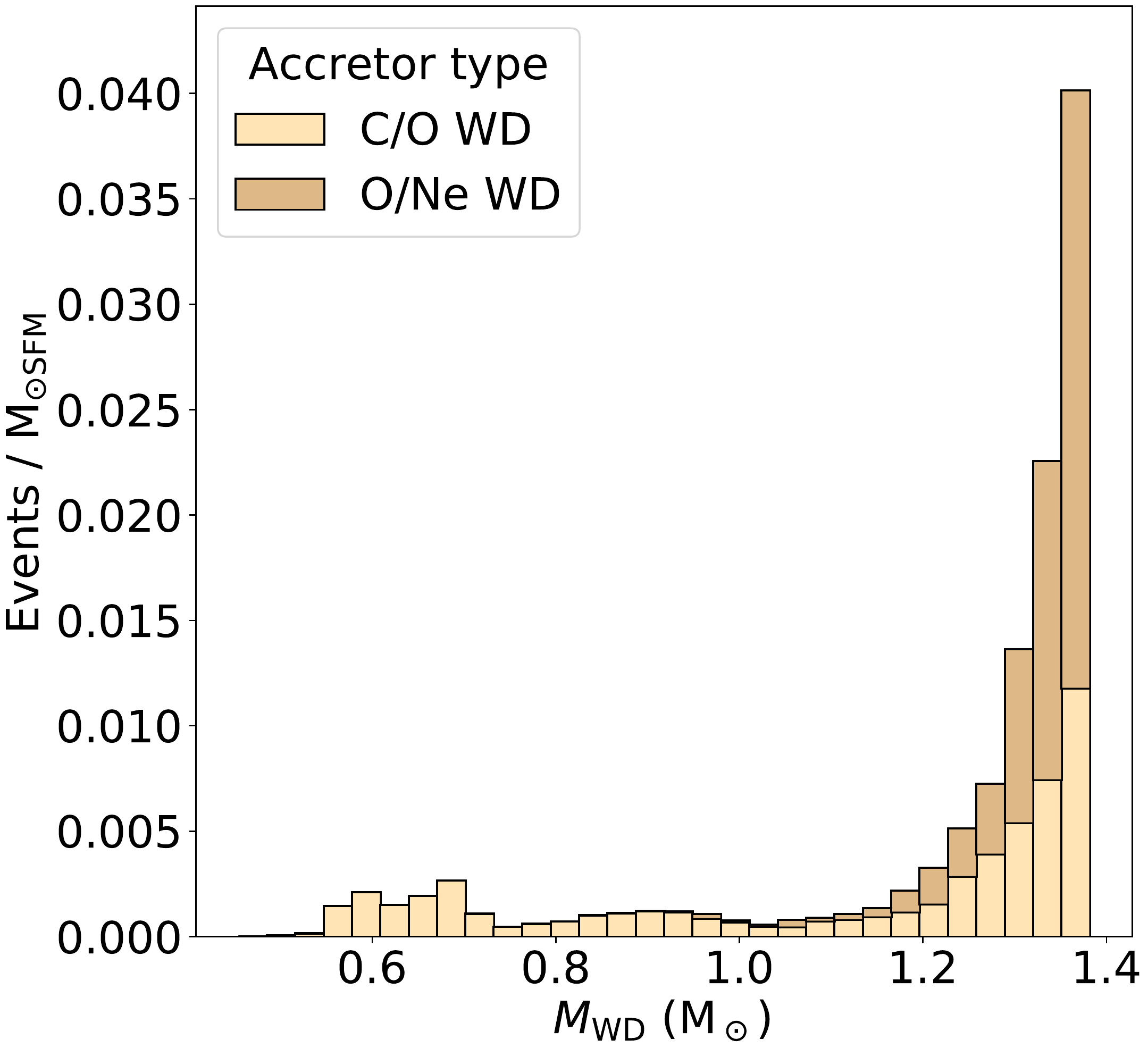}
    \caption{He Novae, including H donors}
\end{subfigure}
     \centering
     \caption{Distribution of white dwarf masses at the time of each nova eruption, coloured by the accretor and donor stellar type. The distributions of both H and He novae are found to heavily favour high mass white dwarfs. He novae from H donors are dominant for \Mwd<0.8 M\solar; otherwise, HeMS donors dominate.}
     \label{fig:histnovam1}
 \end{figure}

Figure \ref{fig:histnovam1} shows the distribution of WD masses at the time of nova eruption for H and He novae.

The vast majority of H novae occur on rarer, high mass WDs, with the distribution strongly peaked as the WD mass approaches \Mchand. In these most extreme WDs, the greatly increased surface gravity reduces the critical ignition mass of the surface H layer to the extent that the number of novae per system overcomes the IMF bias towards lower mass WD systems. Most of these novae are driven by accretion from evolved FGB and TPAGB donor stars; the combination of the higher accretion rates from these more evolved donors and the high WD mass creates ideal conditions for short-recurrence time novae, allowing high event counts despite the relatively short lived nature of the donor stars. A smaller secondary peak around 0.6 M\solar\ is also observed, approximately aligning with the peak in the observed mass distribution of single WD systems \citep{giammichele2012,tremblay2016}. The composition of the accretor WDs transitions from C/O to O/Ne around 0.9 M\solar.

He nova events are also dominated by accreting WDs at the highest masses. In fact, for He novae the skew of the distribution is even more pronounced. Unlike the case of H novae, however, C/O WD accretors continue to contribute even at the highest WD masses. This is due to the far higher propensity of He-nova systems to grow the C/O accretor to high mass, and is responsible for the blurred C/O-O/Ne divide seen in Figure \ref{fig:histnovam1}. It should be noted that we allow for quiescent C burning events to occur in WDs when the He accretion rate is greater than $2.05\times 10^{-6}$ M\solarperyr\ \citep{wang2017}, converting a C/O core to a O/Ne core. Neglecting this physics, the blurring becomes even more pronounced, increasing the fraction of He novae on C/O WDs with masses approaching \Mchand\ by approximately 15 per cent.

For WD masses less than 0.8 M\solar, almost all He novae are produced by H donor stars. The remainder of the parameter space is dominated by HeMS donor stars.

Finally, the distributions in Figure \ref{fig:histnovam1} bear little resemblance to the distribution of WD masses currently derived from nova observations, which peaks around 1.1-1.2 M\solar\ \citep{shara2018}. Figure 8 is not expected to reproduce any kind of nova observation, as no convolution of a star formation rate (SFR) history has been applied here. Instead, Figure \ref{fig:histnovam1} represents the distribution of WD masses of all novae produced in the 15 Gyrs we simulate after star burst. A conclusive comparison with the observed distribution WD masses for novae requires accounting for changes in metallicity, which allows for a more accurate model of the formation history of a given environment (e.g., the Galaxy). This we leave to a future work which will deal specifically with the effect of metallicity on nova populations.

\subsubsection{White Dwarf Accretion Rate}

\begin{figure}
     \centering
\begin{subfigure}{0.8\columnwidth}
    \centering
    \includegraphics[width=\textwidth=1]{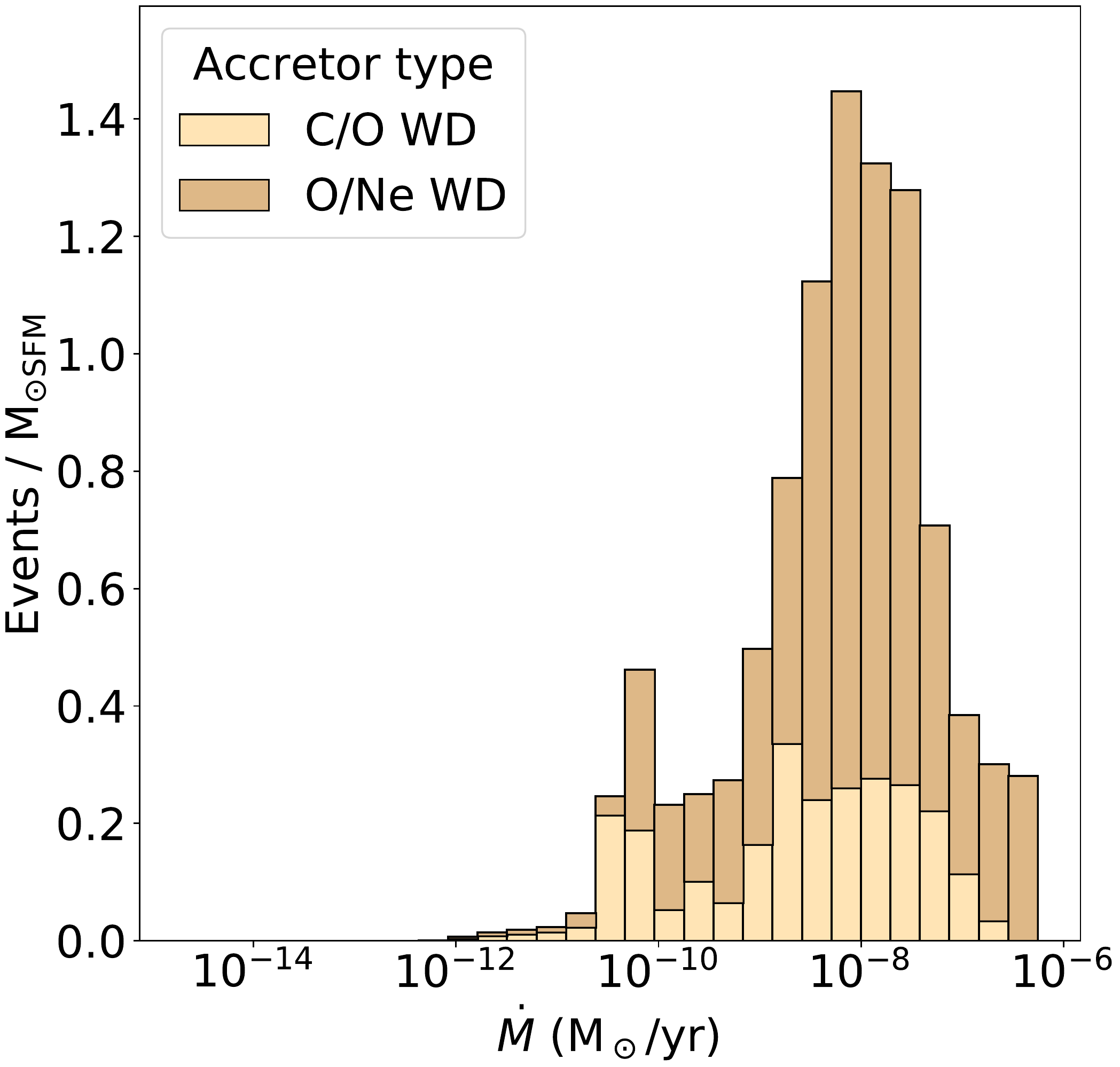}
    \caption{H Novae}
\end{subfigure}

\begin{subfigure}{0.8\columnwidth}
    \centering
    \includegraphics[width=\textwidth=1]{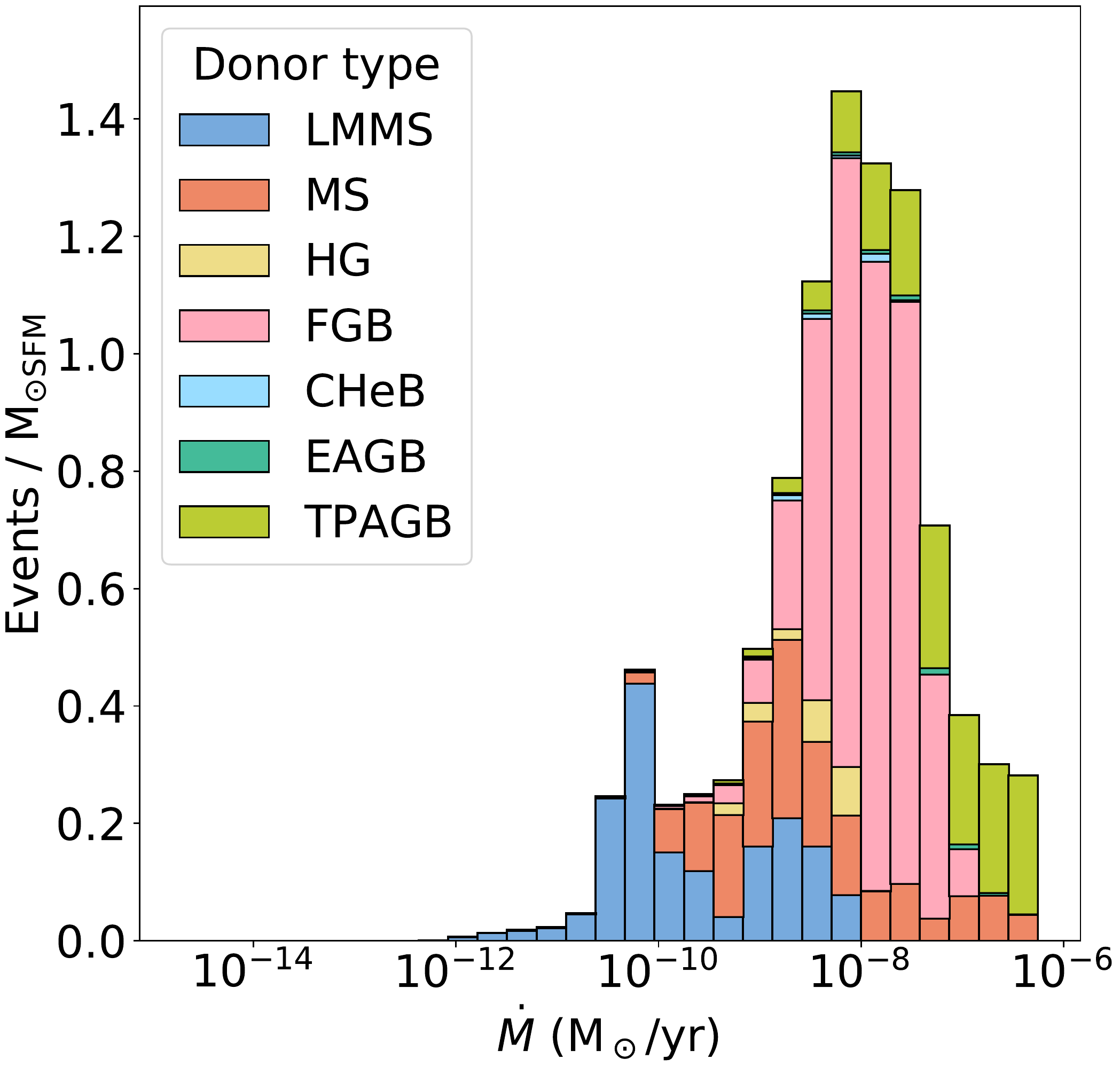}
    \caption{H Novae}
\end{subfigure}

\begin{subfigure}{0.8\columnwidth}
    \centering
    \includegraphics[width=\textwidth=1]{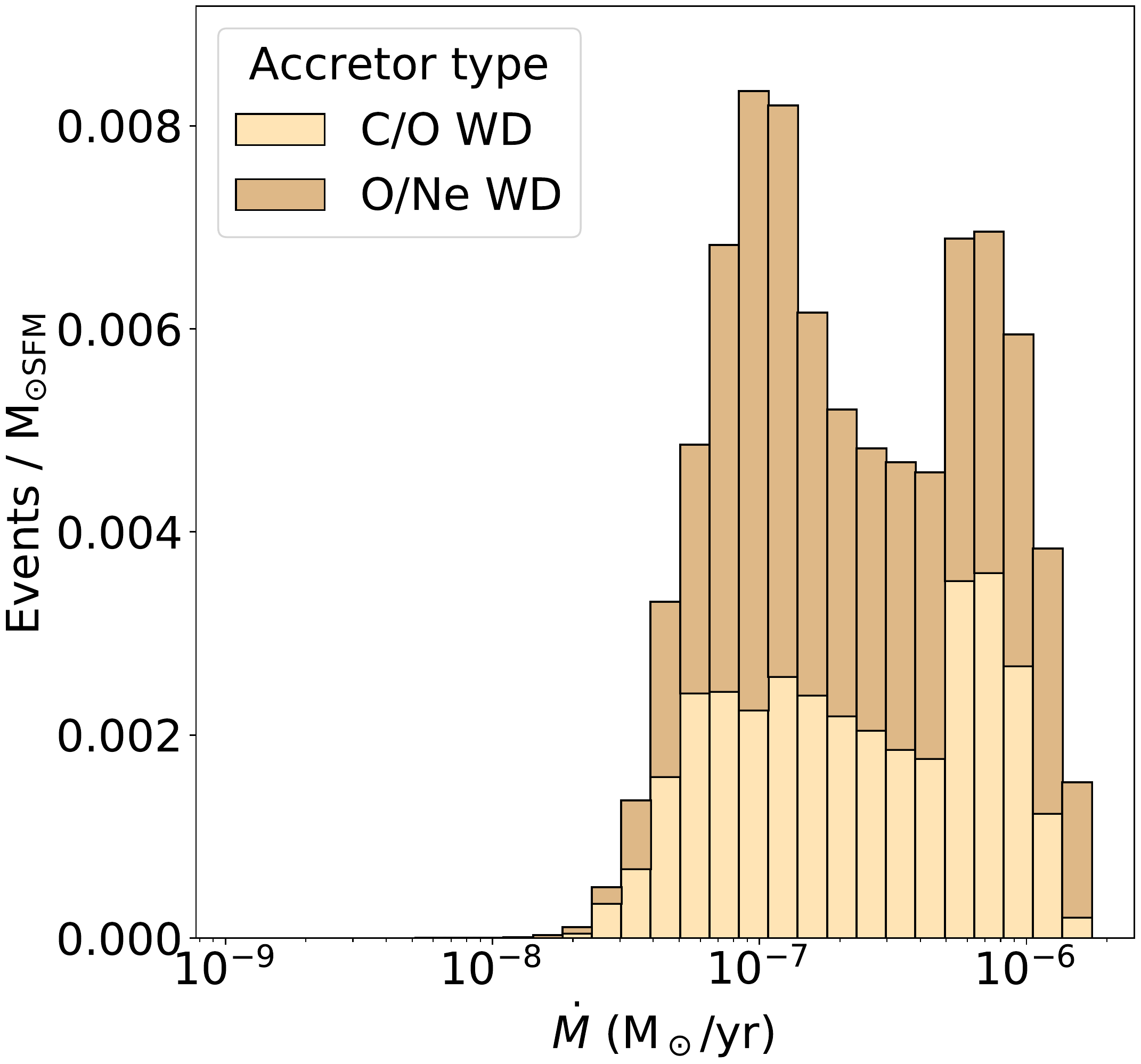}
    \caption{He Novae, excluding H donors}
\end{subfigure}
\caption{Distribution of the mass accretion rates at the time of each nova eruption, coloured by the accretor and donor stellar types. Giant donors dominate the high-accretion rate H-nova systems, while the lowest accretion rate systems are driven by low-mass main sequence stars.}
\label{fig:mdotnovahist}
\end{figure}

Figure \ref{fig:mdotnovahist} shows the distributions of the mass accretion rate immediately prior to the nova eruption.
As expected, in H novae FGB and TPAGB donor stars are responsible for the majority of the highest accretion rate events while LMMS and MS donor stars are responsible for nova systems with lower accretion rates. There exists a weak preference for C/O WDs in low accretion rate systems and O/Ne WDs in high accretion rate systems. The distribution peaks at accretion rates around $10^{-8}$ M\solarperyr, driven by the FGB and TPAGB donor star systems, with a smaller peak present around $10^{-10}$ M\solarperyr\ driven by LMMS and MS donor stars. 

He novae exhibit considerably less spread in the distribution of accretion rates, driven by the increased difficulty in igniting He. While both H- and He-nova systems are able to produce novae at accretion rates up to approximately $10^{-6}$ M\solarperyr, the distribution for He novae tails off by $10^{-8}$ M\solarperyr, compared to the longer tail present for H novae which dies out around $10^{-12}$ M\solarperyr. The event counts for He novae exhibit twin peaks at $10^{-7}$ M\solarperyr and $10^{-6}$ M\solarperyr. The inclusion of H donor channels of He novae does not significantly affect the distribution shown in Figure \ref{fig:mdotnovahist}; we show the distribution excluding He donors due to the presence of a small number of low \Mdot\ outliers which compromise the presentation of the figure.
 
\subsection{Nova Recurrence Times}
\label{sec:rectimes}
\begin{figure}
     \centering
\begin{subfigure}{0.8\columnwidth}
    \centering
    \includegraphics[width=\textwidth=1]{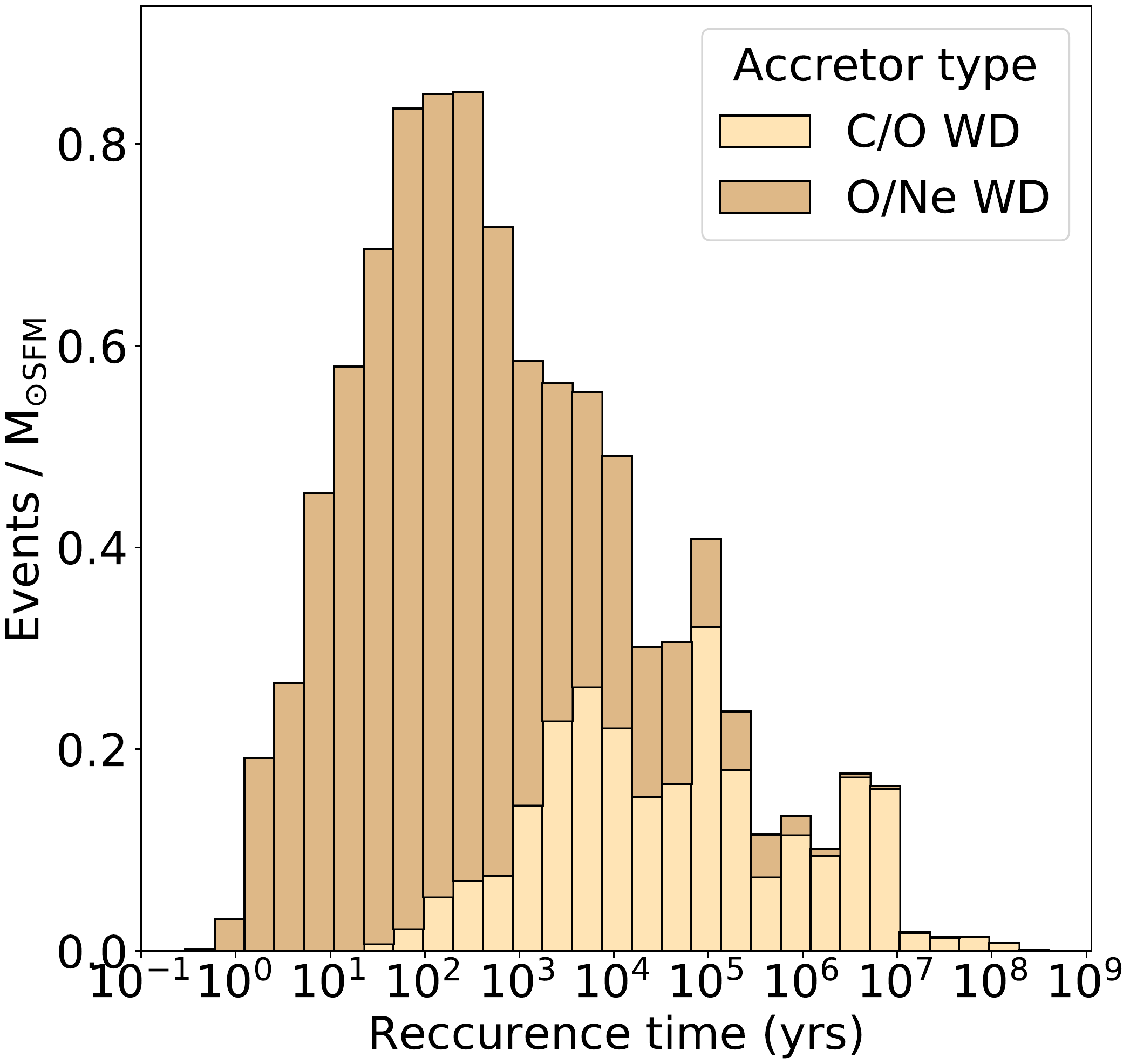}
    \caption{H Novae}
\end{subfigure}
\begin{subfigure}{0.8\columnwidth}
    \centering
    \includegraphics[width=\textwidth=1]{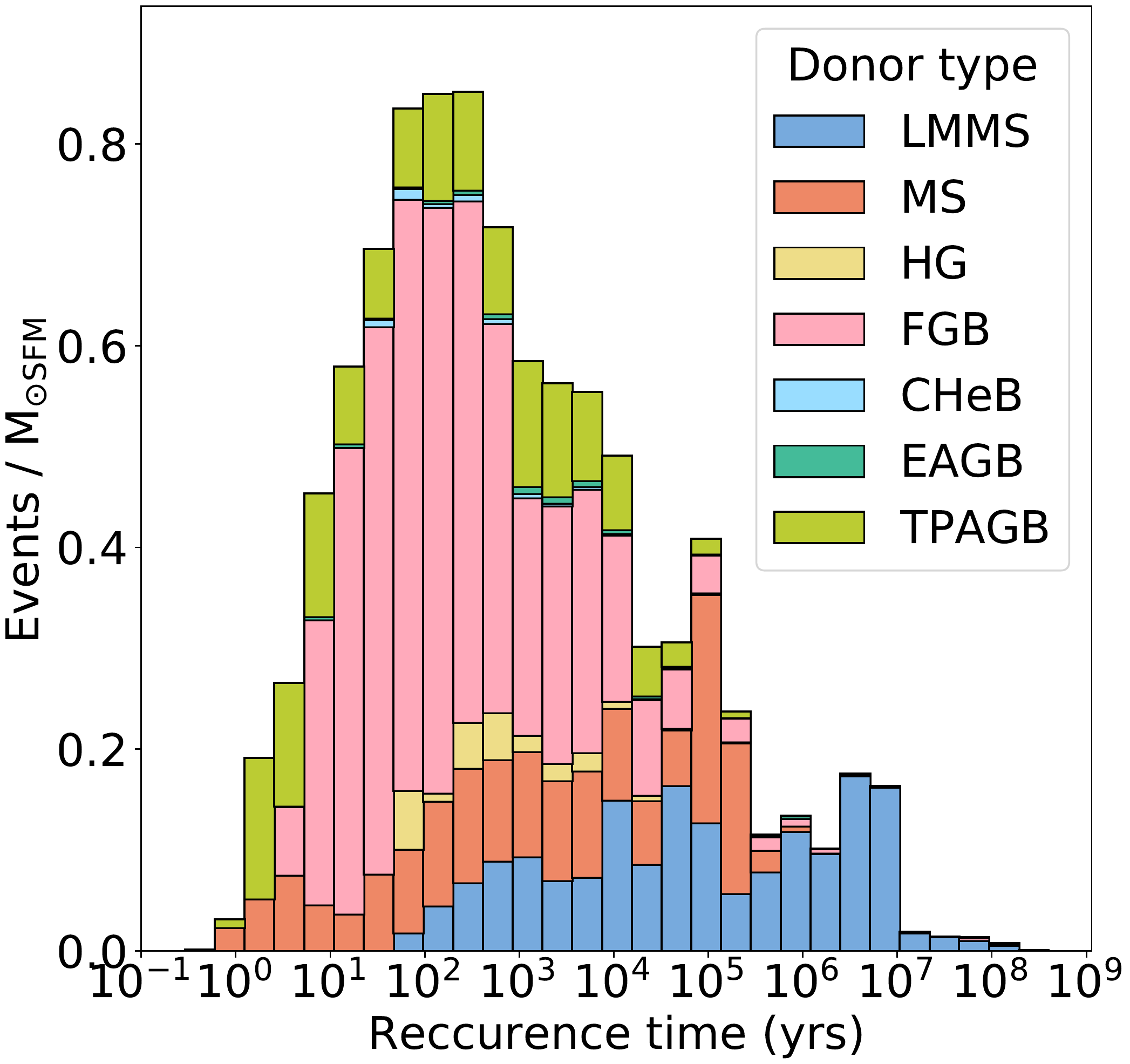}
    \caption{H Novae}
\end{subfigure}
\begin{subfigure}{0.8\columnwidth}
    \centering
    \includegraphics[width=\textwidth=1]{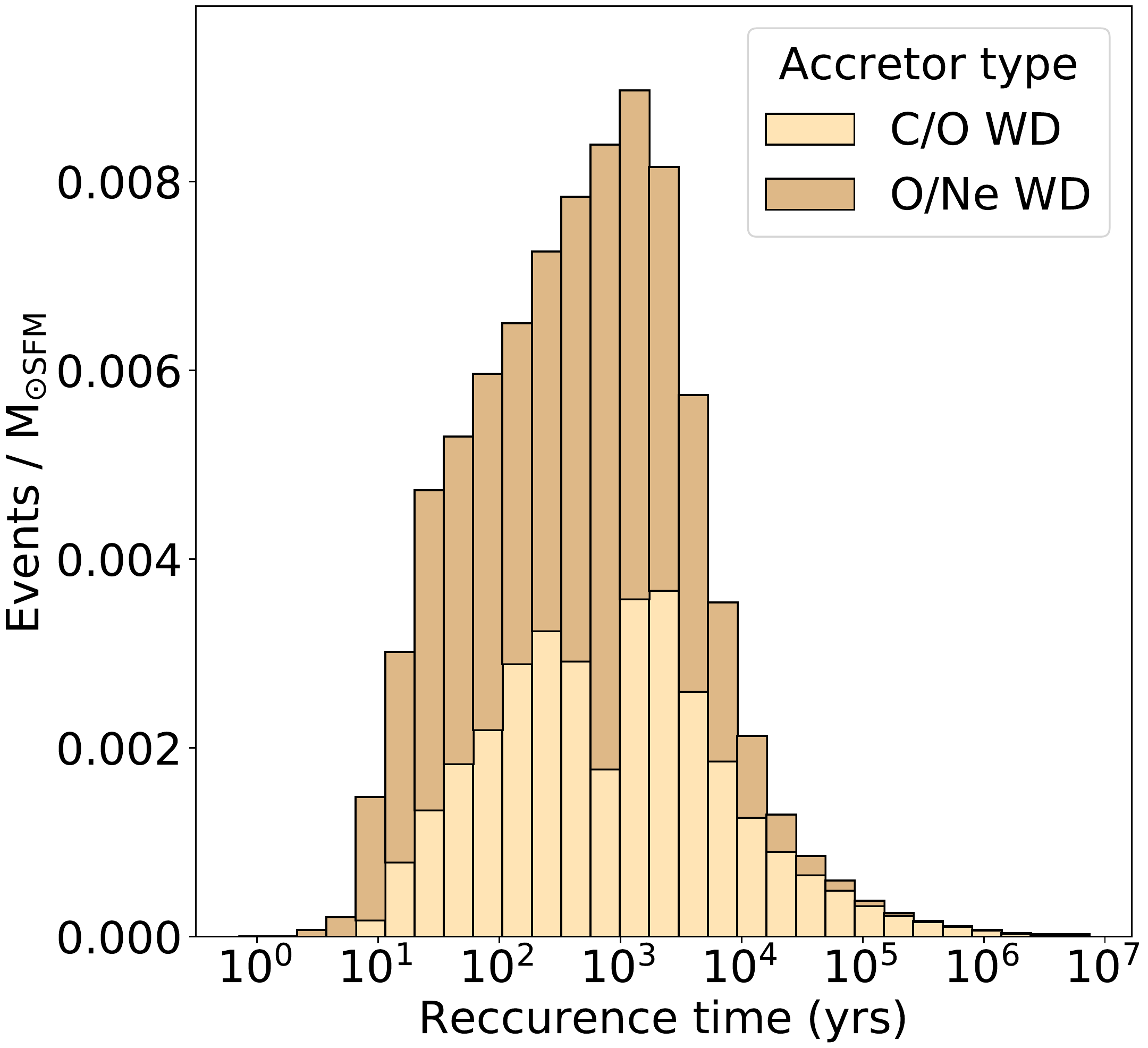}
    \caption{He Novae, excluding H donors}
\end{subfigure}
\caption{Recurrence time distribution at the time of each nova eruption, coloured by the accretor and donor stellar types. H novae with recurrence times shorter than 100 years are mostly produced by O/Ne white dwarfs accreting from giant donor stars.}
\label{fig:histreccurencetimes}
\end{figure}

The recurrence time is defined as the time between subsequent nova eruptions, and is distinct from the observational definition of recurrent novae. It is common to class observed nova systems as recurrent if they have been actually observed to erupt more than once. In practice, only systems with `true' recurrence times shorter than approximately 100 years have a chance to be classified as recurrent, with the remainder (making up the vast majority of nova observations) classed as classical novae. In this work we refer to a nova being `recurrent' if it has a recurrence time shorter than 100 years. Figure \ref{fig:histreccurencetimes} shows the distribution of recurrence times for H and He novae.

We find a huge spread in recurrence times for H novae, with the shortest interval between novae calculated at approximately 100 days, while the longest is found to be almost 400 Myrs. These extremes of the distribution are powered by near-\Mchand\ WDs rapidly accreting material from giant donor stars and WDs accreting material at extremely low accretion rates through stellar wind accretion and Roche lobe filling LMMS stars, respectively.

The distribution is fairly broad, peaking around 100-1000 yr with almost all H novae occurring with recurrence times between 1 year and 10 Myrs. H-nova systems with short recurrence times tend to favour O/Ne WDs, while the long recurrence time novae favour lower mass C/O WD accretors. The distribution of the donor types shows FGB and TPAGB donors dominating lower ($<10^4$ yrs) recurrence time systems and LMMS and MS donor stars dominating the high ($>10^4$ yrs) recurrence time systems.

\begin{figure*}
\centering

\begin{subfigure}{0.95\columnwidth}
    \centering
    \includegraphics[width=\textwidth]{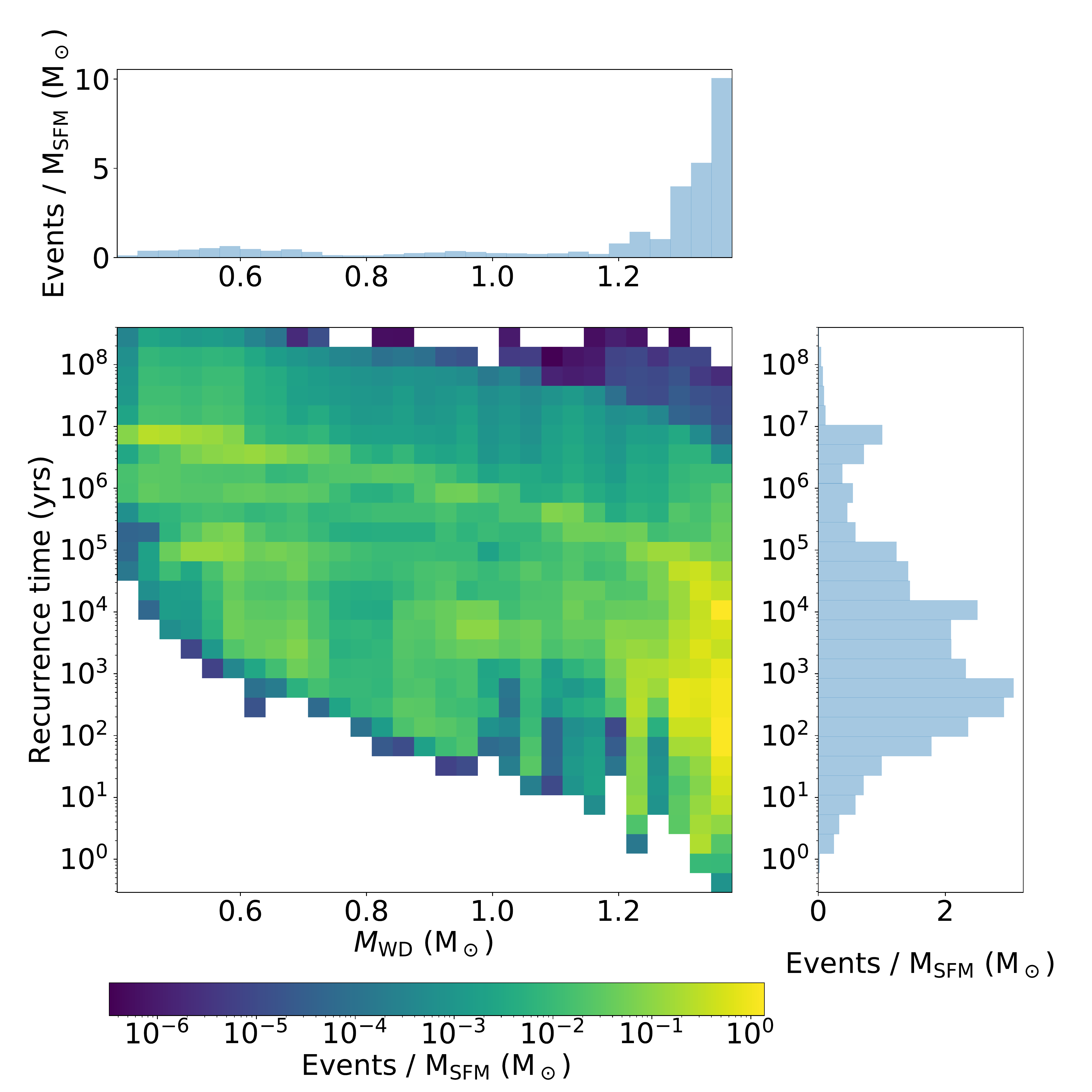}
     \caption{H novae: white dwarf mass vs recurrence time}
\end{subfigure}%
\begin{subfigure}{0.95\columnwidth}
    \centering
    \includegraphics[width=\textwidth]{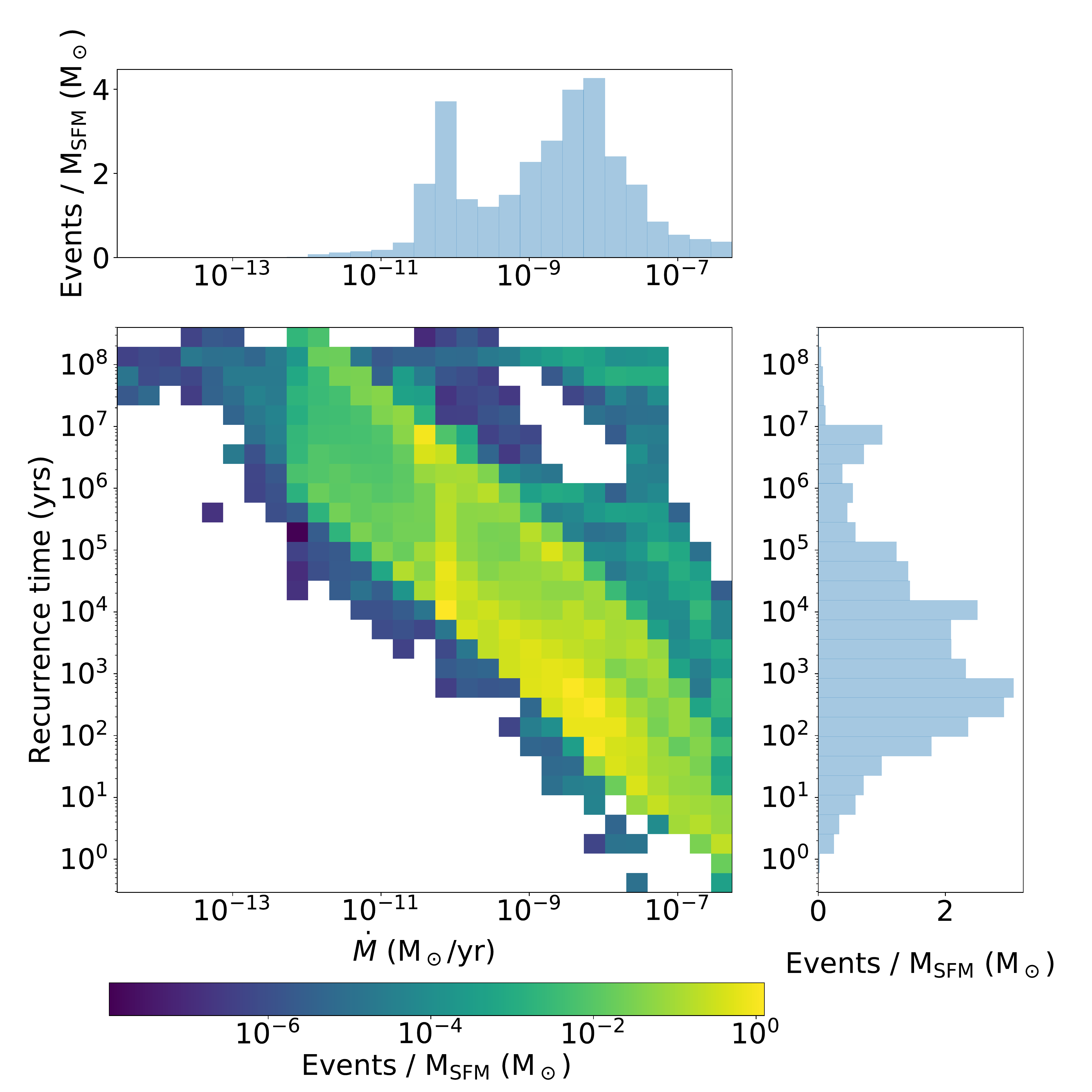}
     \caption{H novae: accretion rate vs recurrence time}
\end{subfigure}
\begin{subfigure}{0.95\columnwidth}
    \centering
    \includegraphics[width=\textwidth]{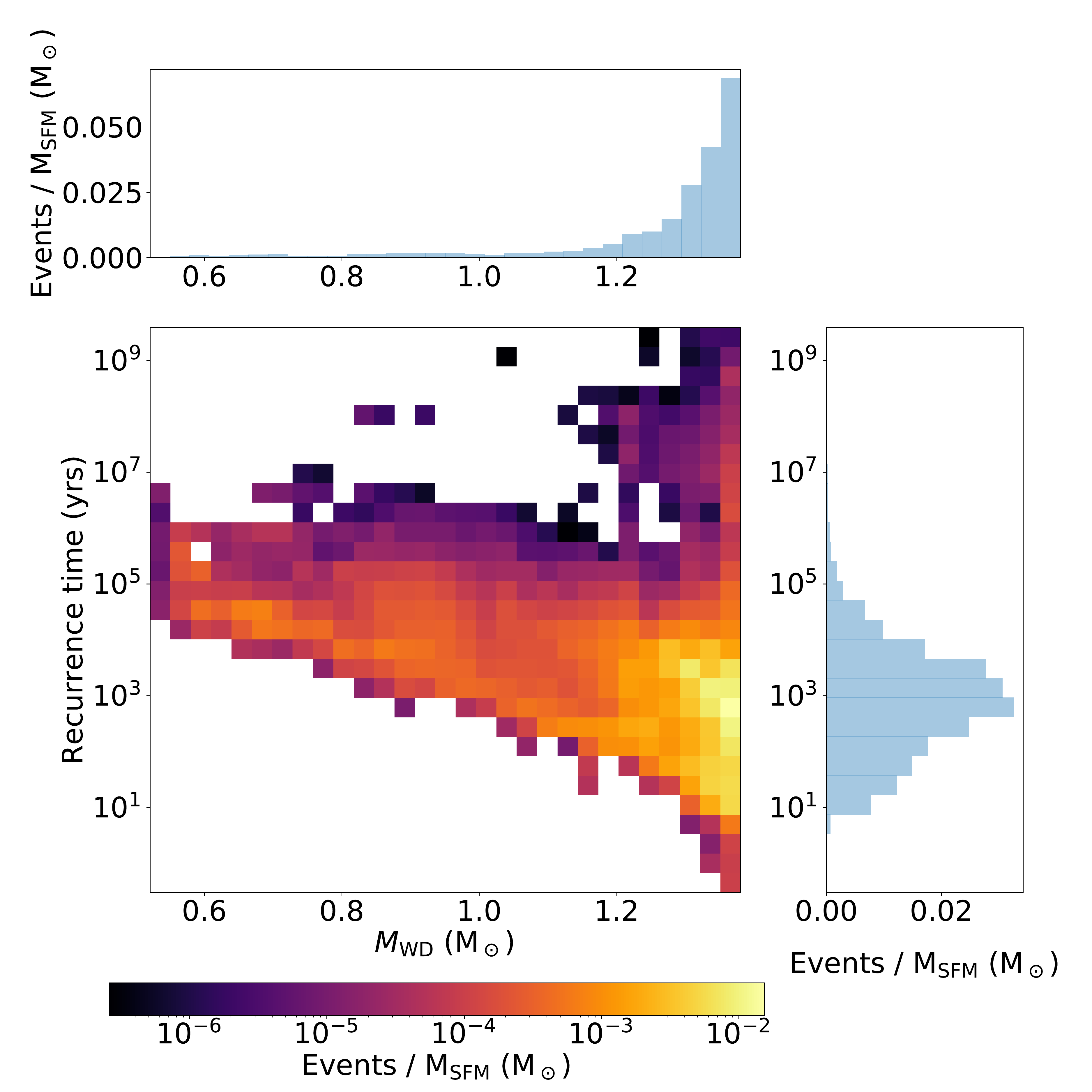}
    \caption{He novae$^{\rm *}$: white dwarf mass vs recurrence time}
\end{subfigure}%
\begin{subfigure}{0.95\columnwidth}
    \centering
    \includegraphics[width=\textwidth]{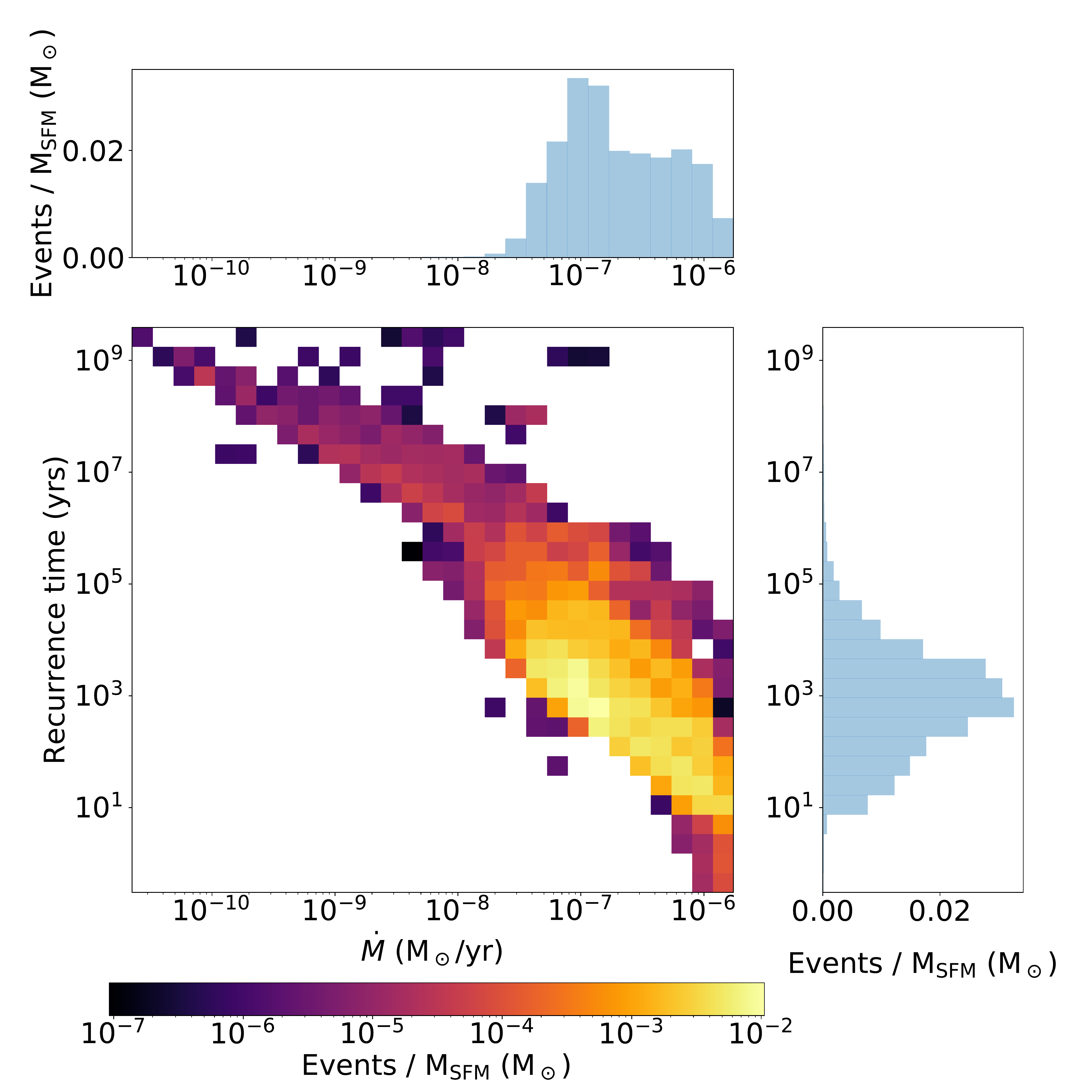}
    \caption{He novae$^{\rm *}$: accretion rate vs recurrence time}
\end{subfigure}

\caption{Distributions of the recurrence time as a function of the white dwarf mass and the accretion rate. The white dwarf mass and accretion rate are sampled at the time of the first nova of each nova pair. The recurrence time is more sensitive to the accretion rate than the white dwarf mass. "$^{\rm *}$": Including H donors.}
\label{fig:hist2Drectime}
\end{figure*}

We find that the recurrence time is more tightly correlated with the mass accretion rate than the mass of the WD, as shown in Figure \ref{fig:hist2Drectime}. Increasing the mass of the WD reduces recurrence times as the critical ignition mass reduces (see Figure \ref{fig:hist2DmwdvsdmcritH}), while the accretion rate has a two-fold effect: increasing \Mdot\ both reduces the critical ignition mass and the time it takes to reach the critical ignition mass. This behaviour is mirrored in the He nova distributions.

Figures \ref{fig:histreccurencetimes} and \ref{fig:hist2Drectime} allow us to make some interesting predictions about the host systems of recurrent H novae. Based on our simulations, we expect recurrent novae to originate almost exclusively from O/Ne WD accretors, particularly recurrent novae with recurrence times $\lesssim10$ yrs. Further, we also expect that most, but not all, of these systems should have giant donor stars. The most rapidly recurring nova systems should be limited to the most massive WDs, with masses very close to \Mchand, and recurrent novae should heavily favour such WDs. However, we also find that less massive WDs as low as approximately 0.6 M\solar\ are expected to be feasible (but rare) sites of recurrent novae. Finally, we predict that recurrent novae should be almost non-existent at accretion rates less than approximately $10^{-9}$ M\solarperyr, and almost all novae from systems accreting at rates greater than approximately $ 10^{-7}$ M\solarperyr occur in recurrent nova systems.

The distribution of He nova recurrence times is more constrained, with most having recurrence times ranging between between 1 yr and 1 Myr, and peaking around $10^3$ yrs. The bulk of the distribution has a relatively homogeneous divide between C/O and O/Ne WDs, with only the extremes of the distribution showing strong preferences for C/O WDs and O/Ne WDs at the high and low end, respectively.

This is in spite of the fact that Figure \ref{fig:hist2Drectime} shows a slightly tighter relation with mass for He novae relative to H novae; even more so than H novae, the bulk of He novae are driven by massive WDs. If we apply the `hundred year' definition of recurrent novae to the distribution of He novae, we find that only WDs greater than 1.1 M\solar\ and accretion rates greater than $10^{-7}$ M\solarperyr\ are capable of driving such events, far more constraining than in H novae. It should be recalled however that there exists only a single confirmed He nova eruption, V445 Puppis (2000) \citep{kato2000,lynch2001,ashok2003, kato2008}, perhaps making the distinction between classical and recurrent He novae somewhat egregious. We discuss our He nova findings in the context of the V445 Puppis observation in Section \ref{sec:discussion}.

\subsection{Nova Rates}
\label{sec:novarates}
\subsubsection{Delay-time distributions}

\begin{figure*}
     \centering
\begin{subfigure}{0.9\columnwidth}
    \centering
    \includegraphics[height=0.9\textwidth]{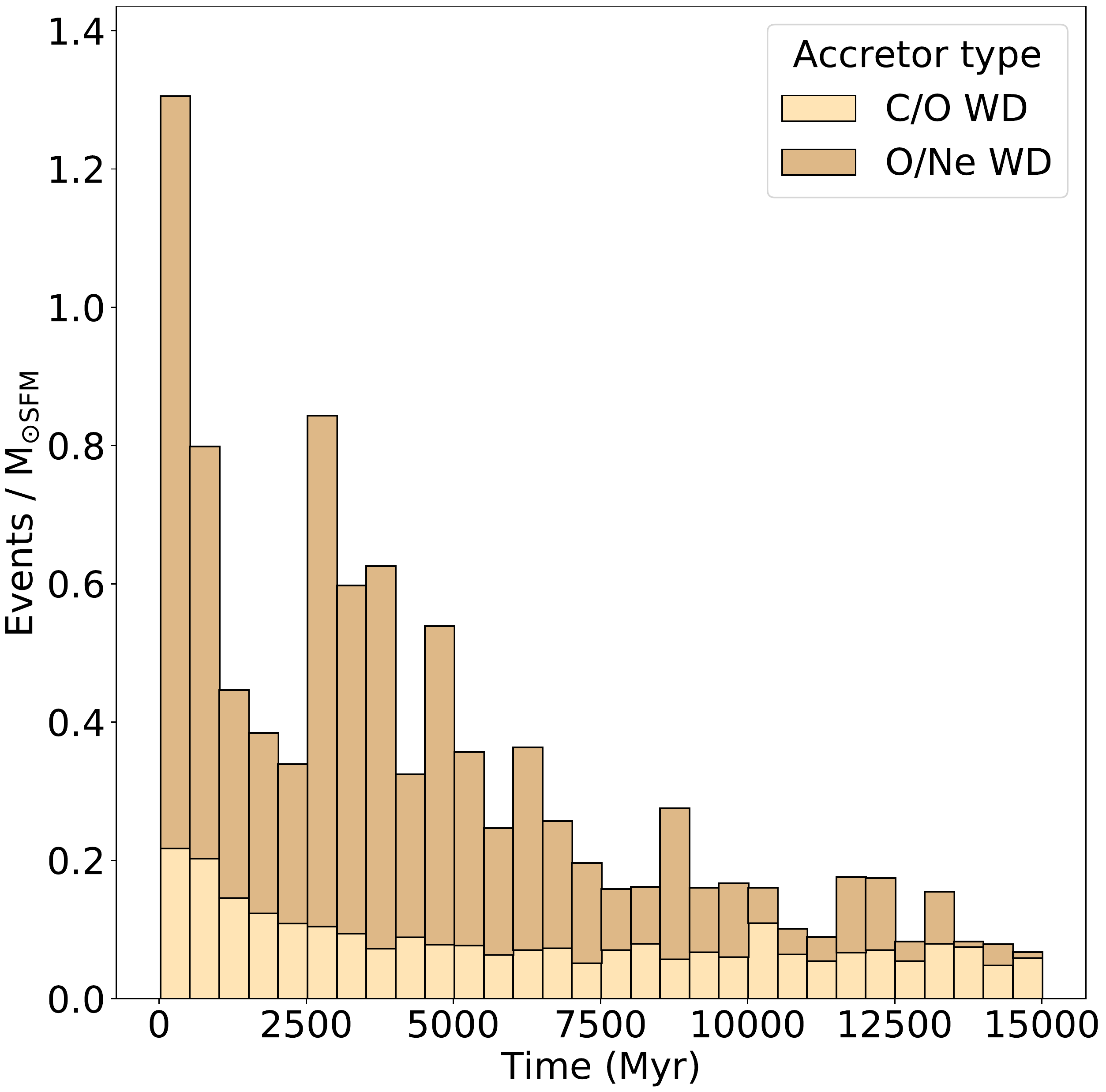}
    \caption{H Novae}
\end{subfigure}%
\begin{subfigure}{0.9\columnwidth}
    \centering
    \includegraphics[height=0.9\textwidth]{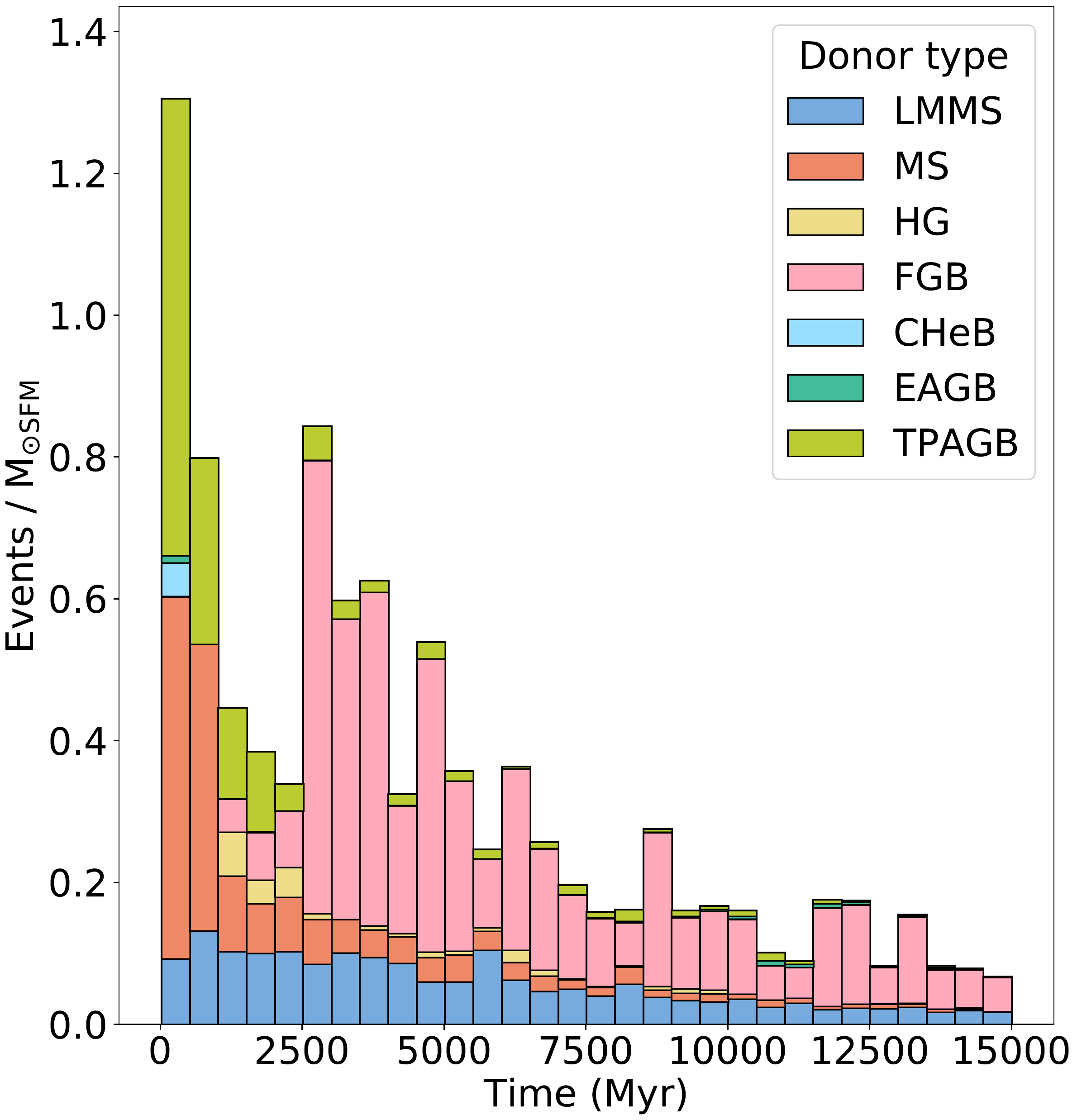}
    \caption{H Novae}
\end{subfigure}

\begin{subfigure}{0.9\columnwidth}
    \centering
    \includegraphics[height=0.9\textwidth]{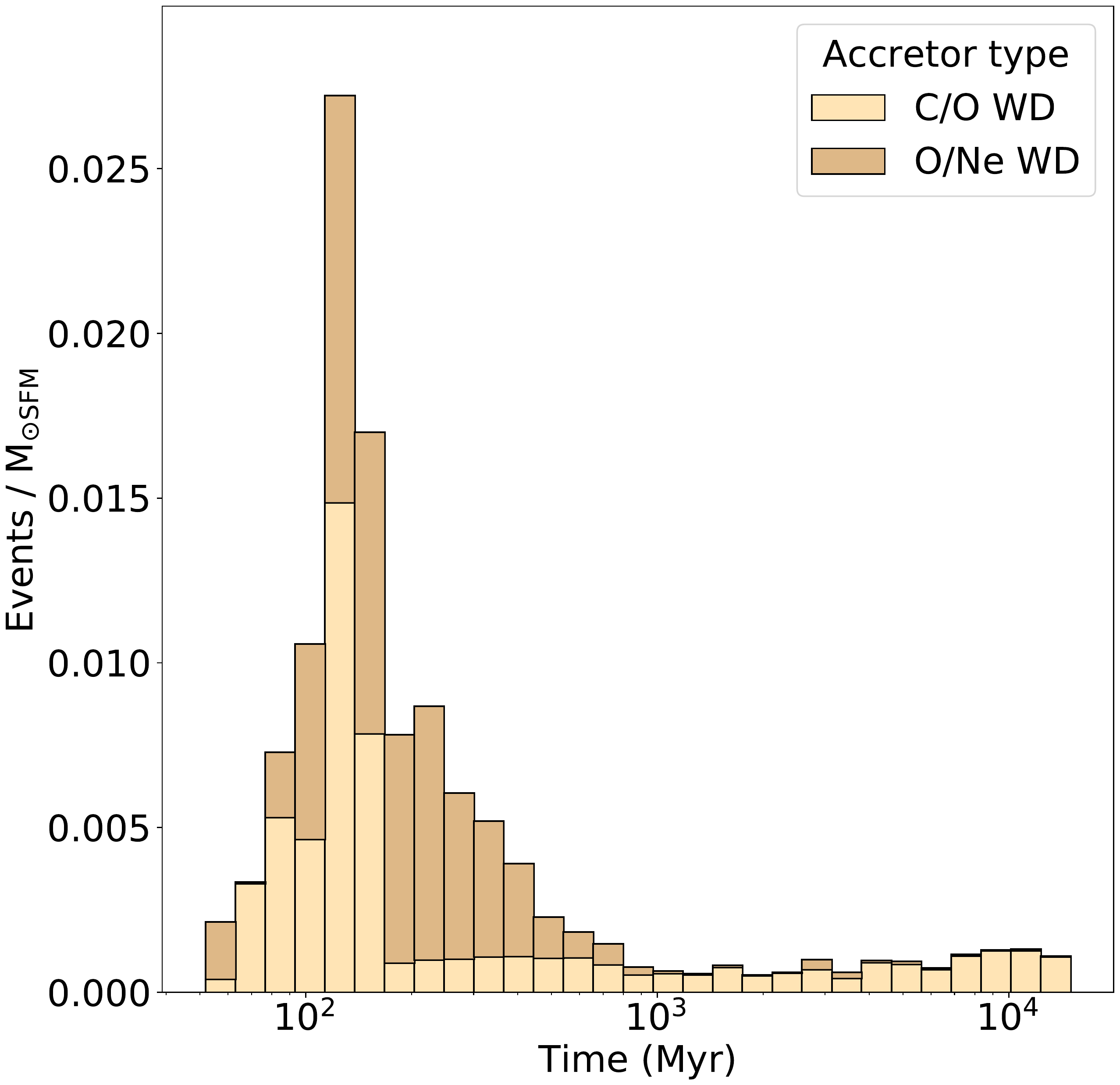}
    \caption{He Novae, including H donors.}
\end{subfigure}%
\begin{subfigure}{0.9\columnwidth}
    \centering
    \includegraphics[height=0.9\textwidth]{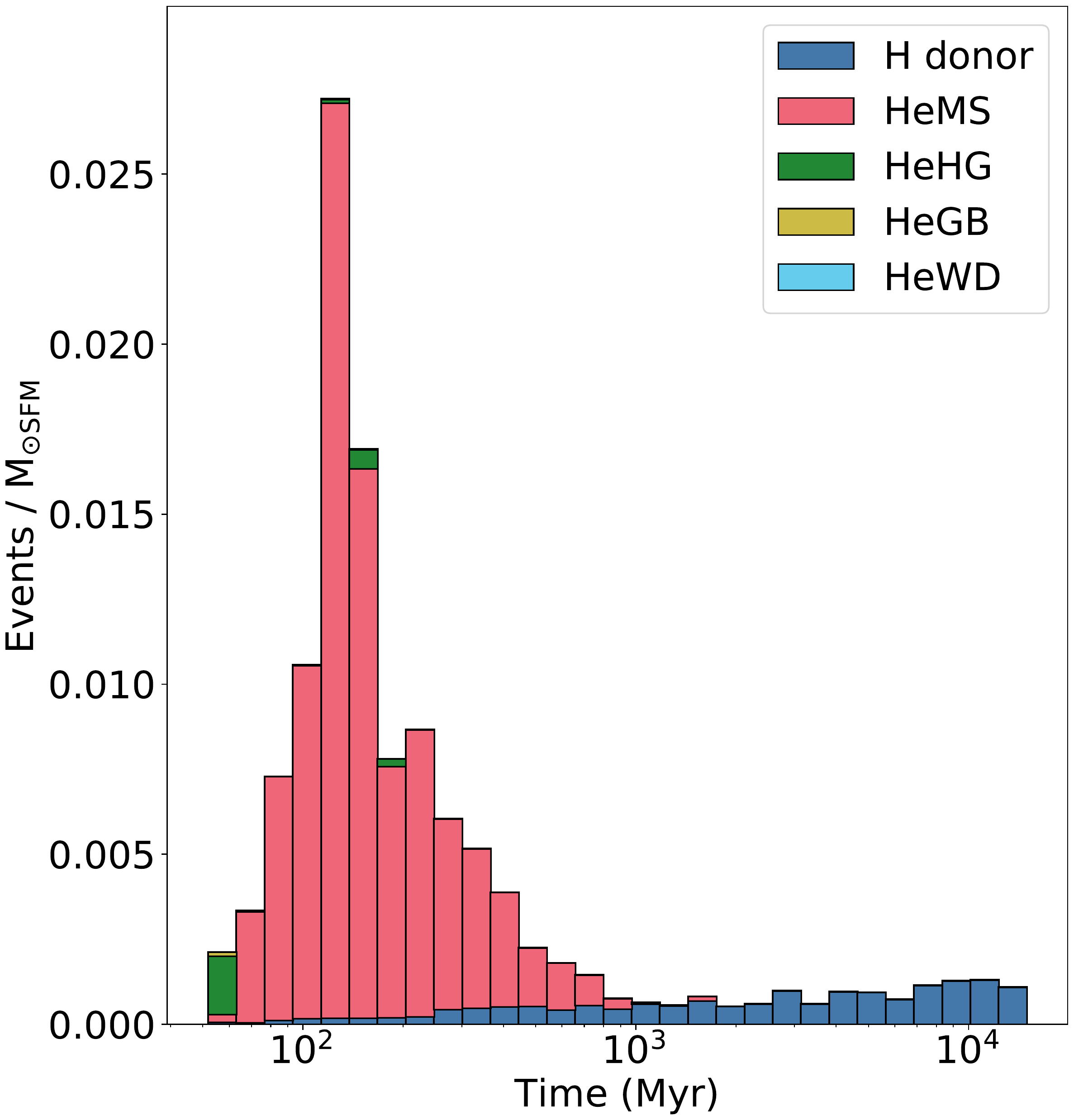}
    \caption{He Novae, including H donors.}
\end{subfigure}
\caption{Delay-time distributions, coloured by the accretor and donor stellar types. He novae from He donors are quite prompt, with almost no contributions beyond 1 Gyr, although a small number of late-time He novae are produced from systems with H donors. H novae are far less prompt, with significant late-time contributions, fueled primarily by giant donor stars, present throughout the simulation.}
\label{fig:delaytimedist}
\end{figure*}

The delay-time distributions of H and He nova events are shown in Figure \ref{fig:delaytimedist}. Note that this is the distribution of the delay-times to each individual nova event, starting from the birth of the system, and not the delay-time to the first nova eruption of each nova system.

The delay-time distribution of H nova events shows that overall nova counts are mostly driven by O/Ne WD accretors, which dominate at earlier times while remaining significant sites of novae even for systems with delay-times greater than 10 Gyrs. Approximately 65 per cent of H nova events (across all times) are found to occur on O/Ne WD accretors, strikingly similar to He novae, for which around 60 per cent of events occur on O/Ne WDs.

For delay-times $<2$ Gyrs, TPAGB donor systems produce most of the H novae, with notable secondary contributions of MS and LMMS donor systems. Beyond 2 Gyr the contribution of FGB donor systems increases substantially as the large population of systems with $M_{2 \ \rm init}\lesssim 2$  M\solar\ evolve off the main sequence. Compared to more massive stars, these systems enjoy longer lifetimes on the FGB, allowing longer periods of mass transfer, compounding the effect of the IMF favouring the birth of such systems. It is also worth noting that while a 1 M\solar\ star has a main-sequence lifetime of $\approx 10$ Gyr, a 2 M\solar\ star has a main-sequence lifetime of only $\approx 1.8$ Gyr \citep{karakas2014}. Thus the contribution of FGB donor stars from low-mass progenitors ($\approx1-2$ M\solar) is observed in Figure \ref{fig:delaytimedist} to be `smeared' over cosmic time. MS and LMMS contributions decline over cosmic time, increasing the relative importance of FGB systems in ultra-late delay-time systems. 

Finally, Figure \ref{fig:finalbinarystateHnova} shows that FGB donors make up only a tiny fraction of the non-remnant--WD systems compared to LMMS-WD binaries, but Figure \ref{fig:delaytimedist} shows that they are important contributors to the nova rate, particularly in late delay-time H nova events. This out-sized contribution by FGB donor stars is due to the much higher mass transfer rates obtainable in these systems compared to LMMS donors, which causes many more novae to be produced per system.

He novae peak in event counts from 100-150 Myrs, climbing steeply from 50 Myrs before slowly trailing off by around 600 Myrs, with a rapidly decaying tail of late-time events with almost no contributions beyond 1 Gyr. From 1-3 Gyrs, there are a small number of He-accreting systems originating from high initial mass ratio ($q_{\rm init}\gtrsim 0.85)$ binaries of relatively low initial mass ($M_{\rm 1,2\ init}\approx 1-1.8$ M\solar), which typically undergo very few He novae ($<10$ per system). Other binaries, also with low initial masses but lower mass ratios, are able to transfer significant mass to the secondary. In these cases the secondary's MS mass becomes so high that it evolves too quickly to create such late delay-time events.

Beyond 1 Gyr, H donor contributions dominate. This is noteworthy, as it implies that in old stellar populations He novae should be expected to be driven almost exclusively by H accretion.
At all other times He nova contributions to the overall transient rate are dominated by HeMS donors. The accretor composition, interestingly, shows a trend opposite that of H novae, with C/O accretors being more important for the most prompt He novae.

\subsubsection{Estimated nova rate history for M31}

Delay-time distributions such as those shown in Figure \ref{fig:delaytimedist} can be numerically convolved with an arbitrary star formation history by breaking the history up into discrete time bins containing the mass of stars formed in each bin. Knowing the contributions to event rates at all delay-times from a unit-mass burst of star formation, the event rate history is computed by superimposing delay-time distributions normalised according to the mass of stars formed in each bin. In this way, the event rate at each time in the history of a given environment becomes the sum of the contributions of every preceding time bin, each of which is modelled as a burst of star formation.

\begin{figure}
\centering
\begin{subfigure}{0.8\columnwidth}
    \centering
    \includegraphics[width=\textwidth]{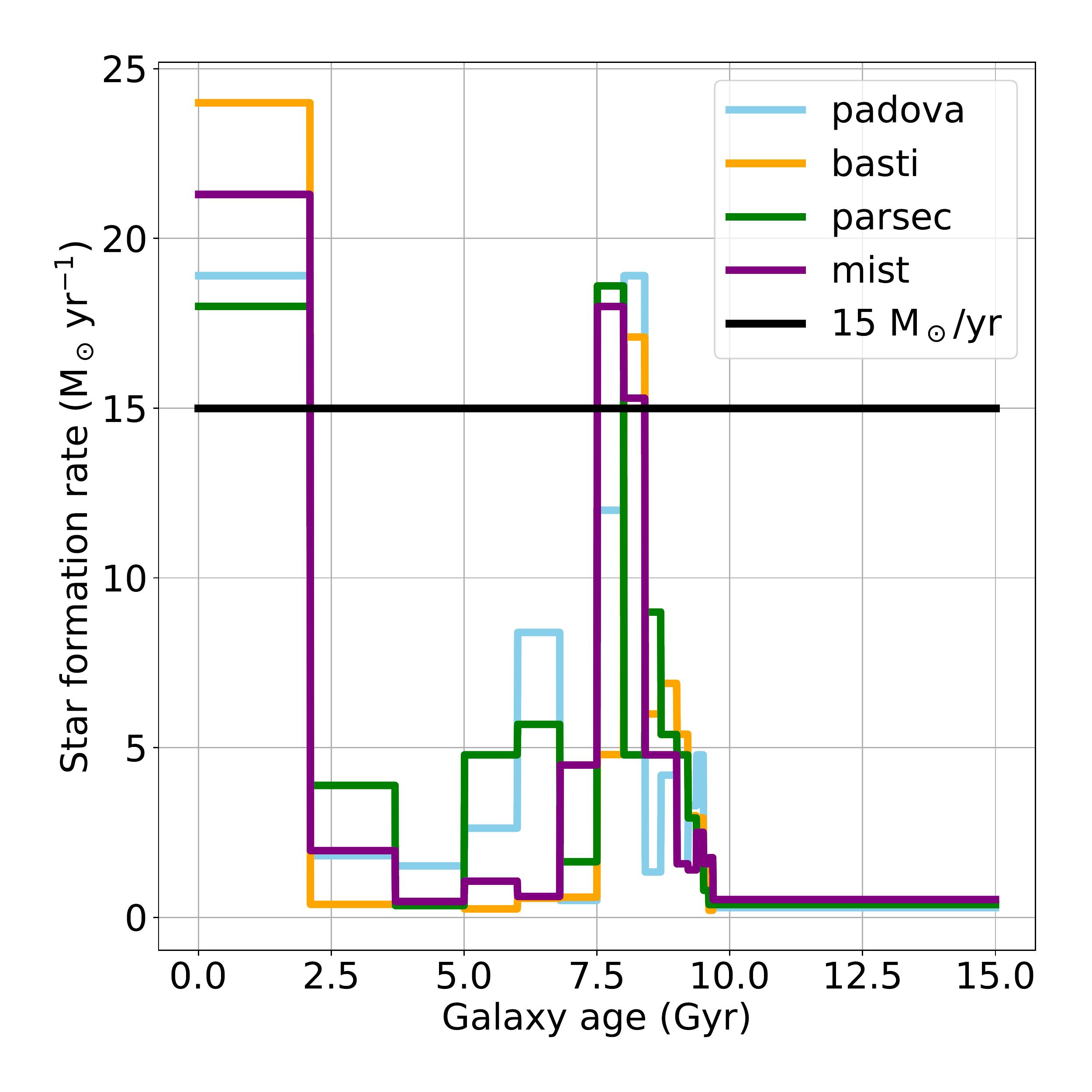}
    \caption{Star formation rate}
\end{subfigure}
\begin{subfigure}{0.8\columnwidth}
    \centering
    \includegraphics[width=\textwidth]{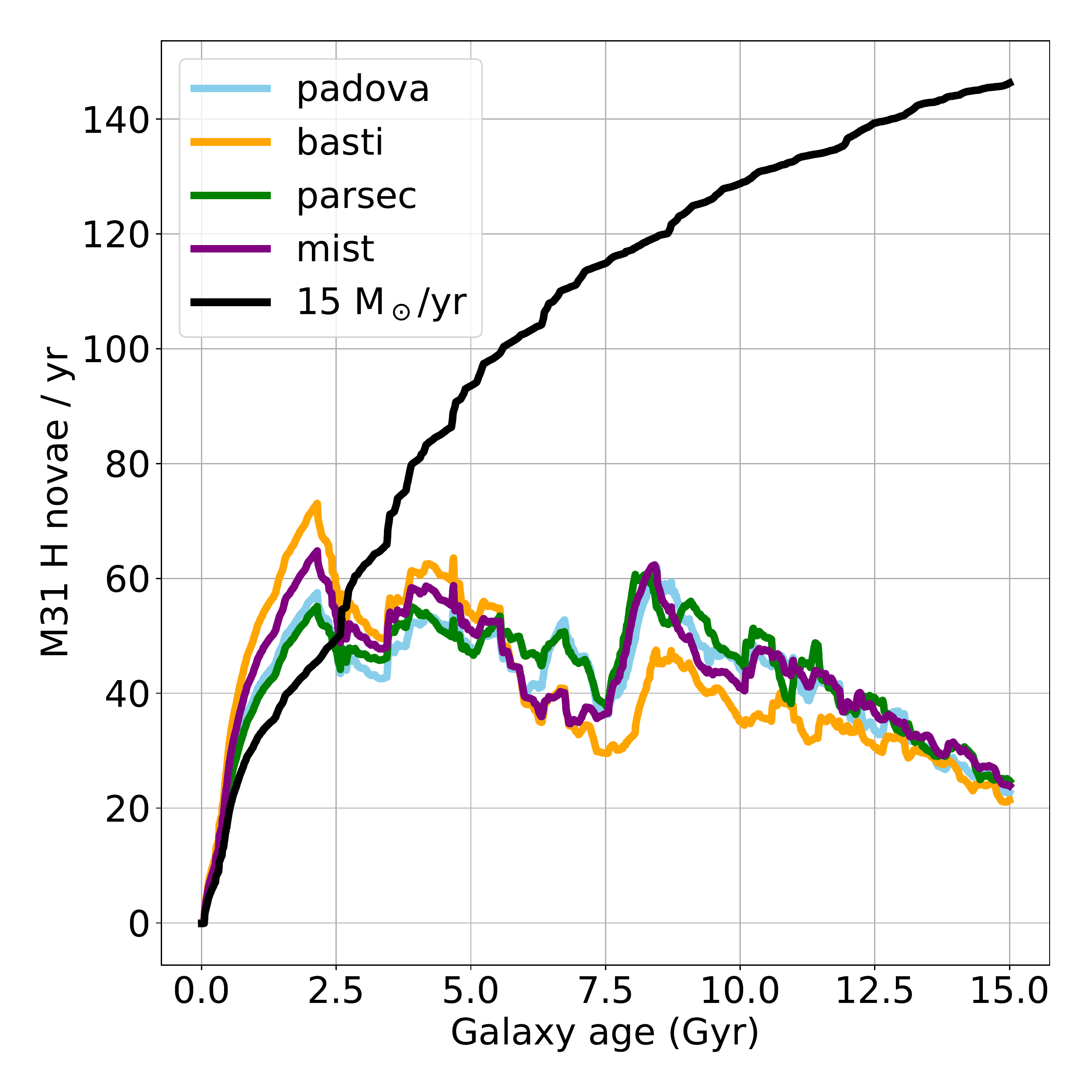}
    \caption{H nova rates}
\end{subfigure}
\begin{subfigure}{0.8\columnwidth}
    \centering
    \includegraphics[width=\textwidth]{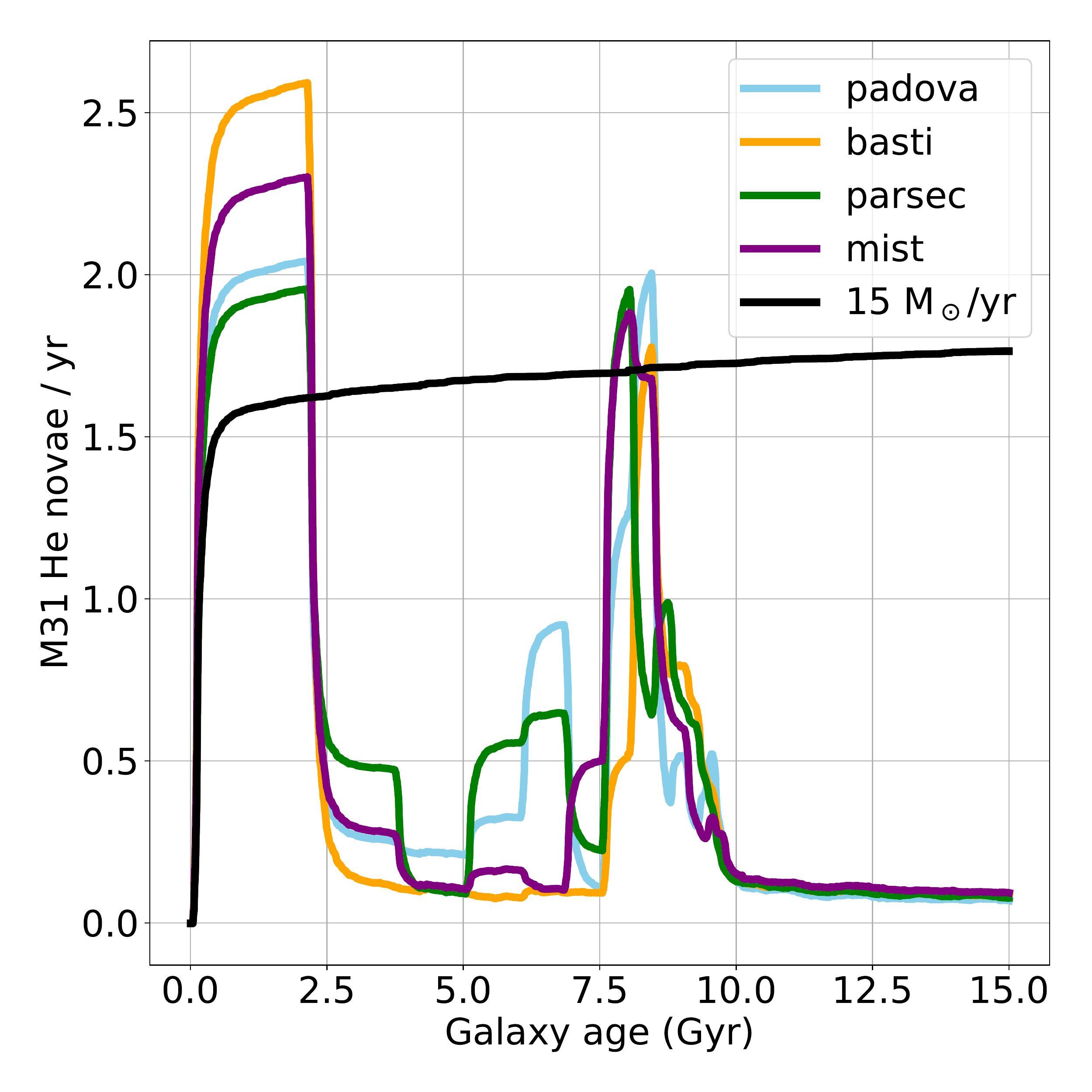}
    \caption{He nova rates}
\end{subfigure}
\caption{Adopted star formation rate (SFR) (a) and predicted nova rate (b,c) histories for M31.  
Each history is labelled according to the set of stellar tracks (Padova, Basti, Parsec, MIST) fit to the colour-magnitude diagrams from the Panchromatic Hubble Andromeda Treasury (PHAT) by \protect \cite{williams2017} to obtain the SFR history. The current age of M31 is assumed to be close to 10 Gyr. Data beyond 10 Gyr is `looking forward' from the present day, with the SFR assumed to remain equal to its current estimated value.}
\label{fig:m31rates}
\end{figure}

Figure \ref{fig:m31rates} shows the predicted H and He nova rates at different times in M31 (Andromeda) using several different star formation histories. The black line is the result for a constant SFR of 15 M\solarperyr, a rate representative of the average SFR of M31 over its assumed 10 Gyr age, with the remaining lines representing more sophisticated models which make use of star formation histories extracted from Figure 11 in \cite{williams2017}. These models are labelled according to the underlying sets of stellar models fit by \cite{williams2017} to colour-magnitude diagrams from the Panchromatic Hubble Andromeda Treasury (PHAT) in order to obtain the SFR history.

We find that H novae track the star formation history poorly because they are distributed across a wide range of delay-times. However, the delay-times of most He novae are so short that their rate (were it known) provides a relatively direct measure of the recent local star formation history. Of course, the short delay-times of these systems also mean that virtually no information about the star formation history at early times can be gleaned from the event rate of He novae. Conversely, the extremely long delay-time systems present for some H-nova systems makes them a potentially useful probe of early star formation history.

It is apparent that there exist significant deviations between the complex, if uncertain, star formation models of \cite{williams2017} and the constant SFR assumption. Given the widespread practice of employing constant SFRs to approximate spiral galaxies, this should be cause for concern for studies attempting to derive absolute rates. The importance of the adopted star formation history when calculating rates is assessed in detail in the context of double compact object mergers in several recent works \cite[e.g.,][]{chruslinska2019,neijssel2019,olejak2020,broekgaarden2021}.
 
We estimate that the current rate of H novae in M31 is approximately $41 \pm 4$ events / yr based on the \cite{williams2017} star formation histories, and $0.14\pm0.015$ for He novae. M31 appears to currently be experiencing a lull in star formation \citep{rahmani2016,williams2017}, impacting estimates of the current rate of He novae in particular. According to the \cite{williams2017} SFR histories, this lull has lasted around 1 Gyr; thus, our models predict that He novae from H donors should be non-negligible when considering He nova rates in M31's current epoch, despite their low rates on a per-star burst basis.

The current nova rate in M31 is discussed further in the context of current observational estimates and other theoretical works in Section \ref{sec:discussion}.

\section{Discussion}
\label{sec:discussion}

The methods employed in this work represent a purely theoretical approach to calculating nova rates. We do not attempt to optimise parameters to best reproduce either the observed nova rate in M31 or any rates calculated in previous theoretical works, but instead have built our standard model based on the merit of the individual physical sub-models and parameters where possible. It is beyond the scope of this work to fully explore the physical parameter space surrounding these events. In particular, we choose to defer addressing the effect of metallicity to a future work where it can be given the attention it deserves.

However, we do estimate the importance of highly uncertain CE physics on predictions of the current nova rate. 

\subsection{Uncertainties in estimates due to CE physics}

\begin{figure*}
\centering
\begin{subfigure}{0.7\columnwidth}
    \centering
    \includegraphics[width=\textwidth]{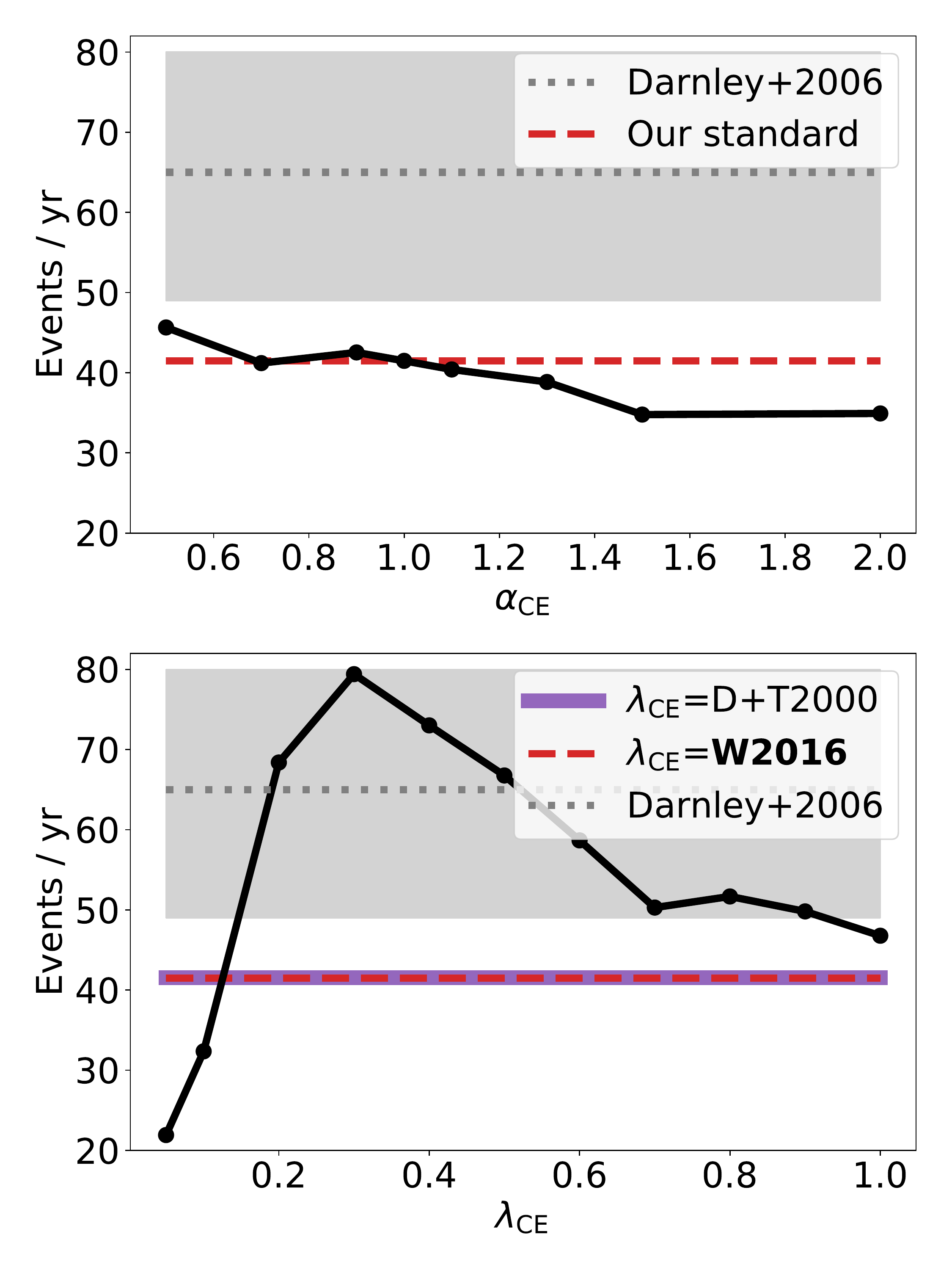}
    \caption{H Novae}
\end{subfigure}%
\begin{subfigure}{0.7\columnwidth}
    \centering
    \includegraphics[width=\textwidth]{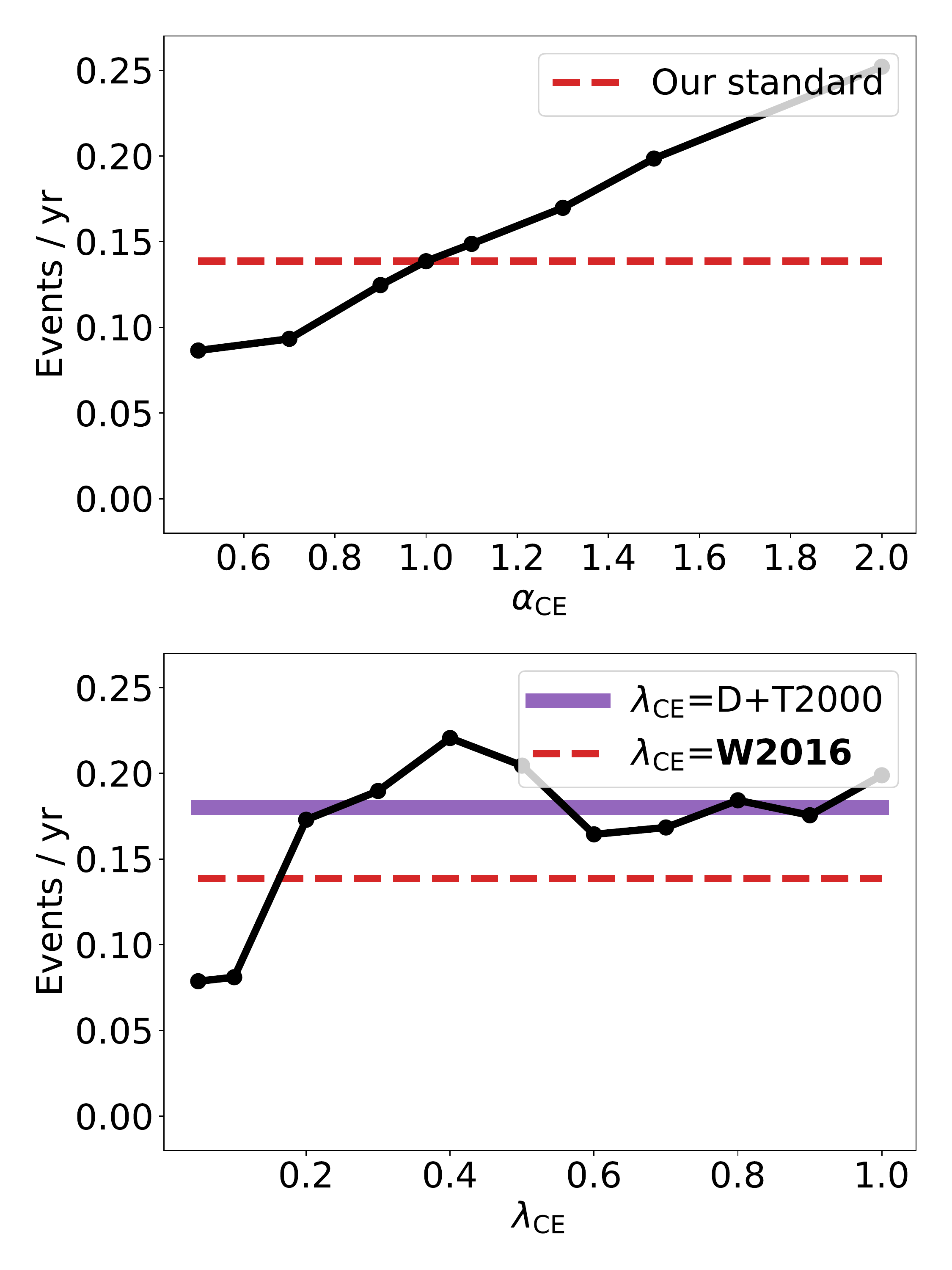}
    \caption{He Novae, including H donors}
\end{subfigure}%
\begin{subfigure}{0.7\columnwidth}
    \centering
    \includegraphics[width=\textwidth]{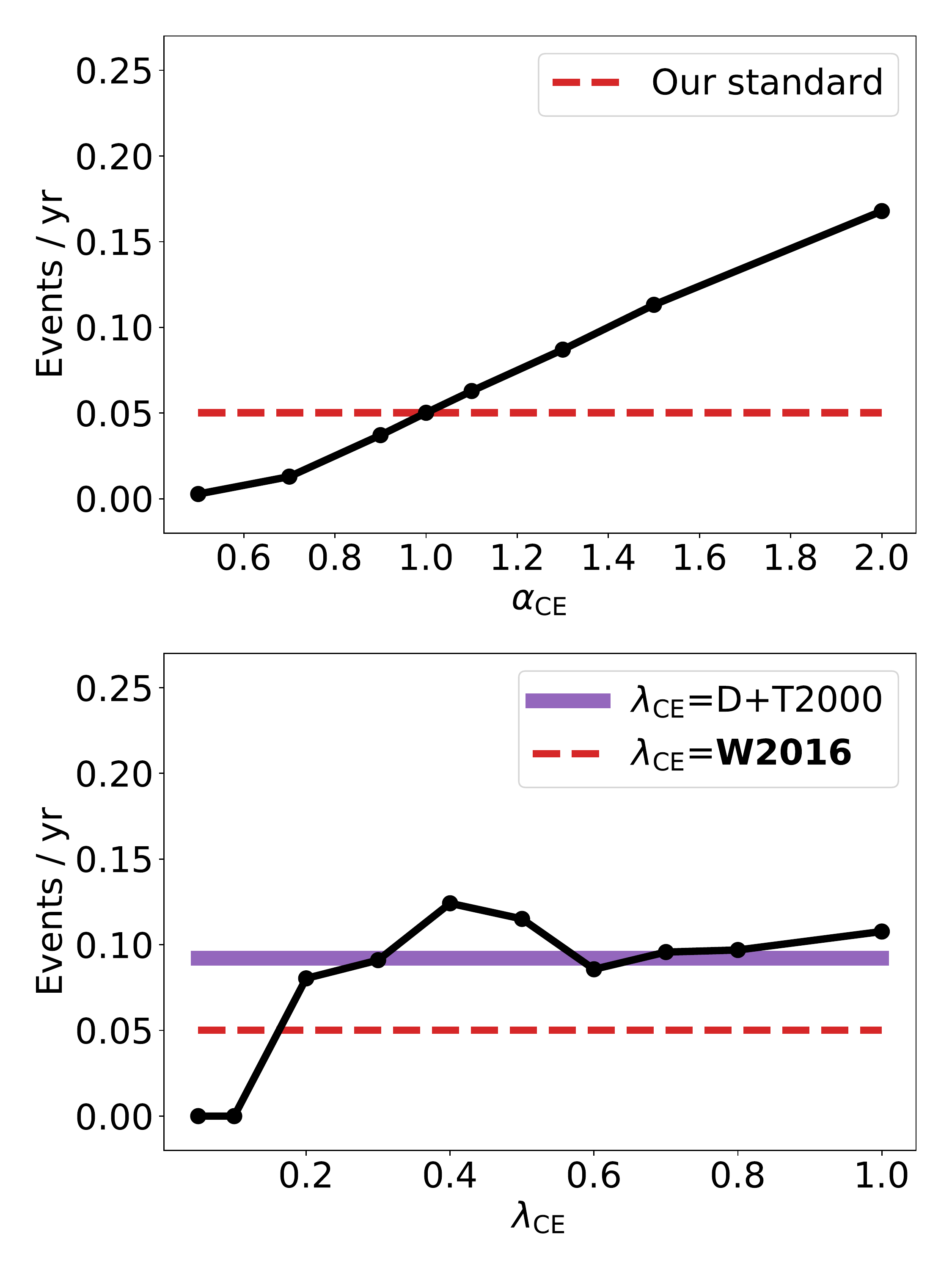}
    \caption{He Novae, excluding H donors}
\end{subfigure}
\caption{The mean current H and He nova rates for M31, varying the common-envelope parameters $\alpha_{\rm CE}$ and $\lambda_{\rm CE}$. The grey shaded region represents the stated uncertainty in the observed nova rate in M31 by \protect \cite{darnley2004,darnley2006}. The H nova rate shows significant variation with the choice of $\lambda_{\rm CE}$. The state-of-the-art W2016 \protect \citep{wang2016binding} model used in this work for $\lambda_{\rm CE}$ results in an underprediction of the nova rate for almost every value of $\alpha_{\rm CE}$ tested.}
\label{fig:m31ratesCEplots}
\end{figure*}

M31 presents the best available laboratory for studying nova rates due to its proximity and near face-on inclination to the Galaxy, and because it is possible to estimate systematic effects and completeness very accurately. The most current observational nova rate is estimated to be $65^{+15}_{-16}$ events / yr \citep{darnley2004,darnley2006}, based on data from the UPAT-AGAPE microlensing survey. Previous analyses have estimated nova rates between 20 and 50 events per year \citep{hubble1929,arp1956,capaccioli1989,shafter2001} for M31. More recent theoretical work by \cite{chen2016} and \cite{soraisam2016} predict an event rate of approximately 100 events / yr when accounting for the absence of novae with speed classes < 10 days \citep{darnley2019} in previous works. In their recent review, \cite{dellavalle2020} question the basis for this correction and propose an `educated guess' of $40^{+20}_{-10}$ events / yr. With the upcoming commencement of the Legacy Survey of Space and Time (LSST) at the Vera C. Rubin Observatory, it is to be hoped that the speed class issue shall soon be resolved and improved constraints placed on nova rates.

Figure \ref{fig:m31ratesCEplots} presents the average current nova rate, derived according to the \cite{williams2017} SFRs, for different values of $\alpha_{\rm CE}$ and $\lambda_{\rm CE}$ \citep[see][for a review of the $\alpha - \lambda$ formalism for the CE phase]{demarco2011alpha}. We vary $\alpha_{\rm CE}$ while keeping $\lambda_{\rm CE}$ set to the model used in this work \citep{wang2016binding}, and likewise vary $\lambda_{\rm CE}$ while keeping $\alpha_{\rm CE}=1$.

As $\alpha_{\rm CE}$ increases, orbital energy is transferred more efficiently to the envelope during CE events. This has the effect of reducing the post-CE hardness of a given binary as it becomes easier to eject the envelope. This can affect the mass transfer rate during subsequent novae, but crucially also alters the populations of systems that survive the CE event and go on to produce novae. 

It is not obvious a priori what the aggregate effect on the nova rate will be. Reducing CE effectiveness in hardening a binary can potentially result in more novae, as `new' nova systems are introduced from regions of the parameter space that previously resulted in mergers, or fewer novae as previously viable nova systems become less productive sites of novae due to increased post-CE orbital separations.

We find that there is an anticorrelation between $\alpha_{\rm CE}$ and the predicted H nova rate. This anticorrelation implies that when increasing $\alpha_{\rm CE}$, the reduction in efficacy of existing nova systems, caused by reduced mass transfer rates due to the wider post-CE binaries, outweighs the benefits of any new nova systems formed. We also find that for all values of $\alpha_{\rm CE}$ tested, the current nova rate in M31 is underpredicted.

A higher $\lambda_{\rm CE}$ corresponds to an envelope that has a lower binding energy and therefore is easier to eject. Increasing $\lambda_{\rm CE}$ therefore reduces the efficacy of CE events, producing a wider post-CE binary but increasing the probability that the binary survives a given CE interaction. An envelope that is more difficult to eject (low $\lambda_{\rm CE}$) results in increased hardening of the binary, similar to the effect of reducing $\alpha_{\rm CE}$.

The relationship between $\lambda_{\rm CE}$ and the event rate is more complex than that of $\alpha_{\rm CE}$, and is found to have a far stronger effect on the estimated H nova rate. We find that all constant values of $\lambda_{\rm CE}$ greater than 0.1 are inconsistent with current observations. It is evident that when increasing $\lambda_{\rm CE}$ beyond 0.3, the effect of introducing new nova systems which previously did not survive a CE interaction and go on to produce novae is outweighed, at least when $\alpha_{\rm CE}=1$, by the effect of increased post-CE orbital separations. Below 0.3 the converse is true, with extremely low values of $\lambda_{\rm CE}$ reducing the rate significantly.

In addition to a set of constant values of $\lambda_{\rm CE}$, we also run cases using more sophisticated methods for calculating $\lambda_{\rm CE}$, using models from \cite{wang2016binding} (our standard physics) and \cite{dewi2000}. Both these models actually predict the same current rate, underpredicting current estimate for the M31 rate by at least 10 novae per year.

Unlike in H novae, a positive relationship exists between $\alpha_{\rm CE}$ and the He nova rate. He-nova systems require far shorter orbital separations than H-nova systems, making them more prone to merging as a result of CE events. For this reason, increasing $\alpha_{\rm CE}$ increases the event rate as systems that previously merged during CE events survive and go on to produce He novae.

The variation of the He nova rate with $\lambda_{\rm CE}$ is complex, with competition between the effects of mergers and post-CE orbital separation determining whether the rate rises or falls. From $\lambda_{\rm CE}$ = 0.2-0.6 the rate peaks, achieving a maximum of 0.12 He novae per year at $\lambda_{\rm CE}$ = 0.4, while for $\lambda_{\rm CE}$ > 0.6 the rate slowly increases with $\lambda_{\rm CE}$. Surprisingly, the rates of H novae arguably show greater variation with the choice of $\lambda_{\rm CE}$ than those of He novae, despite the existence of He novae being fundamentally dependent on CE physics. $\lambda_{\rm CE}<0.1$ produce no He novae at all, as all would-be He-nova systems merge.

As expected due to the current lull in star formation in M31, H donors are shown to be an important component of the He nova rate. The main influence of their inclusion is a shifting of current rate estimates by 0.07 events per year. The uniformity of this shift is due to the fact that most of the He novae produced by this channel are from systems that never undergo common-envelope evolution; thus their contribution is independent of common-envelope physics.

It is clear that both H and He nova rates display significant sensitivity to CE parameters. However, for nova rates to provide a truly useful test of CE physics, more precise estimates for the nova rate are required from both theoretical and observational sources.

\subsection{He novae and V445 Puppis (2000)}

\begin{figure*}
\centering
\includegraphics[width=0.85\textwidth]{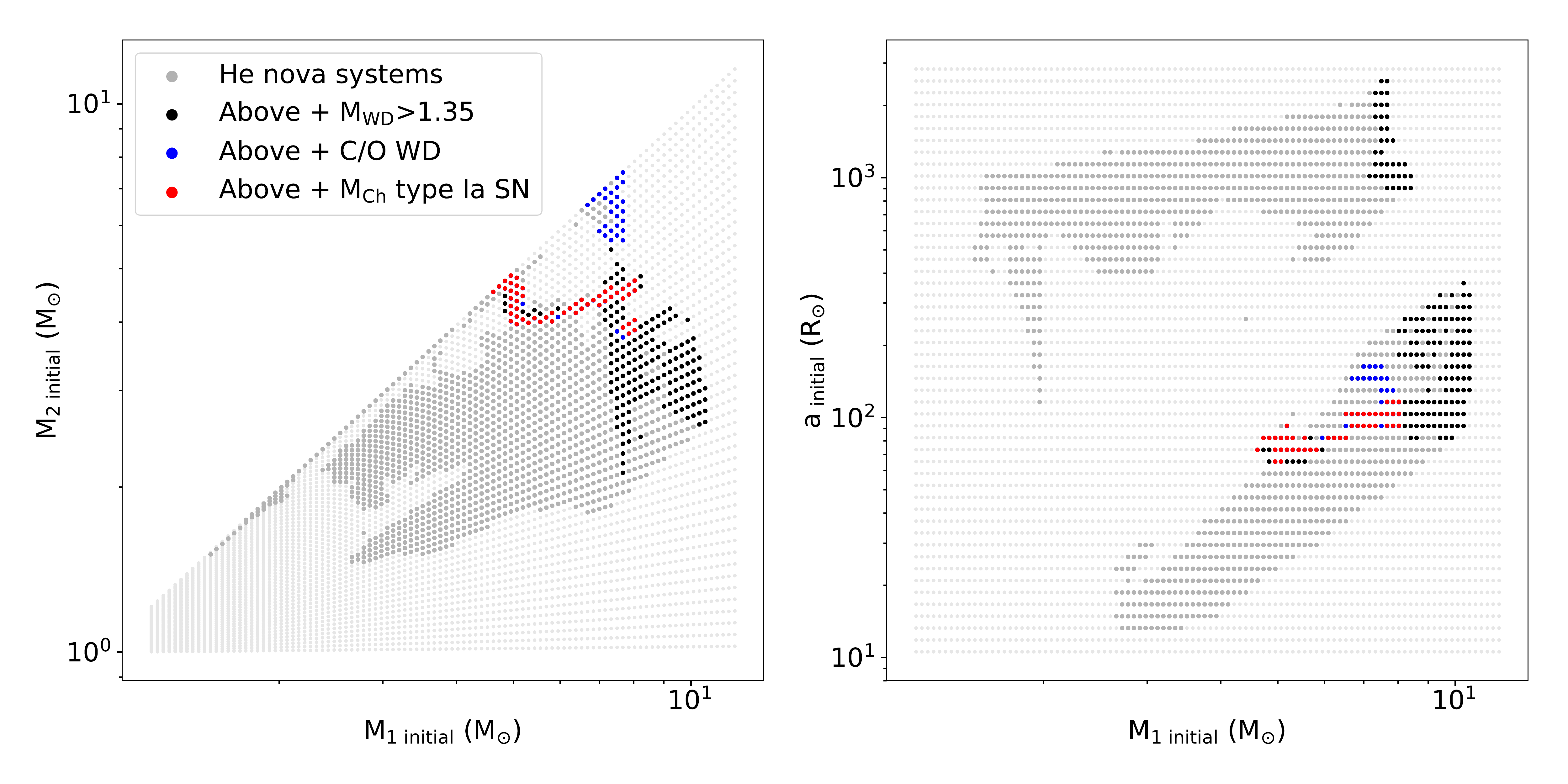}
\caption{Parameter space of initial conditions of He novae, showing different subsets of potential V445 Puppis progenitors. The black data are systems which involve at least one He nova on a white dwarf with \Mwd>1.35 M\solar, the blue are systems which involve at least one He nova on a C/O white dwarf with \Mwd>1.35 M\solar, and the red are the systems which meet both of the preceding criteria and go on to produce a single-degenerate, \Mchand\ type Ia supernova.}
\label{fig:v445pup}
\end{figure*}

There is widespread consensus that the transient V445 Pup was the result of explosive He fusion on the surface of a WD remnant: a He nova \citep{kato2000,lynch2001,ashok2003}.

Critical to this designation was the ability to spectroscopically confirm the absence of H in the spectra. This is an important point, as only a small fraction of existing `nova' observations are spectroscopically confirmed. Thus, from an observational perspective, He novae may be considered confirmed astrophysical events, but with an unconstrained rate. Aside from V445 Pup, four additional potential He nova candidates have been suggested on the basis of an over-abundance of He and the presence of unusual light curves \citep{rosenbush2008}.

In the following discussion, we consider only He-nova systems accreting from He-donor stars. The argument for this restriction is that in a H accreting system, there will almost certainly be some H material present at the time of the He flash which would compromise our ability to positively identify the eruption as a He nova. However, this constraint may be of limited practical importance, as we find that only a small fraction of the total He novae produced per star burst occur under conditions of H accretion, despite almost 95 per cent of all He-nova systems having at least one He nova occur under conditions of H accretion.

Finally, we should highlight that we present the following comparisons primarily for the purpose of comparing with future works. There is a huge amount of the parameter space that remains to be explored; as such a true measure of the uncertainties associated with the numeric results of this work are both large and relatively unknown. Despite these limitations, we believe it is still of interest to briefly discuss our results in the context of the V445 Pup observation.

\cite{kato2008} present a light curve analysis which makes several useful predictions about V445 Pup, chiefly that it involved: 
\begin{enumerate}
    \item \ \ a massive WD accretor ( \Mwd$\gtrsim1.35$ M\solar),
    \item \ an accretion efficiency of approximately 0.5, 
    \item a critical ignition mass $\mathcal{O}(10^{-4})$ M\solar,
    \item and a donor star of approximately 0.8 M\solar.
\end{enumerate}

The apparent absence of Ne in the ejecta implies a C/O WD accretor \citep{woudt2005}, while derived luminosity measurements imply that there remains a rapidly accreting WD within the dusty, bipolar shell of the ejecta \citep{woudt2009}. It should also be noted that in the 20 years since outburst there has been no detection of any subsequent eruptions, placing a hard floor on the recurrence time of the system.

We find that the distribution of accretor masses of He novae peaks strongly towards \Mchand, agreeing well with the \cite{kato2008} estimate for the WD mass of V445 Pup. Further, the distribution of critical ignition masses peaks slightly above $10^{-4}$ M\solar, with this combination of ignition mass and \Mwd\ being by far the most common combination on a per-event basis. However, it should be noted that our critical ignition masses are based on models from \cite{kato2018} which are computed using the same code used in \cite{kato2008} to make the V445 Pup predictions, and therefore the agreement with the critical ignition mass estimate for V445 Pup may be systematic.

A wide range of donor masses are found to be viable, and the estimated donor mass of 0.8 M\solar\ can be considered consistent with our work. However, less massive donor stars (<0.5 M\solar) are significantly more common. We find that approximately 30 per cent of He novae from the most massive ($\gtrsim 1.35$, see Figure \ref{fig:histnovam1}) WDs originate from C/O accretors, making the inferred accretor composition of V445 Pup uncommon but not rare. Neglecting quiescent C burning events from our model, this fraction increases to approximately 45 per cent.

As previously discussed in Section \ref{sec:wdphysics}, there is little agreement in nova efficiency estimates between different detailed modelling works. If we neglect the peak around perfect efficiency events (caused by the large population of He-nova systems which inspiral and merge), we find that an efficiency of $\eta\approx 0.2-0.25$ is most probable for He novae and high efficiency events of $\eta\approx0.5$ are very rare, in tension with the \cite{kato2008} estimate. The accretion efficiency distributions of our nova systems are described in more detail in Appendix \ref{sec:efficiency}.

We find that almost no He nova events occur with sufficiently short recurrence times for us to expect another eruption from V445 Pup in the near future. Although sub-decade He nova recurrence times are possible for the most massive WDs, the number of such short recurrence time events is extremely low. Under the assumption of a massive WD undergoing rapid He accretion (arbitrarily chosen here to mean $\gtrsim 10^{-6}$ M\solarperyr), a second eruption appears possible within the next century and probable within the next millennium.

It should be repeated that, in addition to modelling uncertainties inherent in this work, the recurrence time is highly sensitive to the accretion rate of the system. Although this casts huge uncertainties on any theoretical prediction of the recurrence time of any single system, this does mean that should V445 Pup be observed to erupt again in the (however distant) future, even information as basic as the time that the second eruption occurs could potentially provide a very useful constraint on the underlying physics.

Finally, we present in Figure \ref{fig:v445pup} predictions for the initial system properties of V445 Pup based on our standard model physics. It is apparent that not all C/O WDs that accrete He so close to the \Mchand\ limit go on to produce type Ia supernovae. We calculate the probability of V445 Pup undergoing a \Mchand\ type Ia supernova some time in the future to be 65 per cent if we assume that the accretor is in fact a C/O WD, and 21 per cent otherwise.

The most common alternate fate of these massive, He-accreting C/O WDs is the case where the period of mass transfer from a He-donor is brief, accreting material from a HeGB donor star. This donor star is formed by stripping an evolved star grown massive ($M>10$ M\solar) through accretion of material from the primary. The donor star in this case will undergo a stripped supernova event before the WD is able to accrete sufficient material from its companion, leaving the system in a NS-C/O WD binary. This channel is easily distinguished in Figure \ref{fig:v445pup}, identifiable as the region of blue data where $M_{\rm 2 \ init}$ is the greatest. The donor stars in this channel are quite massive (>1.5 M\solar, although most are 2.5-5 M\solar), and so this channel can potentially be ruled out in consideration of the 0.8 M\solar\ donor estimate from \cite{kato2008}. Doing so increases the fraction of possible V445 Puppis progenitors which undergo type Ia supernova to 91 per cent. Under this assumption, the only alternative fate is for the WD to merge with its donor star either through gravitational-wave driven inspiral or through a CE event upon the donor star evolving off the HeMS. As these outcomes result in a single remnant, we conclude that He nova systems with extremely massive WDs are unlikely to contribute significantly to type Ia supernova rates through double WD (DWD) merger channels. However, \cite{ruiter2013} show that He nova systems with less extreme WDs can contribute to sub-\Mchand\ DWD merger channels of type Ia supernovae.

\section{Conclusion}
\label{sec:conclusion}

In this work, we describe a method for modelling H and He novae in population synthesis codes and implement this method in \binaryc. For the case of a single, solar-metallicity physics set, we present rates and distributions of key physical properties including white dwarf masses, accretion rates, recurrence times, and system birth properties. We also discuss in detail the different evolutionary pathways and considerations governing nova populations, and briefly consider the effect of common-envelope physics on nova rates.

We find that while only a small fraction of H-nova systems undergo a common-envelope phase during their lives, these systems are responsible for the majority of H nova events. These systems arise from initially massive primaries (7--9 M\solar) with lower mass secondaries ($\lesssim 2.5$ M\solar), born at separations between $10^3$--$10^4$ R\solar. The vast majority of He novae are produced by He-accreting systems that have undergone either one or two common-envelope events prior to the first He nova, and both channels produce comparable numbers of He novae per system. We find that H accreting systems make up almost 95 per cent of all He nova systems produced per burst of star formation, but only around 10 per cent of the He novae.

Massive white dwarfs are responsible for most H and He novae, despite the initial mass function favouring lower-mass white dwarfs. This preference for massive white dwarfs is reflected in the distribution of accretor compositions, with approximately 70 per cent of H novae and 55 per cent of He novae occurring on O/Ne WDs.

Significant numbers of H novae are produced from thermally-pulsing asymptotic giant branch donor stars with short delay-times (< 2 Gyrs), while at later times contributions from low-mass main sequence, main sequence, and first giant branch donor stars dominate. The final remnant states of H-nova systems are diverse, dominated by white dwarf binaries but including black holes and neutron stars both in single and binary configurations. He novae are utterly dominated by contributions from He main sequence donor stars at early times, with the vast majority of these He-nova systems ultimately merging with their donor stars through gravitational-wave driven inspiral. For delay-times > 1 Gyr however, He novae which occur during H accretion dominate, fueling a non-negligible late-time tail in the distribution.

The spread in the recurrence times of H novae is vast, from 100 days to 400 Myr, but almost all have recurrence times between a year and 10 Myr and a well defined peak occurs between 100-1000 yr. The He nova distribution has a narrower spread in recurrence times, with most occurring between 1 yr to 1 Myr. The recurrence times of nova systems in general are more tightly correlated with the mass accretion rate than the white dwarf mass. 

We find that H novae with recurrence times shorter than 100 yr should be almost exclusively produced by O/Ne white dwarfs accreting from giant donor stars, but WDs with masses as low as 0.8 M\solar\ may be feasible, if rare, sites of recurrent novae. He novae with recurrence times shorter than 100 yr are limited to white dwarf accretors with masses greater than 1.1 M\solar\ and accretion rates greater than $10^{-7}$ M\solarperyr.

We estimate the current annual rate of H novae in M31 to be $41 \pm 4$, and $0.14\pm0.015$ for He novae, underpredicting the current observational estimate of $65^{+15}_{-16}$ \citep{darnley2004,darnley2006}. Further, we find no values of $\alpha_{\rm CE}$ between 0.5 and 2 that are consistent with the current observational estimate. Varying $\lambda_{\rm CE}$ affects the H nova rate dramatically, with predicted rates varying between 20 and 80 events / yr when $\alpha_{\rm CE}$ = 1. Constant values of $\lambda_{\rm CE}$ greater than 0.1 produce rate estimates that are consistent with current observations while our more sophisticated models, drawn from \cite{wang2016binding} and \cite{dewi2000}, are found to be inconsistent.

Finally, we find that most observation-inferred properties of the He nova V445 Puppis by \cite{kato2008} are consistent with our models, noting that components of this consistency may be systematic due to the inclusion of critical ignition masses for He novae from \cite{kato2018}. Our models suggest that the system is likely to erupt again within the next century, and almost certain to erupt within the next millennium, provided it continues to accrete at a rapid rate. We also find that, provided we accept the assertion of a C/O core \citep[evidenced by the absence of Ne in the spectrum of the ejected material,][]{woudt2005}, the probability of the V445 Puppis system undergoing a Chandrasekhar mass type Ia supernova is $\gtrsim 65$ per cent.

\section*{Acknowledgements}
The authors wish to thank M. Kato and B. Wang for the provision of data used in this work.
A.~R.~C. is supported in part by the Australian Research Council through a Discovery Early Career Researcher Award (DE190100656).
R. ~G. ~I. thanks the STFC for funding, in particular Rutherford fellowship ST/L003910/1 and consolidated grant ST/R000603/1.
A.~J.~R. acknowledges financial support from the Australian Research Council under grant FT170100243.
Parts of this research were supported by the Australian Research Council Centre of Excellence for All Sky Astrophysics in 3 Dimensions (ASTRO 3D), through project number CE170100013.
This research was supported in part by a Monash University Network of Excellence grant (NOE170024).
The authors wish to thank the referee, D. Prialnik, for their comments, which we believe assisted in elevating the scientific value and overall quality of the work.

\section*{Data Availability}

The data underlying this article will be shared on reasonable request to the corresponding author.



\bibliographystyle{mnras}
\bibliography{INbib} 




\appendix

\section{Rate calculations}
\label{sec:normalisationprocess}
As discussed in Section \ref{sec:grids}, the grids computed by \binaryc\ do not reflect physical birth distributions of the input parameters ($M_{\rm 1, init}$, $q_{\rm init}$, and $a_{\rm init}$) which form the grid, but are instead selected for numeric efficiency. Several stages of normalisation are therefore required before physically meaningful rates can be calculated, the process of which is described here. For a more detailed discussion on this subject, the reader is directed to the work of \cite{broekgaarden2019}, the process of which this work closely follows.

We define the rate $R_{\rm sys}$ to be the number of events that would occur on a \textit{per binary system} basis, and is calculated as:

\begin{equation}
    R_{\rm sys}=\frac{1}{N_{\rm systems}} \cdot \sum_{i=1}^{i=N_{\rm systems}} w_{\rm i} * N_{\rm events, i},
\end{equation}
where $ N_{\rm events, i}$ is the number of events in a given system, $N_{\rm systems}$ is the number of systems evolved in a give simulation (see Tables \ref{tab:gridH} and \ref{tab:gridHe}). $w_{\rm i}$ is a weighting factor correcting for discrepancies between the sample distribution\footnote{In \cite{broekgaarden2019}, the sample distribution was denoted $q$; here we instead use the symbol $\xi$ to avoid confusion with the mass ratio $q$.} $\xi_{\rm i}$ and an assumed physical prior distribution $\pi_{\rm i}$:
\begin{equation}
    w_{\rm i}=\frac{\pi_{\rm i}}{\xi_{\rm i}}.
\end{equation}

Under the assumption that each grid variable $M_{1,\rm init}$, $q_{\rm init}$, and $a_{\rm init}$ is independent, $\pi_{\rm i}$ can be written as the product of the probability distribution of each variable:

\begin{equation}
\pi_{\rm i}= P_1(M_{1, \rm init}) \cdot P_2(M_{\rm 1, init}, \ M_{2, \rm init}) \cdot P_{\rm a}(a_{\rm init})
\end{equation}

where $P_1$ is based on the IMF as put forth in \cite{kroupa2001}, and represents the physical distribution of primary masses, $P_2$ is derived from the condition for $\Delta q_{\rm init}=\rm constant$, and $P_{\rm a}$ is derived from the condition for $\Delta \textrm{ln}(a)=\rm constant$. 

The IMF probability distribution $P_1$ is implemented as a three-part broken power law\footnote{We used the package: \url{https://github.com/keflavich/imf} to generate the probability distribution function $P_1$.} with breaks at 0.08 and 0.5 M\solar, a minimum and maximum IMF mass of 0.01 and 150 M\solar\ respectively, and power law indices of 0.3, 1.3, and 2.3.
The other physical priors are defined below.

\begin{align}
P_{\rm a}(a_{\rm init})=
\begin{cases} 
      0 & a_{\rm init}\ \leq a_{\rm min} \\
      \cfrac{1}{\textrm{ln}(a_{\rm max})-\textrm{ln}(a_{\rm min})} & a_{\rm min} < a_{\rm init} < a_{\rm max} \\
      0 & a_{\rm init}\ \geq a_{\rm max} \\
   \end{cases},
\end{align}
where $
    a_{\rm min}= 10 \ \mathrm{R}_\odot , \ a_{\rm max}= 10^6 \ \rm R_\odot.$

\begin{align}
P_2(M_{\rm 1, init} , M_{\rm 2, init})=
\begin{cases} 
      0 & M_{\rm 2, init}\ \leq M_{\rm 2, min} \\
      \cfrac{1}{M_{\rm 1, init}-M_{2,\rm min}} & M_{\rm 2, init} > M_{2,\rm min} \\
\end{cases},
\label{eq:priorM2}
\end{align}  
where $M_{\rm 2, \min} = 0.$
\newline

Similarly, the sample distribution component $\xi_{\rm i}$ can be written as
\begin{equation}
\xi_{\rm i}= P_{1, \rm SD}(M_{1, \rm init}) \cdot P_{2, \rm SD}(M_{\rm 1, init}, \ M_{2, \rm init}) \cdot P_{a, \rm SD}(a_{\rm init}),
\end{equation}

where for our H nova grid $P_{1, \rm SD}$ is the distribution for which $\Delta(M_1) = \rm constant$, $P_{2, \rm SD}$ is the distribution for which $\Delta \textrm{ln}(q) = \rm constant$, and $P_{a, \rm SD}$ is the distribution for which $\Delta \textrm{ln}(a) = \rm constant$. These distributions are defined below for the sample distributions of our H nova grid.

\begin{align}
P_{1, \rm SD}(M_{1, \rm init}) =
\begin{cases} 
0 & M_{1, \rm init}\ \leq M_{1, \rm min\,SD} \\
\multirow{2}{*}{$\cfrac{1}{M_{1, \rm max\,SD}-M_{1, \rm min\,SD}}$} & M_{1, \rm min\,SD} < M_{1,\rm init} \\
& < M_{1,{\rm max\,SD}}\\
0 & M_{1, \rm init}\ \geq M_{1, \rm max\,SD} \\
\end{cases}
\end{align}

where $M_{1, \rm min \, SD} = 0.8 \ \rm M_\odot,\  M_{1, \rm max \, SD} = 20\ \rm M_\odot$.

\begin{align}
P_{2, \rm SD}(M_{2, \rm init}) =
\begin{cases} 
      0 & M_{2\, \rm init}\ \leq M_{2, \rm min\,SD} \\
      \textrm{ln} \Big ( \cfrac{M_{\rm 2, init}}{M_{2, \rm min\,SD}}\Big )^{-1} & M_{\rm 2, init} > M_{2, \rm min\,SD} \\
   \end{cases},
\end{align}

where $M_{2, \rm max \, SD} = 0.1 \ \rm M_\odot.$

\begin{align}
P_{a, \rm SD}(a_{\rm init}) =
\begin{cases} 
      0 & a_{\rm init}\ \leq a_{ \rm min\,SD} \\
      \textrm{ln} \Big ( \cfrac{a_{ \rm max\,SD}}{a_{ \rm min\,SD}}\Big )^{-1} & a_{ \rm min\,SD} < a_{\rm init} < a_{ \rm max\,SD} \\
      0 & a_{\rm init}\ \geq a_{\rm max\,SD} \\
   \end{cases},
\end{align}

where $a_{1, \rm min \, SD} = 3 \ \mathrm{R}_\odot,\  a_{1, \rm max \, SD} = 10^6\ \rm R_\odot.$
\newline
To obtain the number of events per unit star forming mass, we multiply the number of events that occur on average per system ($R_{\rm sys}$) by the number of systems formed per unit star forming mass ($F_\gamma$) when drawn from a distribution according to our assumed physical priors. Then we can write the rate of a given event per unit star forming mass ($R_{\rm sfm}$) as:
\begin{equation}
    R_{\rm sfm}=R_{\rm sys} \cdot F_\gamma.
\end{equation}

As the sample distribution of the grid of initial separations is identical to that of the physical prior, we neglect the initial separation in the calculation. Then we need only  consider the primary and secondary masses, and draw them from the previously defined physical priors to compute $F_\gamma$.

Finally, with the weightings $w_{\rm i}$ accounting for the physical prior distributions and $F_\gamma$ normalising each event per unit star forming mass, we have a weighting which we can apply to each simulated event. This weighting simultaneously accounts for deviations in our sample distribution from an assumed physical prior for the birth distribution of systems and normalising each event per unit star forming mass:
\begin{equation}
    w_{\rm sfm}=w_{\rm i} \cdot F_\gamma
\end{equation}

With this new weighting, we can construct delay-time distributions such as those shown in Figure \ref{fig:delaytimedist}, where the height of each bin represents the number of events per unit star forming mass at time 0 for that time bin. These normalised delay-time distributions can then be used to approximate the event rates in astrophysical environments such as galaxies or star clusters, provided an appropriate star formation history is known. In this work we present galactic rates using delay-time distributions for solar metallicity models only, thus galactic chemical evolution is not accounted for.

To account for the evolution of metallicity over cosmic time, multiple delay-time distributions for different metallicities should be computed in the manner previously described, and then combined with an age-metallicity relation for the desired galaxy or cluster so that the correct metallicity delay-time distribution is employed for the appropriate period of star formation. The computation of such a set of delay-time distributions is beyond the scope of the current work, and it should be noted that observed age-metallicity relations have enormous scatter \citep{casagrande2011,hayden2015}.

\section{Critical ignition mass distributions}
\begin{figure*}
\centering
\begin{subfigure}{0.95\columnwidth}
    \centering
    \includegraphics[width=\textwidth]{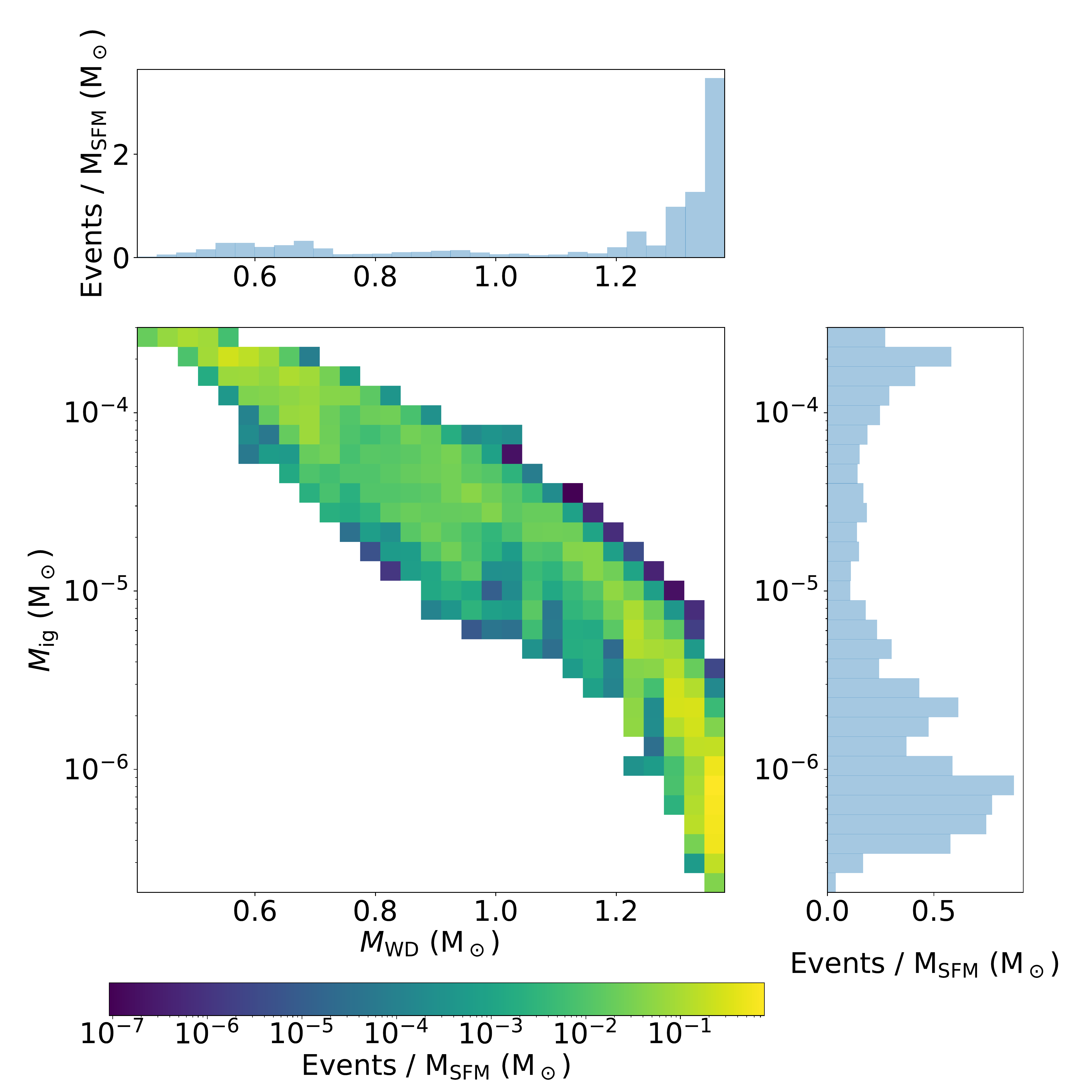}
     \caption{H novae: white dwarf mass vs critical ignition mass}
     \label{fig:hist2DmwdvsdmcritH}
\end{subfigure}%
\begin{subfigure}{0.95\columnwidth}
    \centering
    \includegraphics[width=\textwidth]{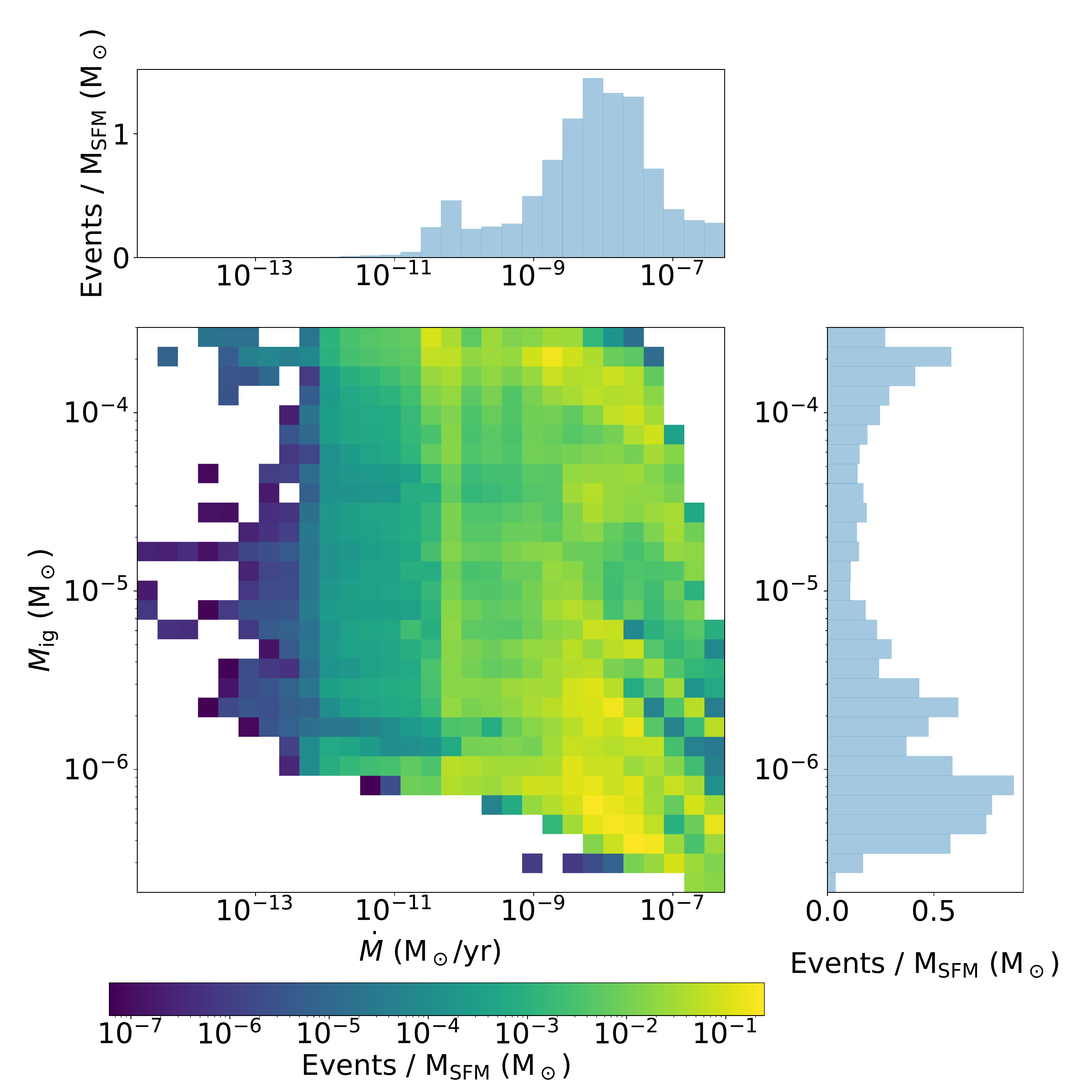}
     \caption{H novae: accretion rate vs critical ignition mass}
     \label{fig:hist2DmdotvsdmcritH}
\end{subfigure}
 
\begin{subfigure}{0.95\columnwidth}
    \centering
    \includegraphics[width=\textwidth]{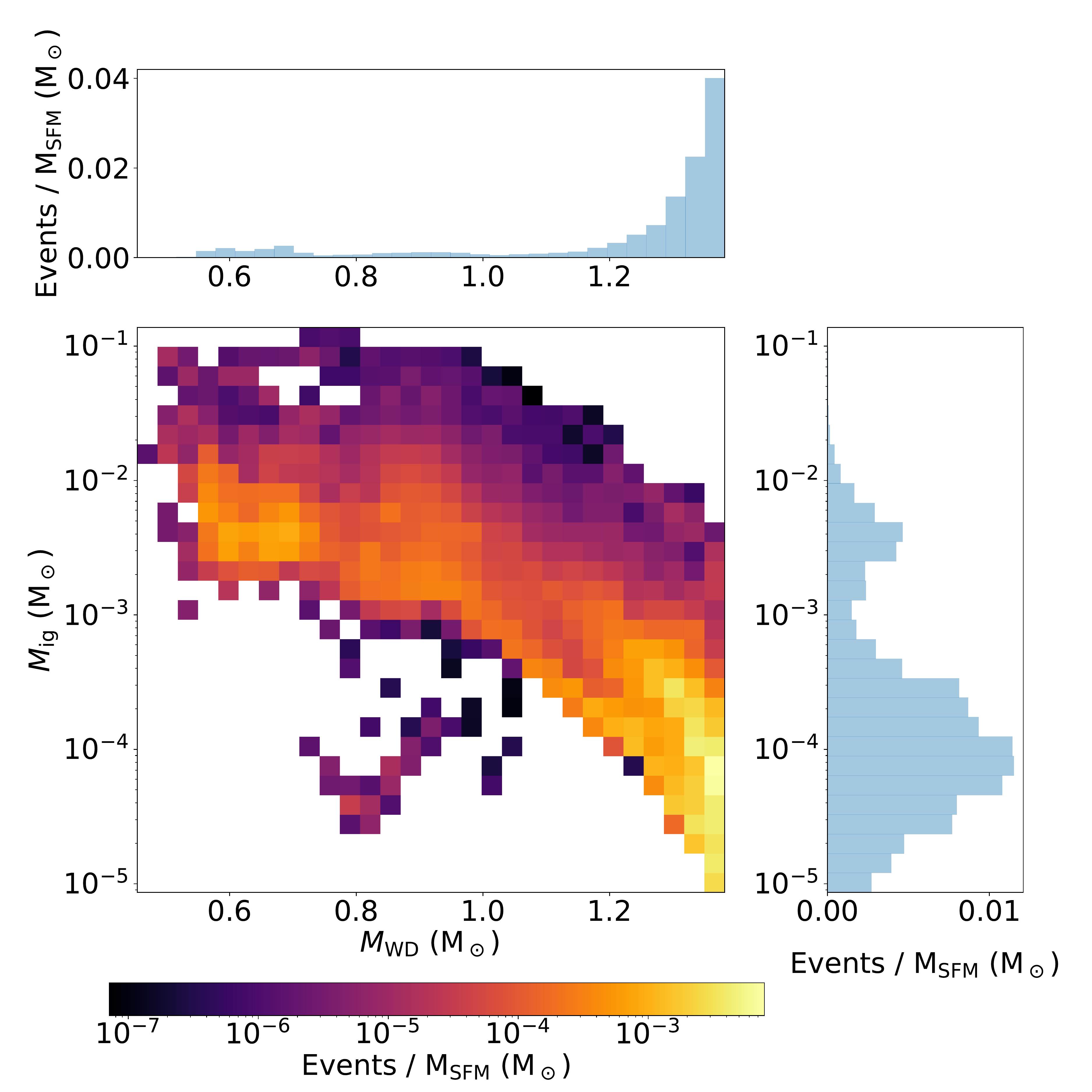}
     \caption{He novae$^{\rm *}$: white dwarf mass vs critical ignition mass}
     \label{fig:hist2DmwdvsdmcritHe}
\end{subfigure}%
\begin{subfigure}{0.95\columnwidth}
    \centering
    \includegraphics[width=\textwidth]{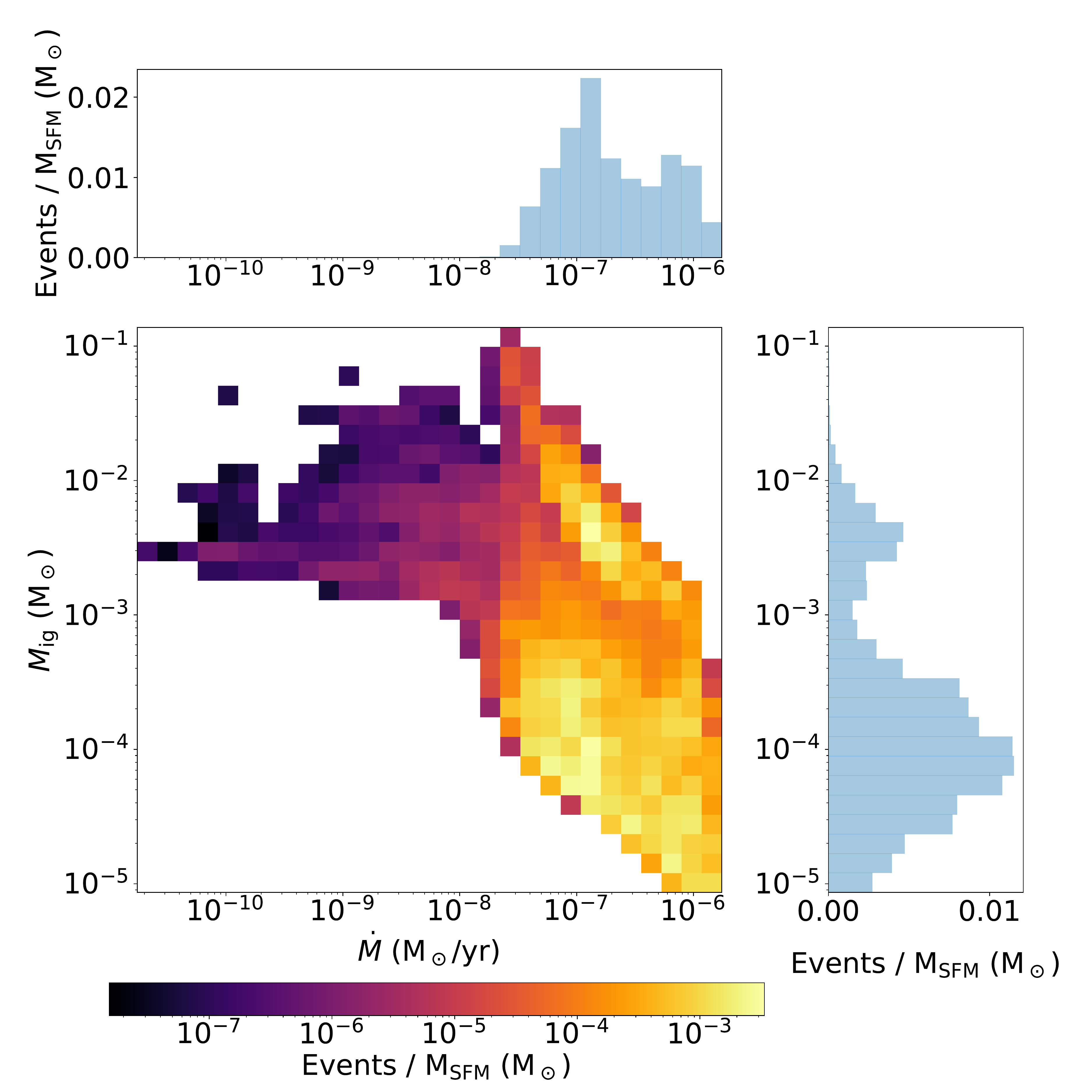}
    \caption{He novae$^{\rm *}$: accretion rate vs critical ignition mass}
    \label{fig:hist2DmdotvsdmcritHe}
\end{subfigure}
\begin{threeparttable}
\begin{tablenotes}
\item[a] Including H donors
\end{tablenotes}
\end{threeparttable}
\caption{Distributions of the critical ignition mass against the white dwarf mass and the accretion rate, all sampled at the time of each nova eruption. The ignition mass is found to depend more strongly on the white dwarf mass than the accretion rate, particularly as the white dwarf mass approaches \Mchand. "$^{\rm *}$": Including H donors.}
\label{fig:hist2Ddmcritplots}
\end{figure*}

The critical ignition mass (\Mig) determines how much material must be accreted prior to nova eruption, directly affecting the rate of novae. In our model we compute \Mig\ as a function of the WD mass and mass accretion rate (see Section \ref{sec:wdphysics}, also Figure \ref{fig:interpfitstuffmig}). The computed distributions of \Mig\ versus \Mwd\ and \Mdot\ are shown in Figure \ref{fig:hist2Ddmcritplots}.

As previously discussed, the expectation is that increasing \Mwd\ leads to lower critical ignition masses, driven by increased compressional heating of the accreted layer. Likewise, increasing \Mdot\ should also lead to lower \Mig\ due to the higher rates of gravothermal energy transfer causing increased heating at the base of the layer.

It is observed in Figures \ref{fig:hist2DmwdvsdmcritH} and \ref{fig:hist2DmdotvsdmcritH} that, for H novae, \Mig\ is extremely sensitive to \Mwd, while being only weakly sensitive to \Mdot. This is quite different to the distribution of \Mig\ for He novae. The He novae distribution is more strongly correlated with \Mwd\ than with \Mdot, but there remains a significant correlation between \Mdot\ and \Mig\ that is only barely discernible for H novae.

The reason for this discrepancy is the more extreme conditions required for the nuclear burning of He compared to H. This results in He novae being more reliant on the gravothermal energy released from the accreted material, the contribution of which scales directly with the accretion rate.

The distribution of critical ignition masses for H novae varies substantially from that of He novae, as expected due to the much greater difficulty of initiating He fusion. He novae are found to have \Mig\ ranging from $\approx 10^{-5} - 0.1$ M\solar \footnote{Double-detonation type Ia supernovae are neglected here. It is possible that the `He novae' with the most massive He shells may manifest as such.}, but most He novae occur with critical ignition masses between $\approx 10^{-5} - 10^{-3}$ M\solar. This is very different to the case for H novae, where the distribution is spread out more evenly throughout between $\approx 10^{-7} \ \mathrm{to} \ 3\times 10^{-4}$ M\solar. 

 \section{Accretion efficiency distributions}
\label{sec:efficiency}
\begin{figure}
     \centering
\begin{subfigure}{0.8\columnwidth}
    \centering
    \includegraphics[width=\textwidth=1]{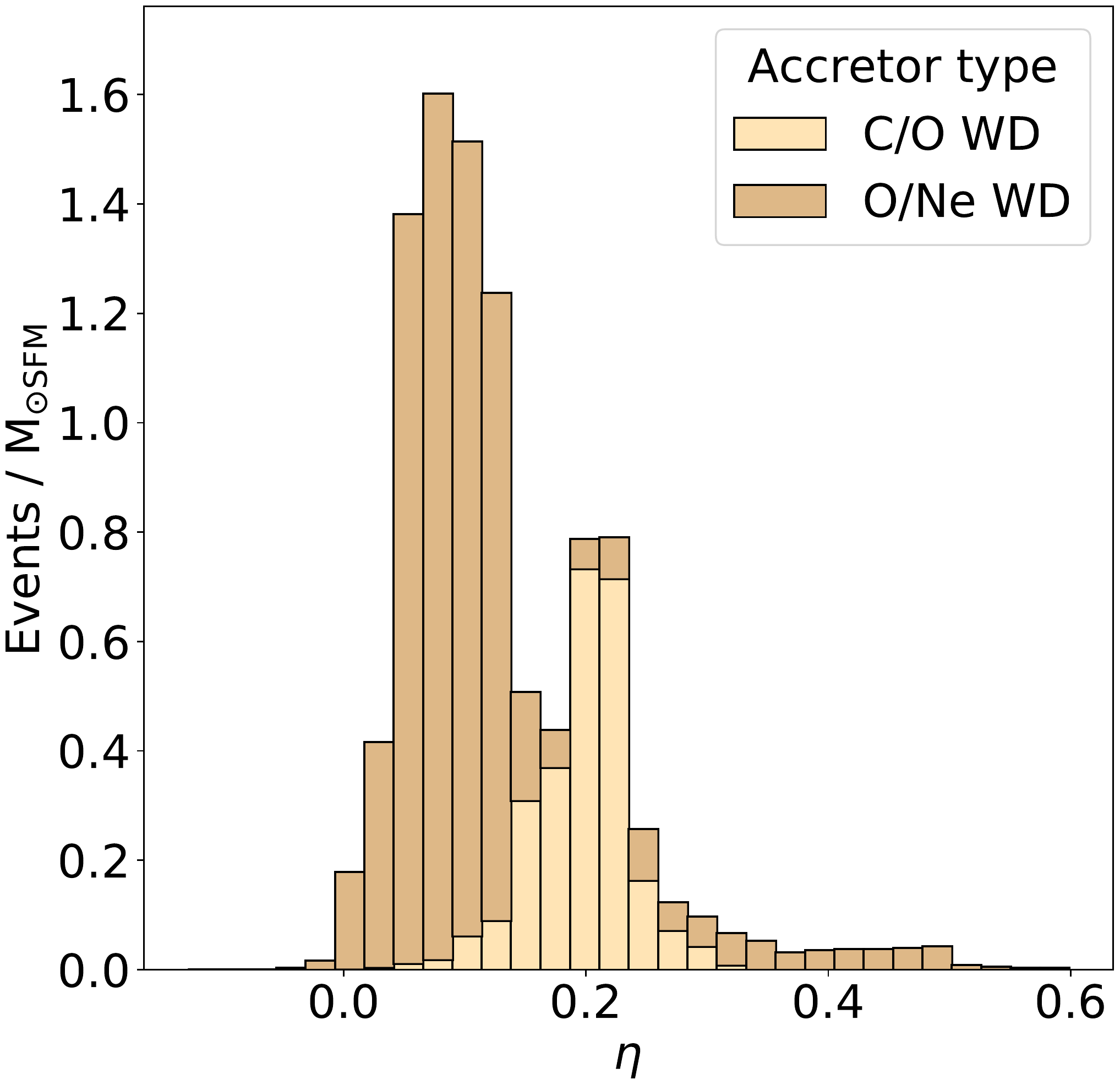}
    \caption{H Novae}
\end{subfigure}

\begin{subfigure}{0.8\columnwidth}
    \centering
    \includegraphics[width=\textwidth=1]{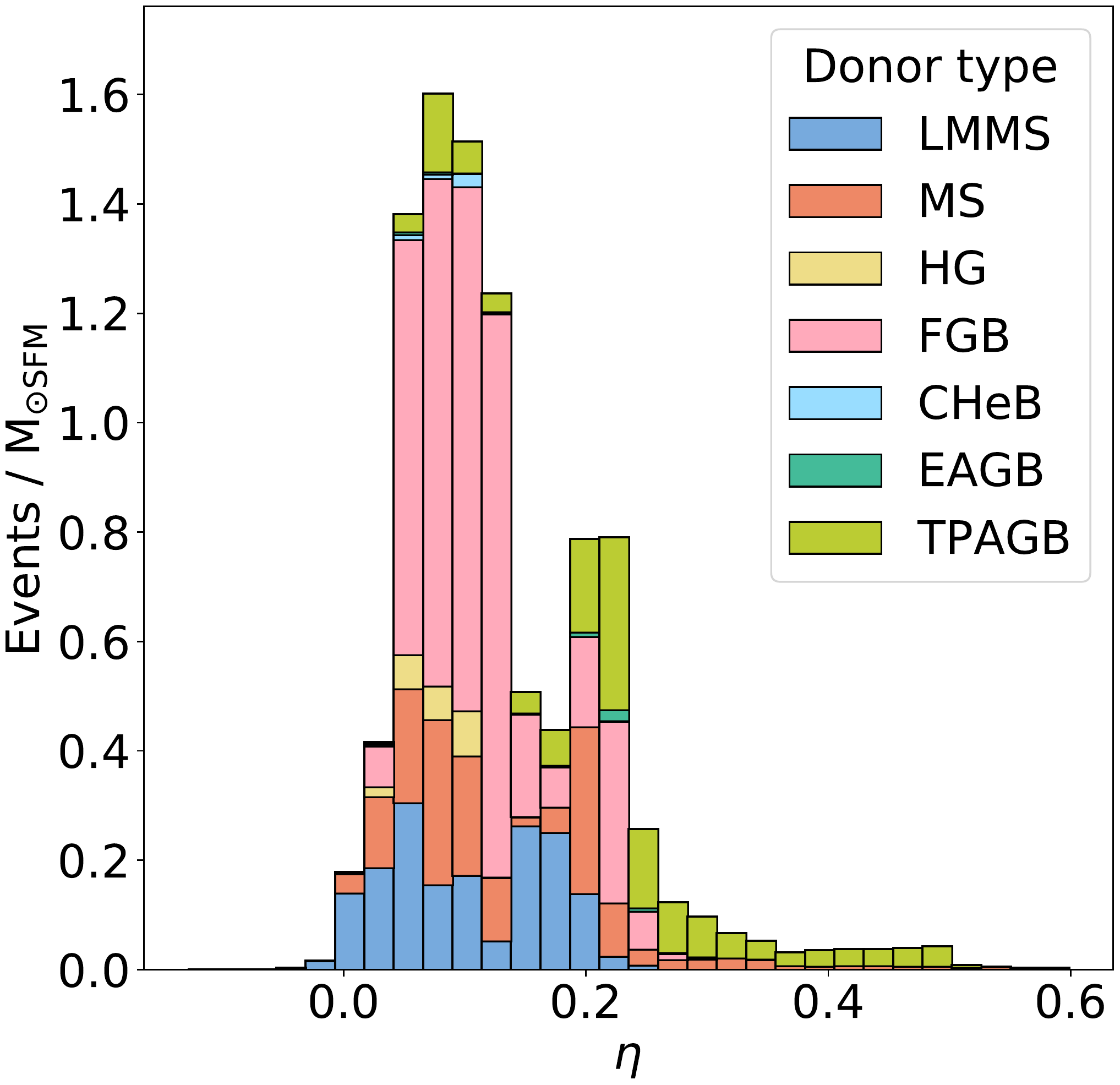}
    \caption{H Novae}
\end{subfigure}

\begin{subfigure}{0.8\columnwidth}
    \centering
    \includegraphics[width=\textwidth=1]{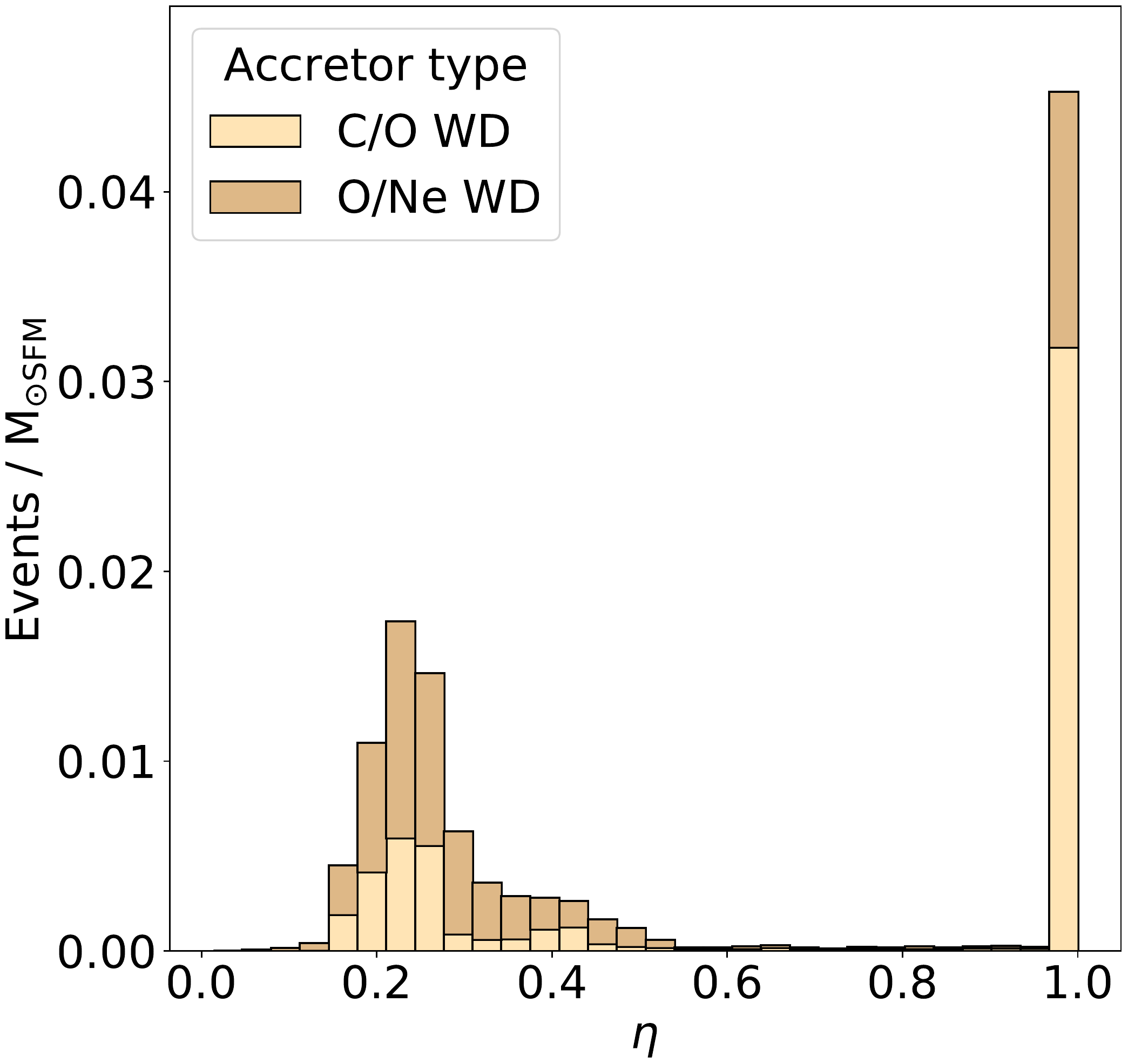}
    \caption{He Novae, including H donors}
\end{subfigure}
 \caption{Accretion efficiency distribution, coloured by the accretor and donor stellar types. The highest accretion efficiencies for H novae are obtained only at the highest accretion rates, driven by giant donor stars overflowing their Roche lobes.}
 \label{fig:histnovaeta}
\end{figure}

\begin{figure}
\centering
\begin{subfigure}{0.95\columnwidth}
    \centering
     \includegraphics[width=\textwidth]{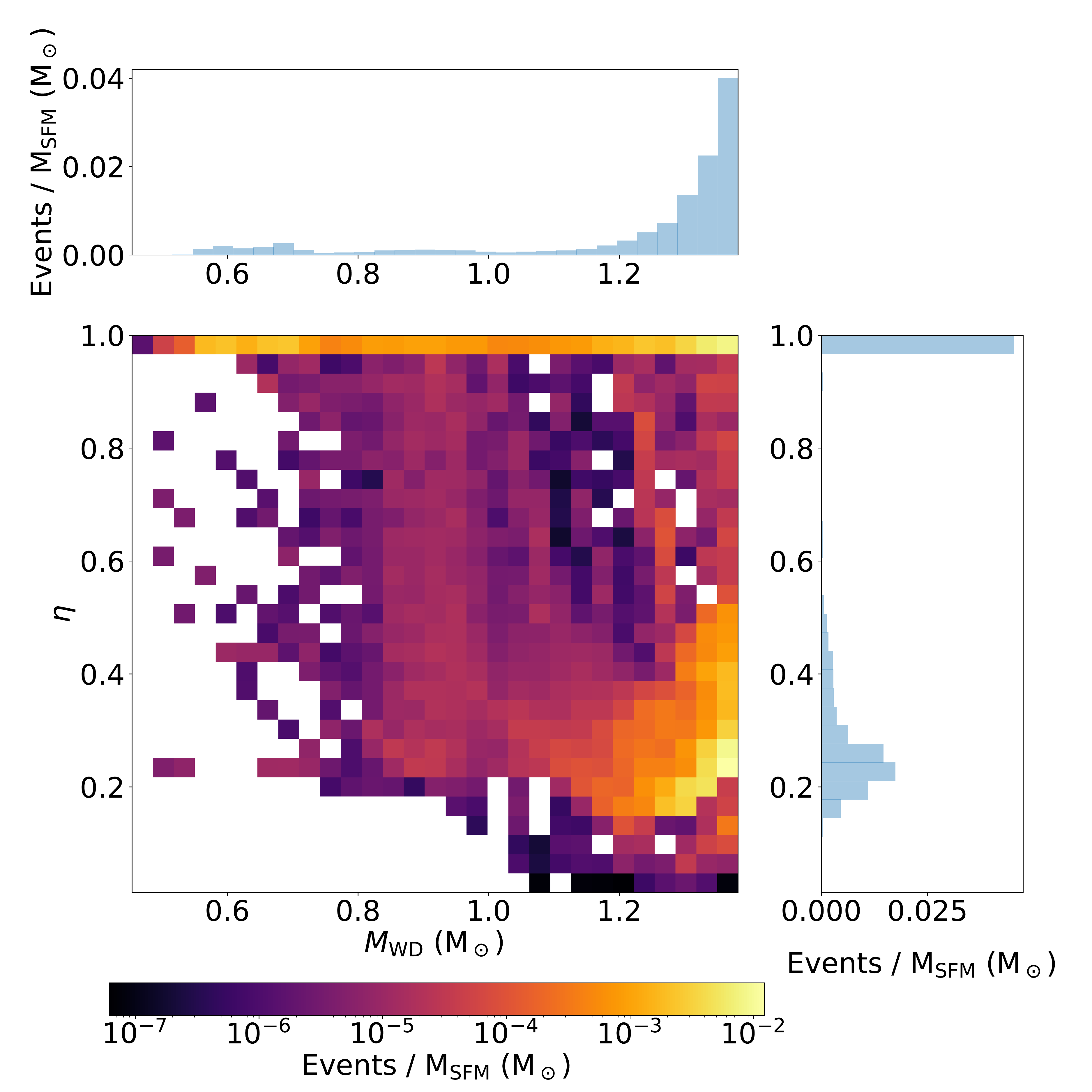}
    \caption{White dwarf mass vs accretion efficiency, including H donors}
\end{subfigure}
\begin{subfigure}{0.95\columnwidth}
    \centering
     \includegraphics[width=\textwidth]{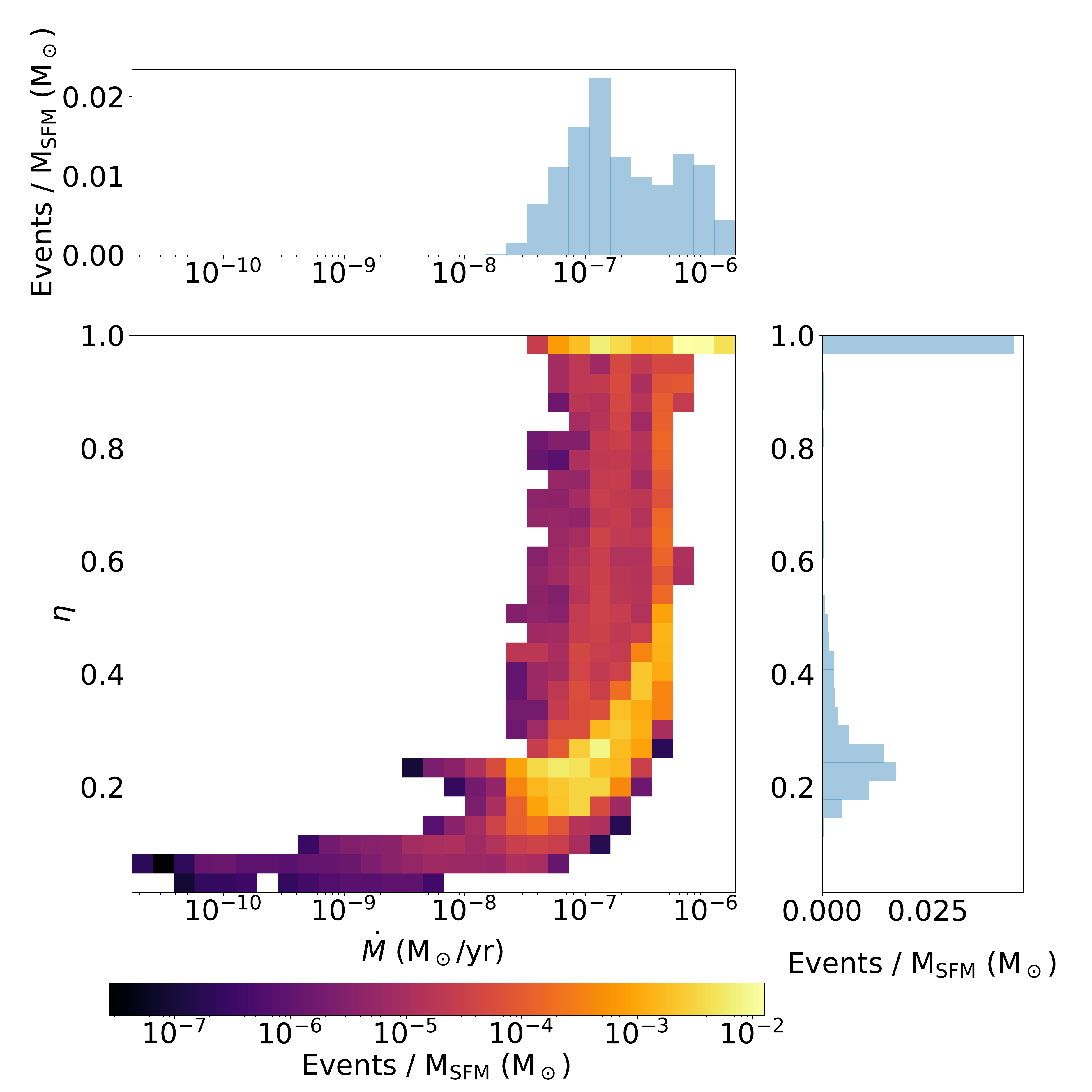}
    \caption{Mass accretion rate vs accretion efficiency, including H donors}
\end{subfigure}
\caption{Distributions of accretion efficiency for He novae against the white dwarf mass and accretion rate. The accretion efficiency is only weakly dependent on the white dwarf mass, but strongly dependent on the mass accretion rate. The large peak around perfect efficiency is driven by the huge number of systems which experience extremely high accretion rates shortly before merging with their He main sequence donor star.}
\label{fig:hist2DmdotvsetaHe}
\end{figure}

The range of permissible $\eta$ according to our fits to data from \cite{wu2017} and \cite{wang2018} is shown in Figure \ref{fig:interpfitstuffeta}. The actual distribution of $\eta$ according to our simulations is shown in Figure \ref{fig:histnovaeta}. Both H and He novae have interesting features in their distributions, the discussion of which provides useful insight into the physics of these events.

First addressing H novae, it can be seen that there are two peaks around $\eta\approx0.1$ and $\eta\approx0.2$, with a significant tail up to $\eta\approx0.6$. Interestingly, while the distribution of nova events from C/O WDs are firmly centered around the peak at $\eta\approx0.2$, O/Ne WD accretors are responsible for both the lower $\eta\approx0.1$ peak and the high $\eta$ tail. The distinction between the two populations of O/Ne WD systems lies in the donor stars. The O/Ne WD systems responsible for the high efficiency tail are those accreting rapidly (\Mdot$\gtrapprox10^{-7}$ M\solarperyr\ from TPAGB donor stars, while those that form the peak around $\eta\approx0.1$ are accreting more slowly (\Mdot$\gtrapprox10^{-8}$ M\solarperyr), primarily from FGB donor stars.

The distribution of He nova efficiencies reveals a distribution peaking around $\eta\approx0.25$, similar to H novae, which drops quickly by $\eta\approx0.15$ and decays more gradually to $\eta\approx0.55$, before a large, dominant peak at $\eta=1$. The distribution shows no significant variation between C/O and O/Ne accretors.

The most interesting feature in the He nova distribution is the dominant peak at $\eta=1$. The reason for its presence is partially revealed in Figure \ref{fig:hist2DmdotvsetaHe}, which shows that this peak is primarily driven by events at the highest mass accretion rates, but is relatively unconstrained in its distribution versus \Mwd. The question becomes: why are there so many events at these exceptionally high accretion rates in the first place? Many of these events are at the limit of what is, in our model, considered the boundary between stable and unstable surface He burning.

The answer lies in the final fates of the He-nova systems, the vast majority of which merge with their donor star. As they inspiral, the rate of mass transfer increases and these systems undergo multiple perfect efficiency eruptions, driven by the substantial reduction in ignition mass caused by the increased accretion rate (see Figure \ref{fig:hist2DmdotvsdmcritHe}).


\bsp	
\label{lastpage}
\end{document}